\documentclass[twocolumn,prl]{revtex4-2}%
\usepackage{graphics,amsmath,amsfonts,amscd,revsymb,latexsym,enumerate,multirow,epsfig}
\usepackage{amsmath}
\usepackage{amsfonts}
\usepackage{amssymb}
\usepackage{color}
\usepackage{graphicx}
\usepackage{mathtools}

\providecommand{\U}[1]{\protect\rule{.1in}{.1in}}
\providecommand{\U}[1]{\protect\rule{.1in}{.1in}}

\newcommand{\ket}[1]{\left|#1\right\rangle}

\graphicspath{{./}}

\begin{document}
\title[]{Theory of superconducting qubits beyond the lumped element approximation}
\author{Ari Mizel}
\email{ari@arimizel.com}
\affiliation{Laboratory for Physical Sciences, 8050 Greenmead Drive, College Park,
Maryland, USA\ 20740}

\begin{abstract}
In the design and investigation of superconducting qubits and related devices, a lumped element circuit model is the standard theoretical approach.  However, many important physical questions lie beyond its scope, e.g. the behavior of circuits with strong Josephson junctions carrying substantial currents and the properties of small superconducting devices.  By performing gauge transformations on self-consistent solutions of the Bogoliubov-de Gennes equations, we develop here a formalism that treats Josephson couplings non-perturbatively.  We apply the formalism to (a) show that Fermi sea effects can contribute to the effective capacitance of small charge qubits; (b) demonstrate an asymmetry in clockwise and counterclockwise current states in small RF squid qubits; and (c) provide a microscopic wavefunction of superconducting Schrodinger cats suitable for computing the number of entangled electrons.  
\end{abstract}
\volumeyear{2021}
\volumenumber{ }
\issuenumber{ }
\eid{ }
\date{\today}




\maketitle

Over the past quarter century, superconducting qubits \cite{Krantz2019} have enjoyed dramatic performance improvements and have attracted growing interest and excitement.  Several remarkable families of superconducting qubits have been invented and investigated \cite{Nakamura1999,*Mooij1999,*Martinis2002,*Vion2002,*Koch2007,*Manucharyan2009}.  To describe these designer quantum systems, a quantum mechanical lumped element (LE) circuit theory is employed very broadly and successfully by researchers \cite{Devoret1997,Girvin2014,Glazman2021}.  However, as the degree of control achieved over superconducting qubits becomes ever more exquisite \cite{Ofek2016,Zhang2021,Somoroff2021,Acharya2022}, a role is emerging for a theoretical description of increased detail and precision.  In this paper, we frame such a description which, among its important features, treats the Josephson coupling between islands non-perturbatively. 

As shown below, this theory can address physical questions that are beyond the scope of LE theory, such as the behavior of strong Josephson couplings supporting substantial currents or the properties of much smaller superconducting qubits.   This description also has the potential to supply answers when ambiguities arise in the application of LE theory (for an example in the case of junctions with Andreev bound states see \cite{Ivanov1999}) and to permit refined calculations of qubit properties such as perturbations of qubit spectroscopy.  This paper presents sample calculations on a charge qubit and on an RF squid qubit, showing consistency with LE theory and also revealing uncharted effects.  We then leverage the microscopic character of our analysis to study the number of electrons in a superconducting ``Schrodinger cat'' \cite{Friedman2000,VanderWal2000,Korsbakken2009,Korsbakken2010}.

LE theory describes a superconducting circuit as a system of distinct islands that are coupled weakly by Josephson junctions.  The justification for this description typically involves treating a tunneling Hamiltonian 
at second order in perturbation theory \cite{Glazman2021,deGennes1966,Ambegaokar1982}.  To go beyond LE theory, and gain new insights, we holistically treat all of the coupled islands of the system as a single superconducting entity.

\begin{figure}[h]
\vspace{-0.5in}
\begin{tabular}{c}
\begin{tabular}{ll} 
(a)\hspace{-0.075in} \begin{tabular}{c}
\includegraphics[width=1.5in]{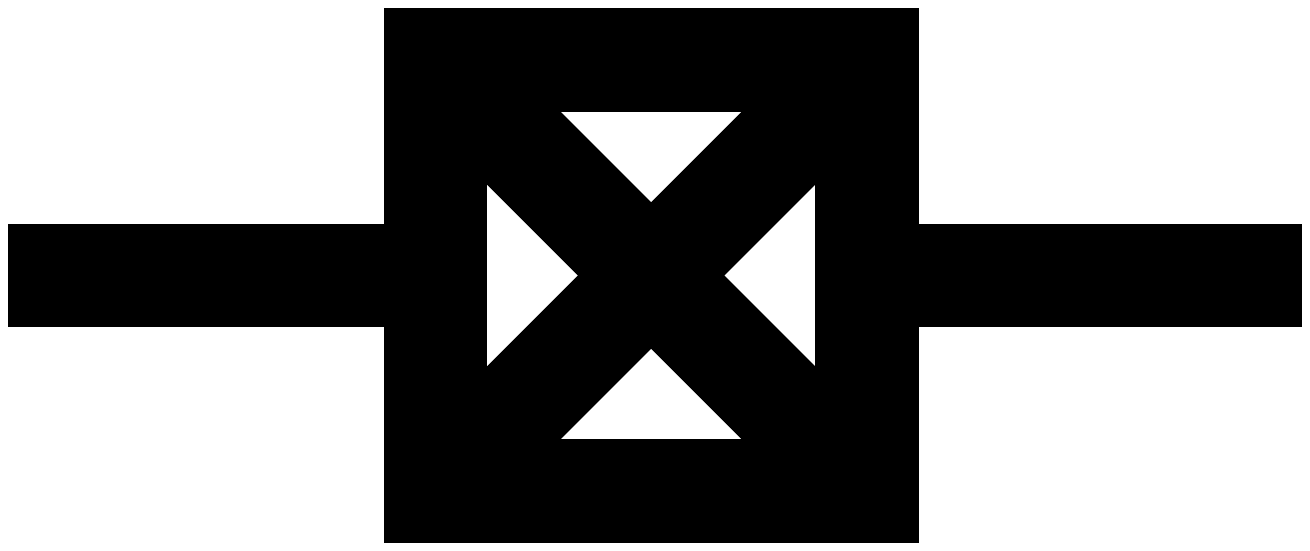}
\end{tabular} &
\hspace{-0.175in} (b) \begin{tabular}{c}
\includegraphics[width=1.4in]{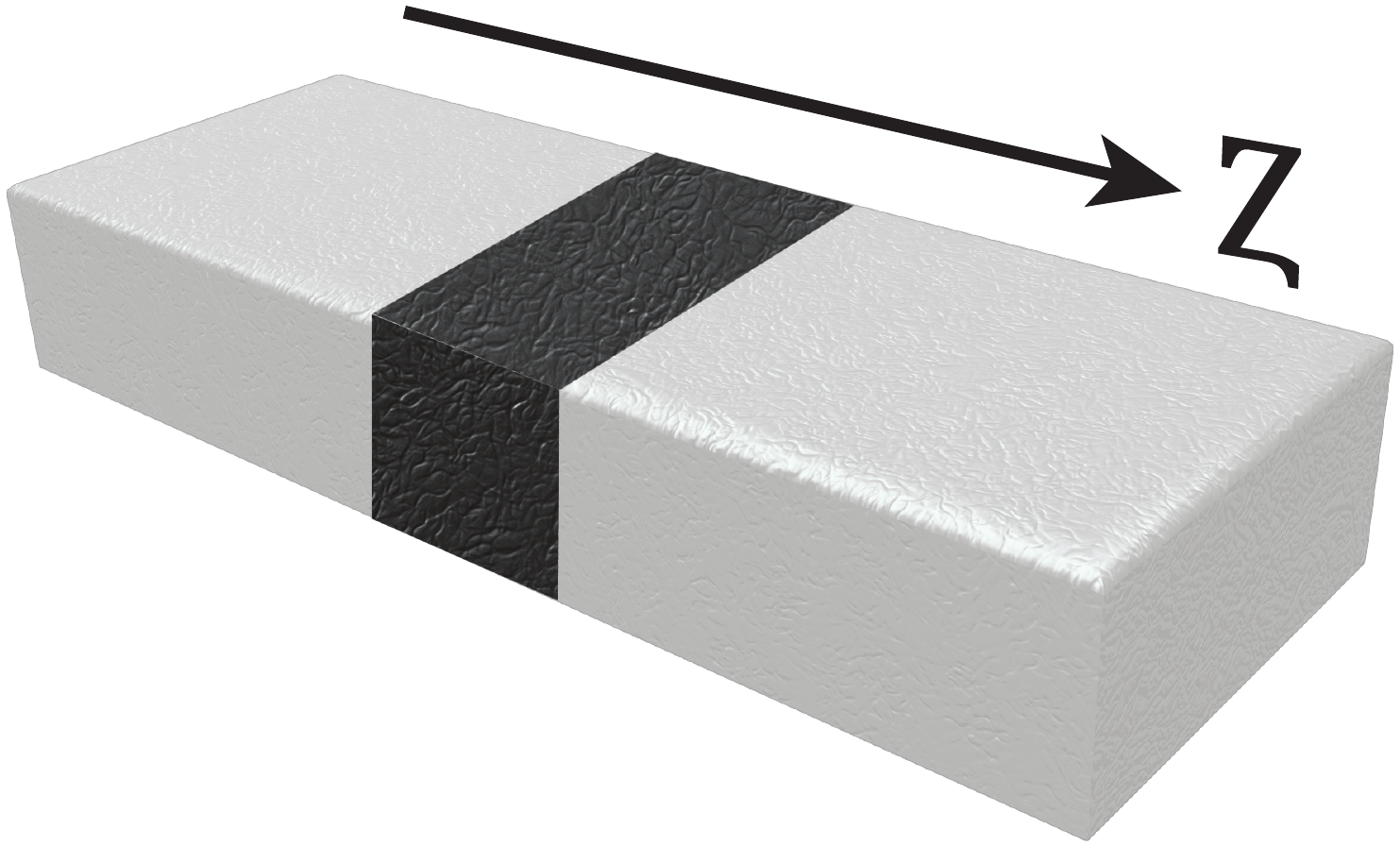}
\end{tabular} 
\end{tabular} \vspace{-0.5in}\\
\end{tabular} 
\begin{tabular}{ll}
(c) \hspace{-0.125in}
\begin{tabular}{c}
\includegraphics[width=1.5in]{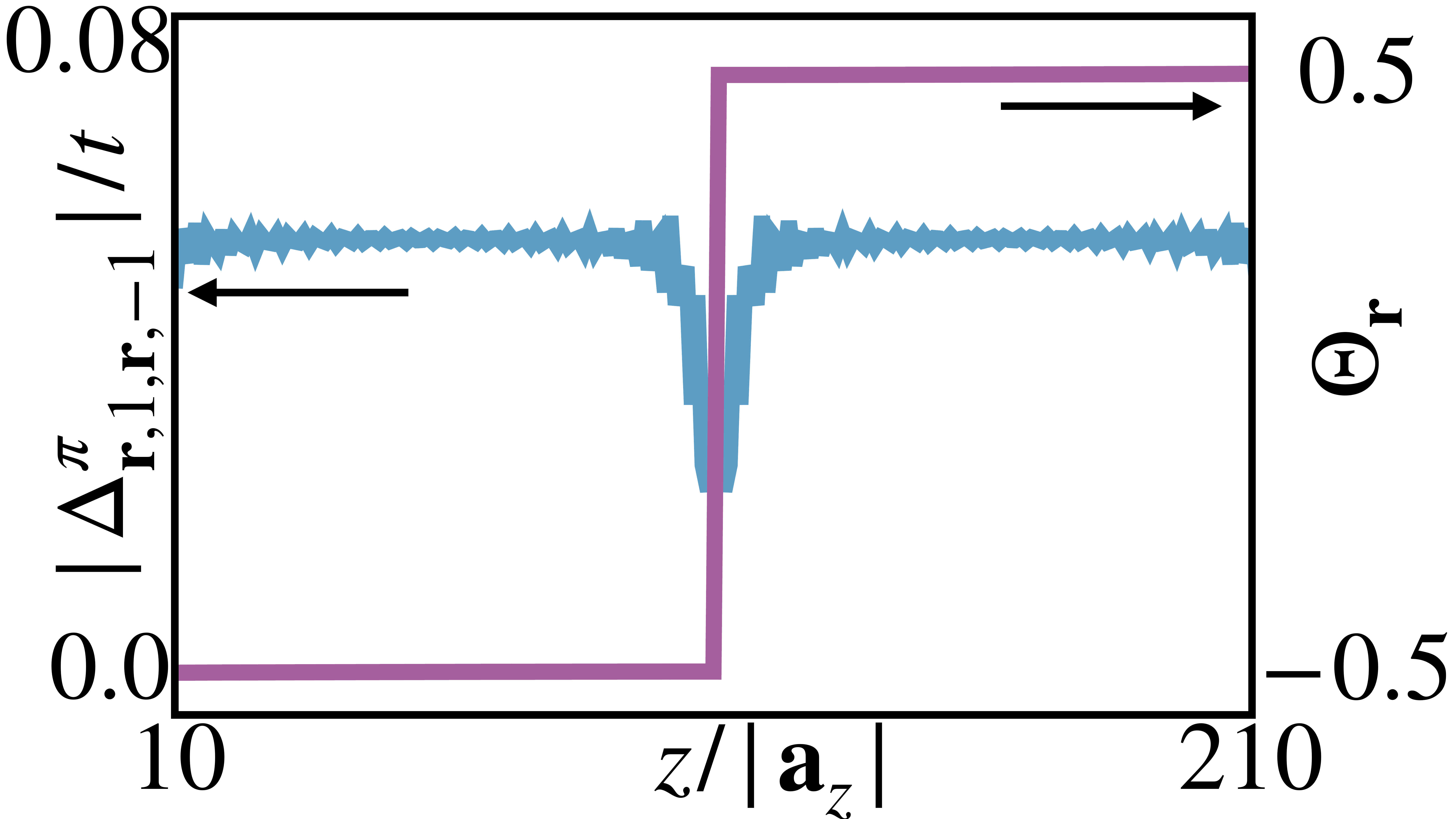}
\end{tabular} & \hspace{-0.125in}
(d)
\begin{tabular}{c}
\includegraphics[width=1.5in]{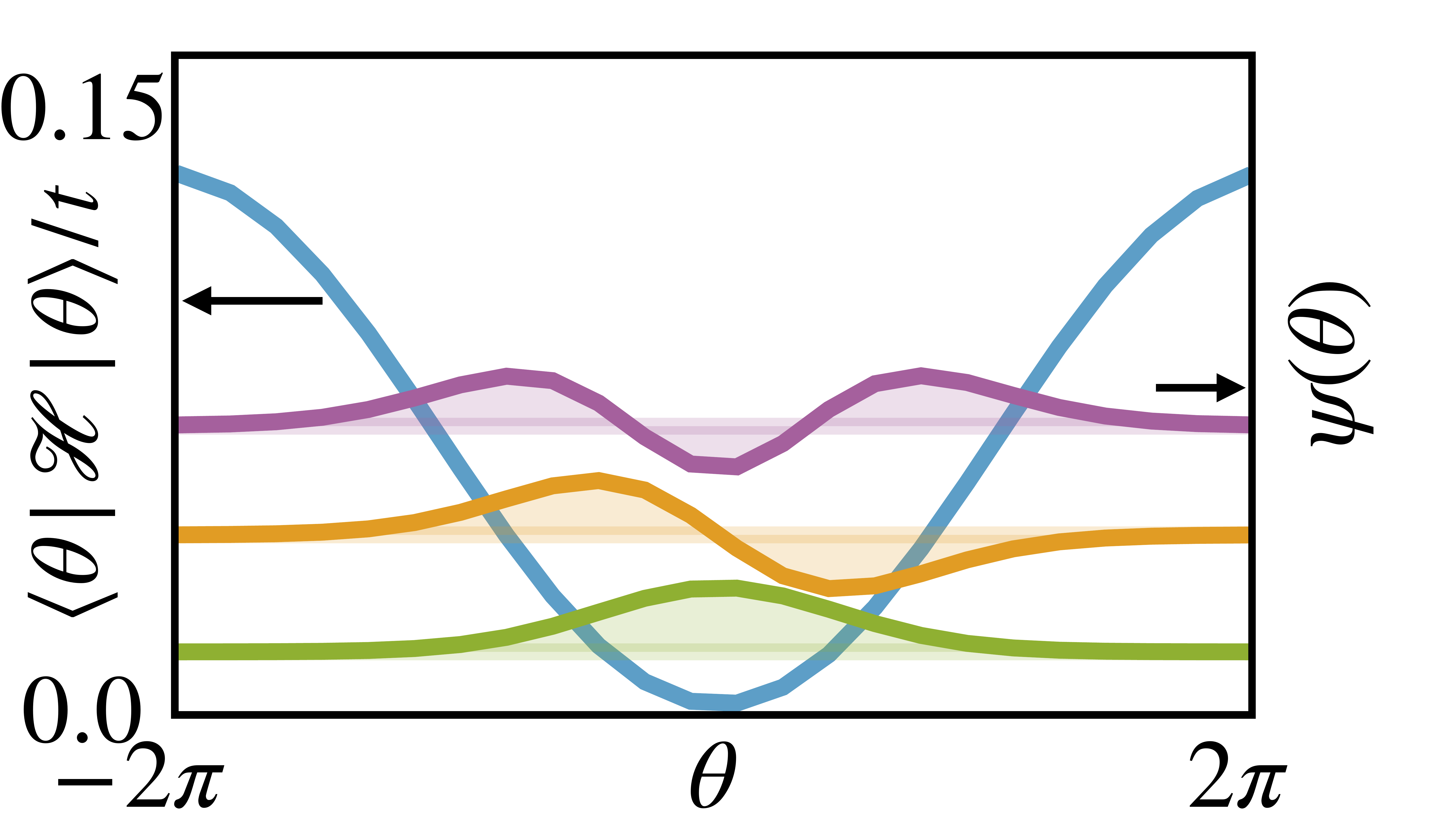}
\end{tabular}
\end{tabular} 
\caption{(a) Circuit diagram of charge qubit.  (b)  Schematic of qubit with Josephson junction in dark region.  (c) Magnitude (blue) of superconducting gap $|\Delta^\pi_{{\mathbf r},\sigma=1,{\mathbf r},\sigma^\prime =-1}|$ depends on $z$ component of ${\mathbf r}$.  Minimum at Josephson junction is evident. Rescaled phase function $\Theta_{\mathbf r}$ (purple). (d) Diagonal elements $\langle \theta|{\mathcal  H}|\theta\rangle$, corresponding to Josephson energy (blue).  Energy eigenstates of (\ref{eq:Hthetathetaprime}) shown in green, orange, and purple.}
\label{fig:chargequbit}
\end{figure}

Assume a microscopic electronic Hamiltonian
\begin{equation}
H = T + P + W
\label{eq:H}
\end{equation}
where the kinetic, potential, and interaction energies are
\begin{align}
& T = - t \sum_{{\mathbf R}}  \sum_{\smash{{\mathbf a}=\pm{\mathbf a}_x,\pm{\mathbf a}_y,\pm{\mathbf a}_z}} \!\!\!\!\!\!\!\!\!\! (t_{{\mathbf r}+{\mathbf a},{\mathbf r}}/t) c^\dagger_{{\mathbf r}+{\mathbf a},\sigma}  c^{ }_{{\mathbf r},\sigma} 
\label{eq:T} \\
&P = \sum_{{\mathbf R}} (v({\mathbf r})-\mu) c^\dagger_{{\mathbf R}}  c^{ }_{{\mathbf R}} \label{eq:P} \\
&W =  \frac{1}{2} \sum_{{\mathbf R},{\mathbf R}^\prime}   W({\mathbf R},{\mathbf R}^\prime) c^\dagger_{{\mathbf R}}c^\dagger_{{\mathbf R}^\prime}c^{ }_{{\mathbf R}^\prime}c^{ }_{{\mathbf R}}.
\label{eq:W}
\end{align}
Here, ${\mathbf R} \equiv ( {\mathbf r},\sigma)$ is a combined position and spin coordinate introduced for notational brevity.  
 The notation is not meant to suggest a tight-binding approximation.  Good descriptions of the low-energy eigenstates of $H$ are obtained by following steps (i)--(iii) below.  For concreteness, we focus on a charge qubit comprising a superconducting grain with a junction region (Fig. \ref{fig:chargequbit}a-b). 

{\bf Step (i)} of our approach defines a set of states that will be superposed to describe the low-energy eigenstates of (\ref{eq:H}).   The choice of states is motivated by the following conventional rationale.  Superconductors have low-energy excited states that allow them to conduct current.  To develop intuition about the form of such current-carrying states, imagine boosting each electron in the ground state by a momentum $\hbar \delta {\mathbf q}$ (henceforth, we set $\hbar = 1$).  Such a boost multiplies the superconducting order parameter
\begin{equation}
\Delta_{{\mathbf R}^\prime,{\mathbf R}} =W({\mathbf R},{\mathbf R}^\prime) \langle c_{\mathbf R} c_{{\mathbf R}^\prime} \rangle 
\end{equation}
by a position-dependent phase $e^{i \delta {\mathbf q} \cdot ({\mathbf r}+ {\mathbf r}^\prime)}$.  This suggests that low-energy excitations can be expressed in terms of phase changes of the superconducting order parameter.

Thus, to describe a charge qubit, we define a set of states $|\theta\rangle$, where $\theta$ denotes the phase drop of the superconducting order parameter.  The LE approximation also works with a set of states $|\theta\rangle$ with dynamics dictated by
\begin{equation}
H_{\text{LE}} = 4 E_C n^2 + E_J (1-\cos\theta)
\label{eq:lumped}
\end{equation}
where $[\theta,n] = i$ and $\theta \in (-\pi,\pi]$.
In our approach, however, we eschew the assumption of an abrupt phase drop and model the junction microscopically.  One consequence is that our $\theta$ is not restricted to $(-\pi,\pi]$. 

To obtain the state $|{\theta=0}\rangle$, we self-consistently solve the BdG equations \cite{deGennes1966,Zhu2016}, collecting the positive-energy eigenstates into matrices $U_{{\mathbf R},{\mathbf K}}$ and $V_{{\mathbf R},{\mathbf K}}$ \cite{SupplementalInformation}.  Here,  ${\mathbf K} = ({\mathbf k},\sigma)$ labels the solutions, with ${\mathbf k}$ denoting some set of quantum numbers but not necessarily the momentum.  

To define $|\theta\rangle$ for other values of $\theta$, one option is to repeat the BdG calculation, fixing the total phase change of the order parameter at $\theta$.  To carry this out, one can return the order parameter phase, after each iteration toward self-consistency, to $-\theta/2$ at one end of the superconducting grain and $\theta/2$ at the other end.  Such a calculation produces an order parameter with a fixed total phase change, but whose local phase change depends upon ${\mathbf r}$ according to the microscopic physics of (\ref{eq:H}): the phase changes rapidly where it is energetically advantageous to do so.  In our case of a charge qubit, the phase will drop rapidly across the Josephson junction.  

Instead of performing this calculation for all $\theta$, here we do so only for $\theta=\pi$.  This both reduces computational effort and leads to results that are more readily compared to the LE approximation.  The computation yields an order parameter $\Delta^\pi_{{\mathbf R},{\mathbf R}^\prime}$ with phase $-\pi/2$ on one side of the system and $\pi/2$ on the other.  We then rescale the phase of the order parameter, defining 
\begin{equation}
\Theta_{\mathbf r} = \text{Arg } \Delta^\pi_{({\mathbf r},1),({\mathbf r},-1)} /\pi
\label{eq:Thetachargequbit}
\end{equation}
which takes the value $-1/2$ on one side of the system and $1/2$ on the other.
 To complete step (i), we define $|\theta\rangle$ as the state obtained by applying a gauge transformation to the self-consistent BdG solution: $U_{{\mathbf R},{\mathbf K}} \rightarrow e^{i \theta \Theta_{\mathbf r}/2} U_{{\mathbf R},{\mathbf K}} $ and $V_{{\mathbf R},{\mathbf K}} \rightarrow e^{-i \theta \Theta_{\mathbf r}/2} V_{{\mathbf R},{\mathbf K}}$.  This definition ensures that the phase of the order parameter of $|\theta\rangle$ indeed changes by $\theta$ across the system.

In {\bf step (ii}) of our approach, each approximate low-energy eigenstate of $H$ is written as a superposition of the states defined in step (i): $|\psi\rangle = \sum_{\theta} \psi(\theta) |\theta\rangle$.

Finally, {\bf step (iii)} entails computing the wavefunction $\psi(\theta)$ by solving the Schrodinger equation
\begin{equation}
\sum_{\theta} \langle \theta | H | \theta^\prime \rangle \psi(\theta^\prime) = E \sum_{\theta} \langle \theta | \theta^\prime \rangle \psi(\theta^\prime).
\label{eq:Hthetathetaprime}
\end{equation}
The overlaps $ \langle \theta | \theta^\prime \rangle$ appear on the right hand side of this equation because the $| \theta \rangle$ do not form an orthonormal basis in general.  The Onishi formula  \cite{Ring2004} states
$ \left|\langle \theta|\theta^\prime\rangle\right|^2 = \mathrm{det }\,\, {\mathcal U}$ where
\begin{align}
&{\mathcal U}_{{\mathbf K},{\mathbf K}^\prime}  =\sum_{\mathbf R}  U^\dagger_{{\mathbf K},{\mathbf R}} e^{-i(\theta - \theta^\prime)\Theta_{\mathbf r}/2} U_{{\mathbf R},{\mathbf K}^\prime} \nonumber \\
 &\hspace{1.1in}+V^\dagger _{{\mathbf K},{\mathbf R}} e^{i(\theta - \theta^\prime)\Theta_{\mathbf r}/2}  V_{{\mathbf R},{\mathbf K}^\prime}.
 \label{eq:Uchargequbit}
\end{align}
The matrix elements $ \langle \theta | H | \theta^\prime \rangle$ of the Hamiltonian (\ref{eq:H}) can be computed using  \cite{Ring2004} 
\begin{equation}
\frac{\langle   \theta^\prime | c^\dagger_{\mathbf{R}_1} c_{\mathbf{R}_2} |\theta \rangle}{\langle\theta^\prime |\theta \rangle} = e^{i \frac{\theta}{2} \Theta_{{\mathbf r}_2}}  (V^* \frac{1}{\mathcal{U}^T}V^T)_{\mathbf{R}_2,\mathbf{R}_1}  e^{-i\frac{\theta^\prime}{2} \Theta_{{\mathbf r}_1}} \label{eq:matrixelements}
 \end{equation}
and a related  formula \cite{SupplementalInformation} for $ \langle  \theta^\prime | c^\dagger_{\mathbf{R}_1}c^\dagger_{\mathbf{R}_2} c_{\mathbf{R}_3}c_{\mathbf{R}_4} |\theta \rangle$.  To solve (\ref{eq:Hthetathetaprime}), we transform to an orthonormal basis with effective Hamiltonian ${\mathcal H}$ \cite{SupplementalInformation}.

This formalism is suitable for first principles or phenomenological computation as well as analytical study.  Here we present an example computation, carrying out steps (i)-(iii) on a model charge qubit formed by a rectangular lattice of $5\times 5 \times 220$ tight-binding sites with a mean occupation of $1012$ electrons \cite{SupplementalInformation}. 
An attractive Hubbard interaction gives rise to superconductivity \cite{Ambegaokar1982}. 
The magnitude of the order parameter $|\Delta^\pi_{({\mathbf r},1),({\mathbf r},-1)}|$ is shown in Fig. \ref{fig:chargequbit}c alongside its rescaled phase (\ref{eq:Thetachargequbit}).

The diagonal matrix elements  $\langle \theta|\mathcal{H}|\theta\rangle$ are shown in Fig.  \ref{fig:chargequbit}d; they correspond to the Josephson term in (\ref{eq:lumped}) and originate from terms in (\ref{eq:T}) associated with hopping across the junction.  The off-diagonal elements of $\langle \theta|\mathcal{H}|\theta^\prime\rangle$ cause transitions from one value of $|\theta\rangle$ to another; they correspond to the capacitive term in (\ref{eq:lumped}).  Based on LE theory, 
one expects this term to arise from (\ref{eq:W}).  However, in our calculations (\ref{eq:W}) makes only a small contribution which even deviates from the quadratic $4E_C n^2$ form; presumably this is because we assume an attractive Hubbard interaction rather than a Coulomb interaction.  Surprisingly, the kinetic energy (\ref{eq:T}) makes the dominant contribution.  Physically, the reason is that transferring $n$ Cooper pairs across our small charge qubit shrinks the Fermi sea of one half of the superconducting grain and grows the Fermi sea of the other.  Expanding the total energy of the Fermi seas in $n$ gives a quadratic term $4E_C n^2$.  It may be possible to amplify and exploit such underexplored capacitance features in device designs.  The eigenstates in Fig. \ref{fig:chargequbit}d take a form in consonance with the LE solution of (\ref{eq:lumped}).  Indeed,  (\ref{eq:lumped}) arises from our formalism in the appropriate abrupt-junction limit \cite{SupplementalInformation}.

\begin{figure}
\begin{tabular}{ll}
(a)\begin{tabular}{c}
\includegraphics[width=1.25in]{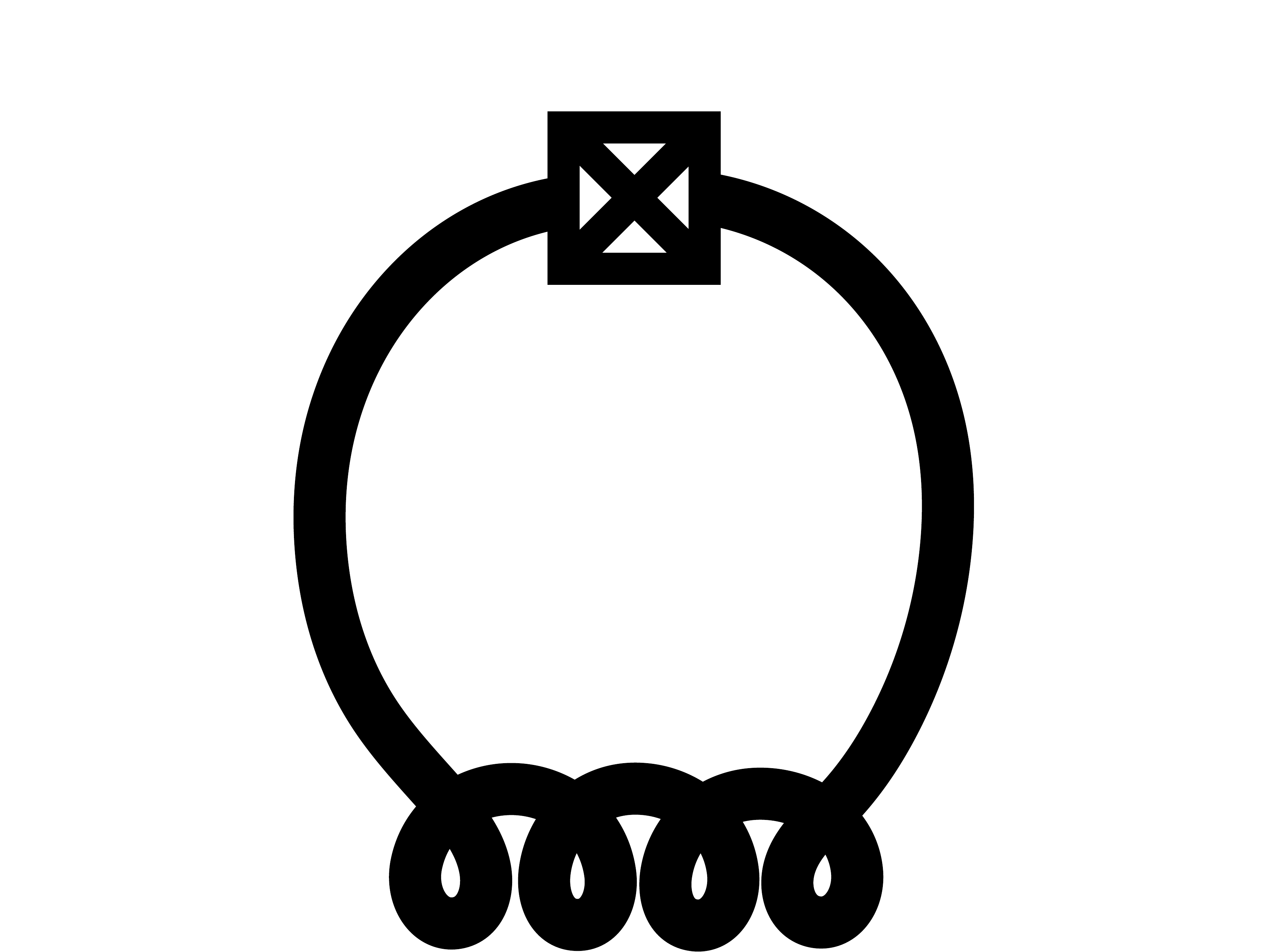}
\end{tabular} & \hspace{-0.15in}
(b) \begin{tabular}{c}
\includegraphics[width=1.35in]{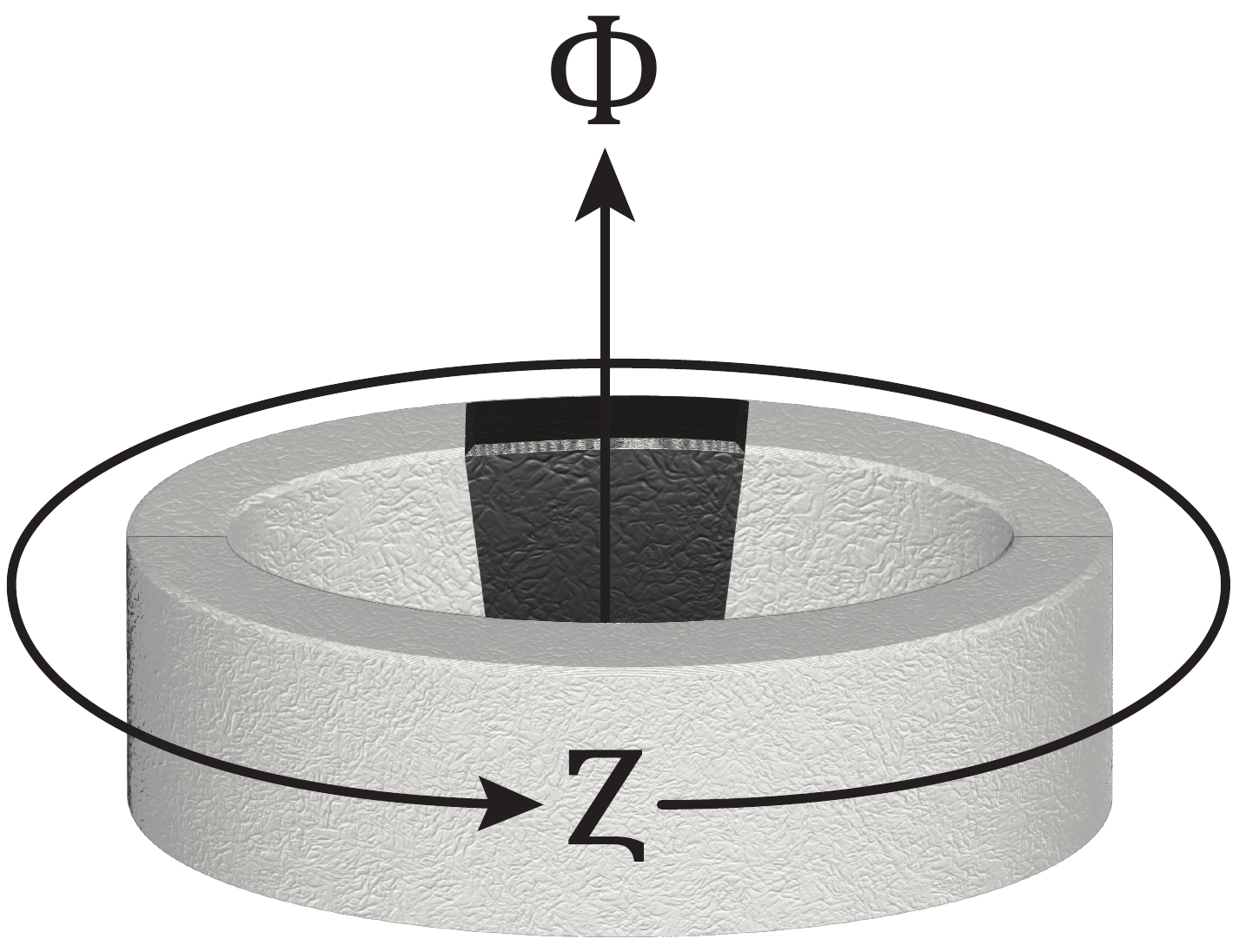}
\end{tabular} \\
(c)\begin{tabular}{c}\hspace{0.05in}
\includegraphics[width=1.5in]{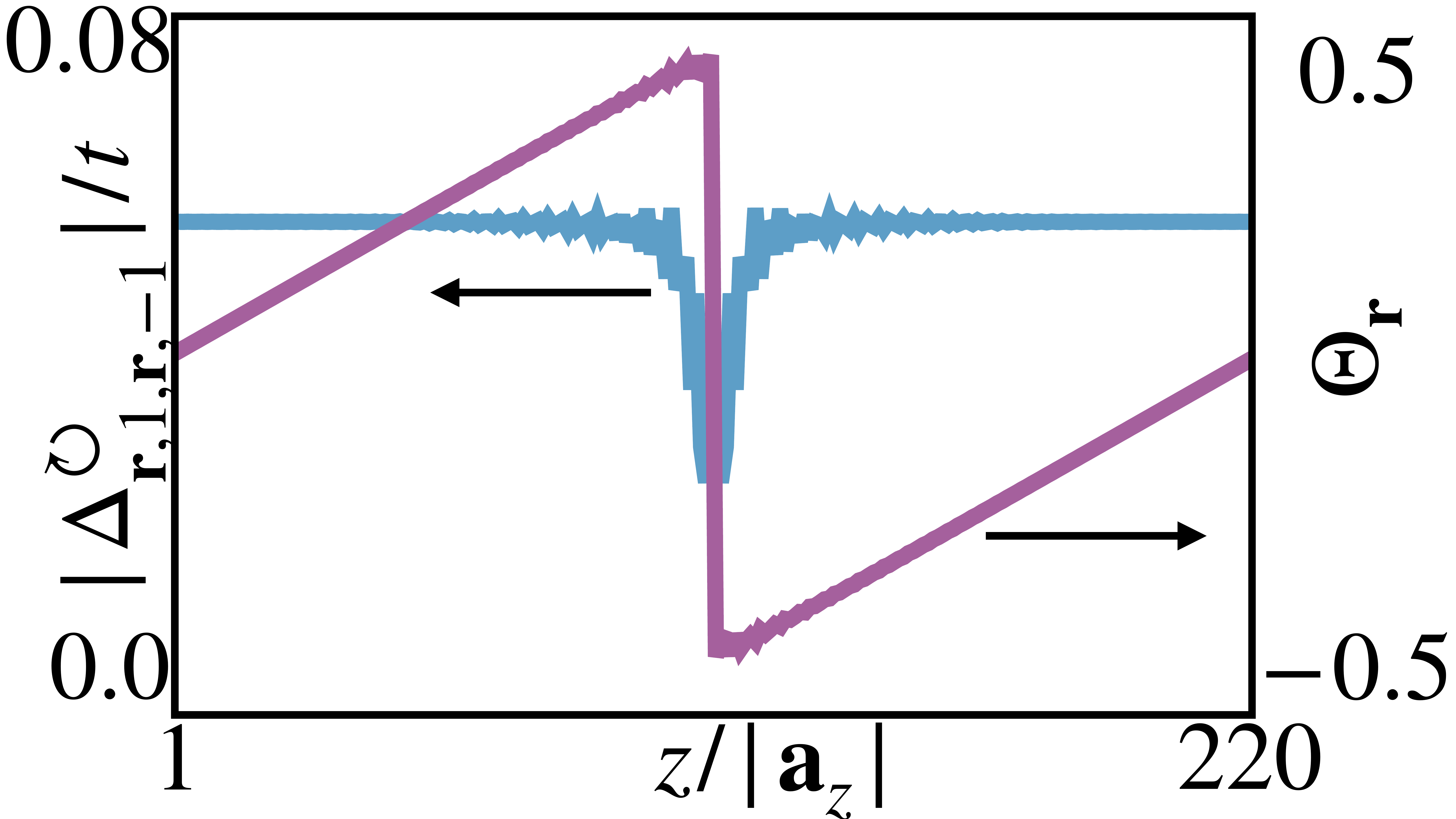}
\end{tabular} & \hspace{-0.15in}
(d) \begin{tabular}{c}
\includegraphics[width=1.5in]{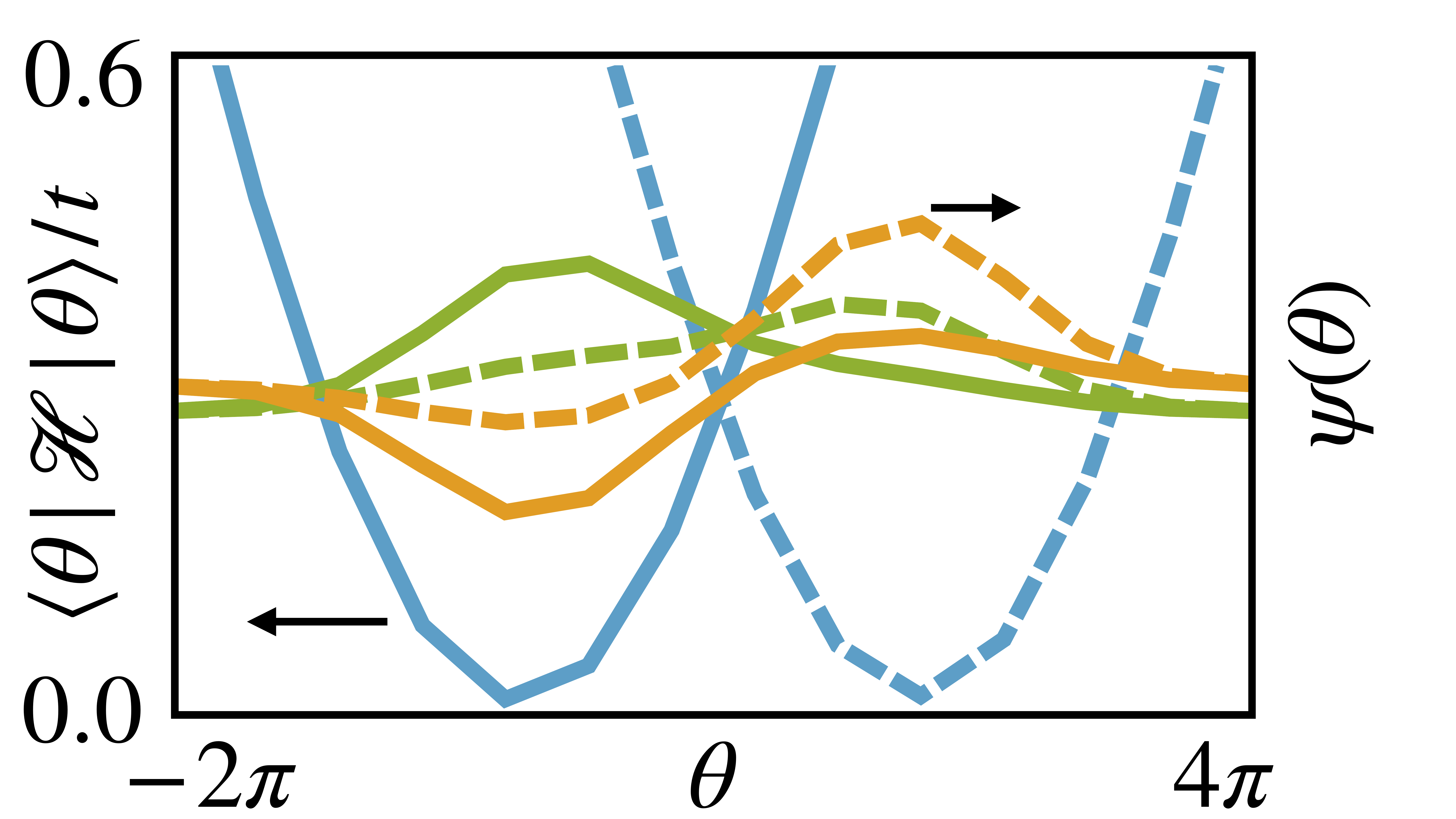}
\end{tabular} \end{tabular}
\caption{(a) Circuit diagram of RF squid qubit.  (b)  Schematic of qubit; magnetic flux $\Phi=h/(4e)$ threads loop. (c) Magnitude (blue) of superconducting gap $|\Delta^\circlearrowright_{{\mathbf r},\sigma=1,{\mathbf r},\sigma^\prime =-1}|$ depends only on $z$ component of ${\mathbf r}$.  Rescaled phase function $\Theta_{\mathbf r}$ (purple).   (d) Diagonal elements ${\langle \circlearrowright,\theta| {\mathcal H}|\!\!\circlearrowright,\theta\rangle}$ (solid) and $\langle \circlearrowleft,\theta| {\mathcal H}|\!\!\circlearrowleft,\theta\rangle$ (dashed) form effective potential.  Ground state (green) and first excited state (yellow) each has a counterclockwise (solid) and clockwise (dashed) component.}
\label{fig:RFsquidqubit}
\end{figure}

We now turn from a charge qubit to a more complicated example: an RF squid qubit threaded by magnetic flux (see Fig. \ref{fig:RFsquidqubit}a-b).  In this case, step (i) involves more than one self-consistent BdG solution: ${|\!\!\circlearrowright,\theta ^\circlearrowright\rangle}$ carries a current circulating clockwise around the loop and ${|\!\!\circlearrowleft,\theta ^\circlearrowleft \rangle}$ carries a current circulating counterclockwise.  (If the Josephson energy of the junction is sufficiently large compared to the inductive energy of the loop, still more self-consistent solutions arise; they are easily included in the analysis if needed.)  The corresponding self-consistent BdG solutions are collected into matrices $U^\circlearrowright_{{\mathbf R},{\mathbf K}}$, $V^\circlearrowright_{{\mathbf R},{\mathbf K}}$ and $U^\circlearrowleft_{{\mathbf R},{\mathbf K}}$, $V^\circlearrowleft_{{\mathbf R},{\mathbf K}}$ respectively.  The associated order parameters are $\Delta^\circlearrowright_{{\mathbf R},{\mathbf R}^\prime}$ and $\Delta^\circlearrowleft_{{\mathbf R},{\mathbf R}^\prime}$.
The parameter $\theta^\circlearrowright$
equals the change in $\text{Arg } \Delta^\circlearrowright_{({\mathbf r},1),({\mathbf r},-1)}$ along the inductance of the loop (i.e. the total phase change of the order parameter around the loop minus the phase change across the junction), and $\theta ^\circlearrowleft$ is defined analogously.

A rescaled phase function equals
\begin{equation}
\Theta_{\mathbf r} = \text{Arg } \Delta ^\circlearrowright_{({\mathbf r},1),({\mathbf r},-1)}/\theta^\circlearrowright .
\label{eq:Theta}
\end{equation} 
Note that (\ref{eq:Theta}) is defined using $\Delta^\circlearrowright_{{\mathbf R},{\mathbf R}^\prime}$ rather than $\Delta^\circlearrowleft_{{\mathbf R},{\mathbf R}^\prime}$; this choice is made so $\Theta_{\mathbf r}$ does not have any vortices that accumulate multiples of $2\pi$ as $\Theta_{\mathbf r}$ proceeds around the qubit loop.  This property is evident in Fig. \ref{fig:RFsquidqubit}c.  If the direction of applied flux were flipped,  $\Delta^\circlearrowleft$ would appear in the definition instead.

In terms of (\ref{eq:Theta}), $\ket{\circlearrowright,\theta}$ is defined as the BdG state obtained by performing a gauge transformation $U^\circlearrowright_{{\mathbf R},{\mathbf K}} \rightarrow e^{i (\theta-\theta^\circlearrowright) \Theta_{\mathbf r}/2} U^\circlearrowright_{{\mathbf R},{\mathbf K}} $ and $V^\circlearrowright_{{\mathbf R},{\mathbf K}} \rightarrow e^{-i (\theta-\theta^\circlearrowright)\Theta_{\mathbf r}/2} V^\circlearrowright_{{\mathbf R},{\mathbf K}}$.  This definition ensures that the phase of the order parameter of $\ket{\circlearrowright,\theta}$ changes by $\theta$ instead of $\theta^\circlearrowright$ along the inductor.  Analogously, $\ket{\circlearrowleft,\theta}$ is defined as the BdG state obtained by performing a gauge transformation $U^\circlearrowleft_{{\mathbf R},{\mathbf K}} \rightarrow e^{i (\theta-\theta^\circlearrowleft) \Theta_{\mathbf r}/2} U^\circlearrowleft_{{\mathbf R},{\mathbf K}} $ and $V^\circlearrowleft_{{\mathbf R},{\mathbf K}} \rightarrow e^{-i (\theta-\theta^\circlearrowleft)\Theta_{\mathbf r}/2} V^\circlearrowleft_{{\mathbf R},{\mathbf K}}$.

In step (ii), we write 
\begin{equation} \ket{\psi} = \sum_\theta \,\, \psi(\circlearrowright,\theta)  \ket{\circlearrowright,\theta} +  \psi(\circlearrowleft,\theta)  \ket{\circlearrowleft,\theta}.
\end{equation}  Step (iii) then requires us to solve
\begin{align}
\sum_{\theta^\prime} & \begin{bmatrix} \langle \circlearrowright,\theta|H|\!\!\circlearrowright,\theta^\prime\rangle & \langle \circlearrowright,\theta|H|\!\!\circlearrowleft,\theta^\prime\rangle \\ \langle \circlearrowleft,\theta|H|\!\!\circlearrowright,\theta^\prime\rangle & \langle \circlearrowleft,\theta|H|\!\!\circlearrowleft,\theta^\prime\rangle \end{bmatrix} \begin{bmatrix}  \psi(\circlearrowright,\theta^\prime) \\  \psi(\circlearrowleft,\theta^\prime) \end{bmatrix} \label{eq:H2by2} \\
& = E \sum_{\theta^\prime} \begin{bmatrix} \langle \circlearrowright,\theta|\circlearrowright,\theta^\prime\rangle & \langle \circlearrowright,\theta\circlearrowleft,\theta^\prime\rangle \\ \langle \circlearrowleft,\theta|\circlearrowright,\theta^\prime\rangle & \langle \circlearrowleft,\theta|\circlearrowleft,\theta^\prime\rangle \end{bmatrix} \begin{bmatrix}  \psi(\circlearrowright,\theta^\prime) \\  \psi(\circlearrowleft,\theta^\prime) \end{bmatrix}. 
 \nonumber
\end{align}
After transforming to an orthonormal basis, we produce a $2 \times 2$ effective Hamiltonian ${\mathcal H}$ \cite{SupplementalInformation}.

We carry out steps (i)-(iii), performing a numerical computation on a model RF squid qubit analogous to the charge qubit computation above \cite{SupplementalInformation}.
Because of the $2\times 2$ structure of the matrix $\mathcal H$, its diagonal  gives rise to the 2 potentials $\langle \circlearrowright, \theta|{\mathcal H}|\!\!\circlearrowright,\theta\rangle$ and $\langle \circlearrowleft, \theta|{\mathcal H}|\!\!\circlearrowleft,\theta\rangle$ depicted in Fig \ref{fig:RFsquidqubit}d; the curve formed by taking the lower potential at each $\theta$ produces an RF squid double-well potential as expected.  Indeed, under suitable conditions discussed in \cite{SupplementalInformation},  ${\mathcal H}$ reduces to the LE Hamiltonian
\begin{equation}
H_{\text{LE}} = 4 E_C n^2 + E_J (1-\cos\theta) + E_L (\theta-\phi)^2/2
\label{eq:lumpedRF}
\end{equation}
where $\theta \in \mathbb{R}$ is unbounded, $[\theta,n] = i$, and $\phi = 2 \pi (2 e/h) \Phi$ is fixed by the applied flux $\Phi$.

The energy eigenstates of $\mathcal H$, shown in Fig \ref{fig:RFsquidqubit}d, accord with those expected within LE theory.  However, they exhibit an asymmetry between the two peaks of each wavefunction.  This effect occurs because the microscopic electronic Hamiltonian  (\ref{eq:H}) does not possess the symmetry at half a superconducting flux quantum $h/4e$ that it does at half an electronic flux quantum $h/2e$: the symmetry between clockwise and counterclockwise superconducting states in the loop emerges only approximately when electrons bind into Cooper pairs.   This effect is absent from LE theory; it becomes significant for mesoscopic loop sizes \cite{Zhu1994} and is captured by our formalism.  

The microscopic character of our theory makes it particularly suitable for investigating Schrodinger cat states in superconducting loops.  Several remarkable experiments \cite{Friedman2000,VanderWal2000} have realized these cats, formed by superposing a clockwise supercurrent and a counterclockwise supercurrent. 
To assess the implications \cite{Leggett1985} of these experiments, it is of central importance to quantify the size of the cat -- the number of entangled particles.  We will focus on \cite{Friedman2000} since it raises the breathtaking possibility that billions of electrons might have been entangled.

Computing the number of entangled electrons, $\Delta N$, is beyond the scope of LE theory \cite{SupplementalInformation}.  
The state-of-the-art calculation is due to Korsbakken et al. \cite{Korsbakken2009,*Korsbakken2010}.
Given a partition of the many-body Schrodinger cat state into two terms $\sqrt{\frac{1}{2}}{|\!\circlearrowright\rangle}+\sqrt{\frac{1}{2}}{|\!\circlearrowleft\rangle}$, they introduced
\begin{equation}
\Delta N = \frac{1}{2} \sum_{\mathbf Q}|\langle\circlearrowright\!| c^\dagger_{\mathbf Q}c_{\mathbf Q}  |\!\circlearrowright\rangle - \langle\circlearrowleft\!| c^\dagger_{\mathbf Q}c_{\mathbf Q}  |\!\circlearrowleft\rangle|,
\label{eq:DeltaN}
\end{equation}
to count how many particles are in different modes in ${|\!\circlearrowright\rangle}$ and ${|\!\circlearrowleft\rangle}$.  Here, ${\mathbf Q}$ labels a state in whichever basis of single-particle electron states maximizes $\Delta N$, which is the basis in which the matrix ${\langle\circlearrowright\!| c^\dagger_{\mathbf Q}c_{{\mathbf Q}^\prime}  |\!\circlearrowright\rangle} - {\langle\circlearrowleft\!| c^\dagger_{\mathbf Q}c_{{\mathbf Q}^\prime}  |\!\circlearrowleft\rangle}$ is diagonal.  Korsbakken et al. evaluated (\ref{eq:DeltaN}) by approximating ${|\!\!\circlearrowright\rangle}$ and ${|\!\!\circlearrowleft\rangle}$ as Galilean-boosted BCS states rather than using LE theory.  They thereby derived the approximation $\Delta N = 3 I L_z/4 e v_F$
 \cite{SupplementalInformation}.  As a check, we evaluate (\ref{eq:DeltaN}) using the RF squid qubit eigenstates that we obtained numerically above.  For these, their approximation differs from (\ref{eq:DeltaN}) by $\sim 30\%$.
 
\begin{figure}
\includegraphics[width=3.0in]{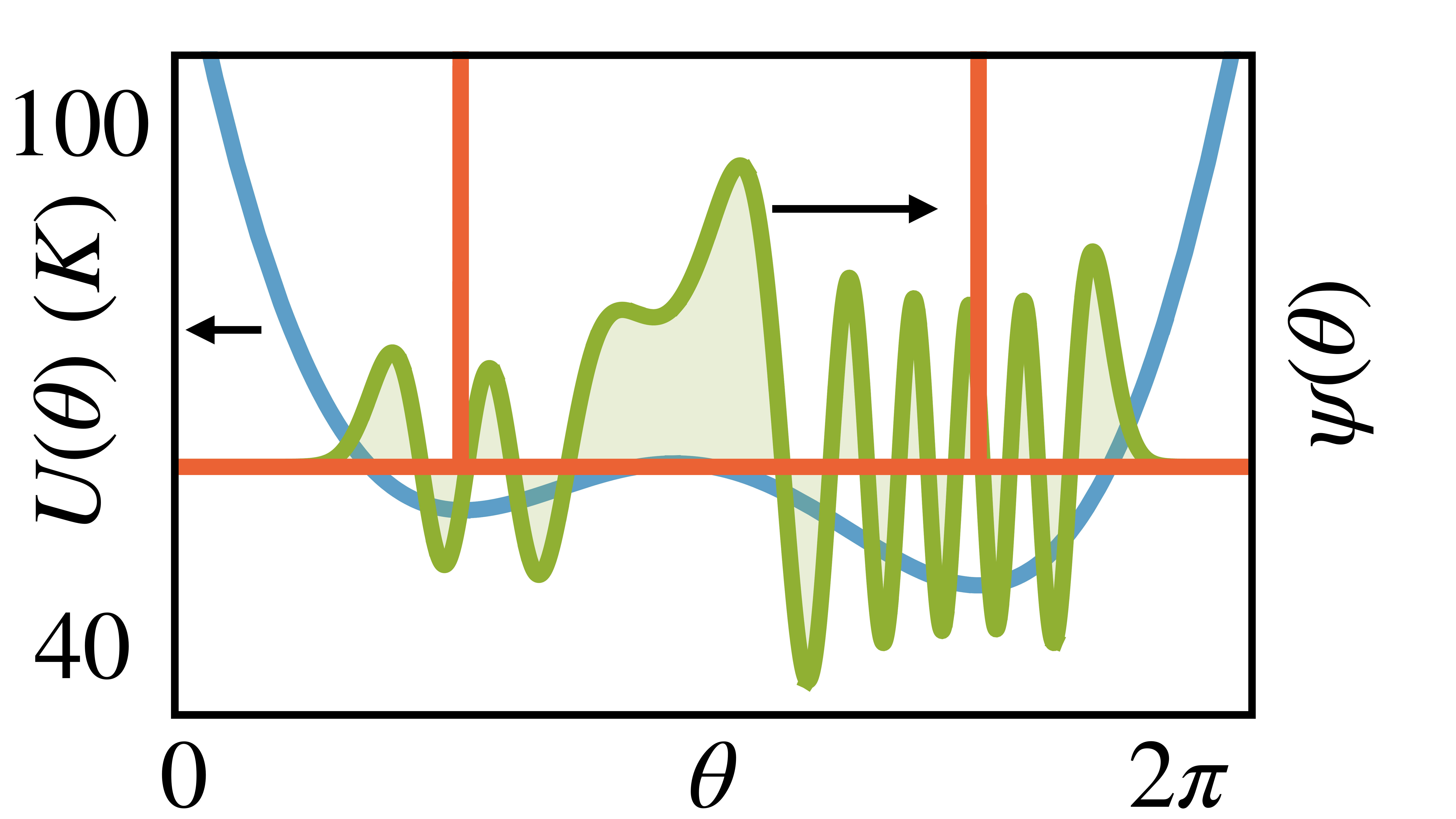}
\caption{LE theory potential $E_J (1-\cos \theta) + \frac{1}{2} E_L  (\theta - \phi)^2$ is shown in blue.  LE wavefunction of Schrodinger cat state of \cite{Friedman2000} is shown in green.  The 2-symmetric-peak form assumed by \cite{Korsbakken2009,Korsbakken2010} is depicted in red.  We use parameters that are suitable for \cite{Friedman2000}: $E_C= 9.0$ mK = $0.19$ GHz, $E_J =76$ K = $1.6$ THz, $E_L= 645$ K = $13.5$ THz, and $\phi = 2 \pi (0.51425)$.}
\label{fig:Friedmanwavefunction}
\end{figure}

 Applying their approximation to the experiment \cite{Friedman2000}, they concluded \cite{Korsbakken2009,*Korsbakken2010} that $\Delta N \sim 3800 - 5750$ electrons. However, note that the wavefunction realized experimentally in \cite{Friedman2000} takes the form of the green curve in Fig. \ref{fig:Friedmanwavefunction} according to LE theory.  It differs dramatically from a Schrodinger cat form $\sqrt{\frac{1}{2}}{|\!\circlearrowright\rangle}+\sqrt{\frac{1}{2}}{|\!\circlearrowleft\rangle}$ and particularly so from the pristine wavefunction assumed by Korsbakken et al. (red curve in Fig. \ref{fig:Friedmanwavefunction}).   A definitive evaluation of the number of entangled electrons in \cite{Friedman2000} would therefore require a tractable alternative to (\ref{eq:DeltaN}) that does not assume a Schrodinger cat form $\sqrt{\frac{1}{2}}{|\!\circlearrowright\rangle}+\sqrt{\frac{1}{2}}{|\!\circlearrowleft\rangle}$.  Unfortunately, defining such a metric is a vexatious problem \cite{Frowis2018} out of the scope of our study.

In conclusion, we have developed a microscopic theory of superconducting qubits that goes beyond the standard LE approximation.  It accords with LE theory both for charge and RF squid qubits and yields several insights inaccessible to LE theory.  Numerous applications of this tool are anticipated as experimental progress allows increasingly precise study of superconducting qubits.

We gratefully thank M. Kruger and I. M. Mandelberg.

\bibliographystyle{apsrev4-1}
\bibliography{../../bibliography/ari_mizel}

\pagebreak

\section{Supplemental Material}

\setcounter{figure}{0}
\renewcommand{\thefigure}{S\arabic{figure}}
\setcounter{equation}{1}

\subsection{Bogoliubov-de Gennes Equations}

This section derives the Bogoliubov-de Gennes (BdG) equations, following pp. 137--145 of \cite{deGennes1966}. 
We begin with the definition of the Bogoliubov transformation

\begin{align}
c_{\mathbf{R}} = \sum_\mathbf{K}  U_{\mathbf{R},\mathbf{K}}\alpha_\mathbf{K} + V^* _{\mathbf{R},\mathbf{K}}\alpha^\dagger_\mathbf{K}  \nonumber \\
c^\dagger_{\mathbf{R}} = \sum_\mathbf{K}  U^*_{\mathbf{R},\mathbf{K}} \alpha^\dagger_\mathbf{K} + V_{\mathbf{R},\mathbf{K}}\alpha_\mathbf{K}.
 \tag{S\arabic{equation}}
\label{eq:Bogoliubov}
 \end{align}
Imposing anticommutation relations, we obtain
\stepcounter{equation}
\begin{equation}
\{c_{\mathbf{R}},c_{\mathbf{R}^\prime}\} =  V^* _{\mathbf{R},\mathbf{K}} U_{\mathbf{R}^\prime,\mathbf{K}} + U_{\mathbf{R},\mathbf{K}} V^* _{\mathbf{R}^\prime,\mathbf{K}} = 0
 \tag{S\arabic{equation}}\end{equation}
which becomes
\stepcounter{equation}
\begin{equation}
V^*U^T+UV^\dagger = 0
 \tag{S\arabic{equation}}
 \label{eq:UVplusVU}
 \end{equation}
as a matrix equation.
Similarly,
\stepcounter{equation}
\begin{equation}
\{c_{\mathbf{R}},c^\dagger_{\mathbf{R}^\prime}\} =  U _{\mathbf{R},\mathbf{K}} U^*_{\mathbf{R}^\prime,\mathbf{K}} + V^*_{\mathbf{R},\mathbf{K}} V_{\mathbf{R}^\prime,\mathbf{K}} = \delta_{\mathbf{R},\mathbf{R}^\prime}
 \tag{S\arabic{equation}}\end{equation}
where a Kronecker delta appears on the right hand side.  As a matrix equation, this becomes
\stepcounter{equation}
\begin{equation}
UU^\dagger+V^*V^T = I.
\label{eq:UUplusVV}
 \tag{S\arabic{equation}}\end{equation}
 Now,  (\ref{eq:Bogoliubov}) can be written as a matrix equation
 \stepcounter{equation}
\begin{equation}
\begin{bmatrix} 
c_{\mathbf{R}} \\ c^\dagger_{\mathbf{R}}
\end{bmatrix} = \begin{bmatrix} U & V^*\\V & U^* \end{bmatrix} \begin{bmatrix} 
\alpha_{\mathbf{R}} \\ \alpha^\dagger_{\mathbf{R}}
\end{bmatrix}
\label{eq:Bogoliubovmatrix}
 \tag{S\arabic{equation}}.
 \end{equation}
Relations (\ref{eq:UVplusVU}) and (\ref{eq:UUplusVV}) imply 
 \stepcounter{equation}
\begin{equation}
\begin{bmatrix} U & V^*\\V & U^* \end{bmatrix} \begin{bmatrix} U^\dagger & V^\dagger\\V^T & U^T \end{bmatrix} = \begin{bmatrix} I& 0\\0& I \end{bmatrix} \tag{S\arabic{equation}}.
\end{equation}
Since the right inverse of a square matrix is also a left inverse, we also have
 \stepcounter{equation}
\begin{equation}
\begin{bmatrix} U^\dagger & V^\dagger\\V^T & U^T \end{bmatrix} \begin{bmatrix} U & V^*\\V & U^* \end{bmatrix} = \begin{bmatrix} I& 0\\0& I \end{bmatrix} \tag{S\arabic{equation}}.
\label{eq:matrixinverse}
\end{equation}
This equation gives 2 identities; collecting them together with (\ref{eq:UVplusVU}) and (\ref{eq:UUplusVV}) gives a total of 4 identities
 \stepcounter{equation}
\begin{align*}
UU^\dagger+V^*V^T = I & \hspace{0,2in} & U^\dagger U+V^\dagger V = I\\
V^*U^T+UV^\dagger = 0 & \hspace{0,2in} &  V^T U + U^T V = 0.
\tag{S\arabic{equation}}
\label{eq:Bogoliubovidentities}
\end{align*}
Equation (\ref{eq:matrixinverse}) implies that the inverse of the Bogoliubov transformation is
\stepcounter{equation}
\begin{align*}
& \alpha_\mathbf{K}  = \sum_\mathbf{R}  U^*_{\mathbf{R},\mathbf{K}}c_{\mathbf{R}} + V^*_{\mathbf{R},\mathbf{K}}c^\dagger_{\mathbf{R}} \\
& \alpha^\dagger_\mathbf{K}  = \sum_\mathbf{R}  U_{\mathbf{R},\mathbf{K}}c^\dagger_{\mathbf{R}} + V_{\mathbf{R},\mathbf{K}}c_{\mathbf{R}}.
 \tag{S\arabic{equation}}
 \label{eq:Bogoliubovinverse}
 \end{align*}
 
We use the Bogoliubov transformation to find eigenstates of a mean-field approximation to the Hamiltonian.  Starting with 
\stepcounter{equation}
\begin{equation}
 H = \sum_{{\mathbf R},{\mathbf R}^\prime} c^\dagger_{{\mathbf R}^\prime} h_{{\mathbf R}^\prime,{\mathbf R}} c_{\mathbf R} + \frac{1}{2}W({\mathbf R}^\prime,{\mathbf R}) c^\dagger_{{\mathbf R}^\prime} c^\dagger_{\mathbf R} c_{\mathbf R} c_{{\mathbf R}^\prime},
 \tag{S\arabic{equation}}\end{equation}
we make a mean-field approximation
\stepcounter{equation}
\begin{align*}
H^\prime &=  \sum_{{\mathbf R},{\mathbf R}^\prime} c^\dagger_{{\mathbf R}^\prime} h_{{\mathbf R}^\prime,{\mathbf R}} c_{\mathbf R} +  \frac{1}{2}W({\mathbf R}^\prime,{\mathbf R}) \times  \tag{S\arabic{equation}}\\
& \Big(c^\dagger_{{\mathbf R}^\prime} c_{{\mathbf R}^\prime} \langle c^\dagger_{\mathbf R} c_{\mathbf R} \rangle + \langle c^\dagger_{{\mathbf R}^\prime} c_{{\mathbf R}^\prime} \rangle c^\dagger_{\mathbf R} c_{\mathbf R} - \langle c^\dagger_{{\mathbf R}^\prime} c_{{\mathbf R}^\prime} \rangle \langle c^\dagger_{\mathbf R} c_{\mathbf R} \rangle \\
&- c^\dagger_{{\mathbf R}^\prime} c_{\mathbf R} \langle c^\dagger_{\mathbf R}  c_{{\mathbf R}^\prime}\rangle -\langle c^\dagger_{{\mathbf R}^\prime}  c_{\mathbf R} \rangle c^\dagger_{\mathbf R} c_{{\mathbf R}^\prime} + \langle c^\dagger_{{\mathbf R}^\prime}  c_{\mathbf R} \rangle \langle c^\dagger_{\mathbf R} c_{{\mathbf R}^\prime} \rangle\\
&+c^\dagger_{{\mathbf R}^\prime} c^\dagger_{\mathbf R} \langle c_{\mathbf R} c_{{\mathbf R}^\prime} \rangle +\langle c^\dagger_{{\mathbf R}^\prime} c^\dagger_{\mathbf R} \rangle c_{\mathbf R} c_{{\mathbf R}^\prime} - \langle c^\dagger_{{\mathbf R}^\prime} c^\dagger_{\mathbf R} \rangle \langle c_{\mathbf R} c_{{\mathbf R}^\prime}\rangle   \Big). 
\end{align*}
In each of the three lines in parentheses, we have paired the first operator $c^\dagger_{{\mathbf R}^\prime}$ in the interaction term $\frac{1}{2}W({\mathbf R}^\prime,{\mathbf R}) c^\dagger_{{\mathbf R}^\prime} c^\dagger_{\mathbf R} c_{\mathbf R} c_{{\mathbf R}^\prime}$ one of the remaining 3 operators ($c^\dagger_{{\mathbf R}^\prime}$ is paired with $c_{{\mathbf R}^\prime}$  in the first line then $c_{\mathbf R}$ in the second line then $c^\dagger_{\mathbf R}$ in the third line).  Within each line, we applied the  approximation
\begin{align*}
ab & = [(a-\langle a \rangle) + \langle a \rangle][(b-\langle b \rangle) + \langle b \rangle] \\
& \approx (a-\langle a \rangle) \langle b \rangle + \langle a \rangle (b-\langle b \rangle) + \langle a \rangle \langle b \rangle\\
& = a  \langle b \rangle +  \langle a \rangle b -  \langle a \rangle \langle b \rangle
\end{align*}
taking $a$ to be the product of $c^\dagger_{{\mathbf R}^\prime}$ and its pair while taking $b$ to be the product of the remaining 2 operators.  The approximation assumes $(a-\langle a \rangle) (b-\langle b \rangle)$ is the product of 2 small quantities and can be neglected.
It is convenient to set
\stepcounter{equation}
\begin{align*}
H^\prime & \equiv \sum_{{\mathbf R},{\mathbf R}^\prime} c^\dagger_{{\mathbf R}^\prime} h^\prime_{{\mathbf R}^\prime,{\mathbf R}} c_{\mathbf R} \\
& + \frac{ \Delta_{{\mathbf R}^\prime,{\mathbf R}} }{2} c^\dagger_{{\mathbf R}^\prime} c^\dagger_{\mathbf R}+\frac{( \Delta_{{\mathbf R}^\prime,{\mathbf R}})^*}{2}  c_{\mathbf R} c_{{\mathbf R}^\prime} + \text{const} \tag{S\arabic{equation}}
\end{align*}
where 
\stepcounter{equation}
\begin{equation}
\Delta_{{\mathbf R}^\prime,{\mathbf R}} = W({\mathbf R},{\mathbf R}^\prime) \langle c_{\mathbf R} c_{{\mathbf R}^\prime} \rangle
 \tag{S\arabic{equation}}\end{equation}
and 
\stepcounter{equation}
\begin{align*}
& h^\prime_{{\mathbf R}^\prime,{\mathbf R}}  =   h_{{\mathbf R}^\prime,{\mathbf R}} \tag{S\arabic{equation}}\\
 & + \delta_{{\mathbf R}^\prime,{\mathbf R}} \sum_{{\mathbf R}^{\prime\prime}}  W({\mathbf R},{\mathbf R}^{\prime\prime}) \langle c^\dagger_{{\mathbf R}^{\prime\prime}} c_{{\mathbf R}^{\prime\prime}}\rangle - W({\mathbf R},{\mathbf R}^\prime) \langle c^\dagger_{\mathbf R}  c_{{\mathbf R}^\prime}\rangle.
\end{align*}
To obtain this form of $h^\prime_{{\mathbf R}^\prime,{\mathbf R}}$, we have used $W({\mathbf R},{\mathbf R}^\prime) = W({\mathbf R}^\prime,{\mathbf R})$.  Note that the final term of $h^\prime_{{\mathbf R}^\prime,{\mathbf R}}$ is the exchange interaction.

We next demand that the Bogoliubov transformation diagonalize the mean-field Hamiltonian, so that $H^\prime = E_g + \sum_{{\mathbf K}} \epsilon_k \alpha^\dagger_{\mathbf K} \alpha_{\mathbf K}$.  This implies $[ \alpha_{\mathbf K},H^\prime] = \epsilon_k  \alpha_{\mathbf K}$ and $[\alpha^\dagger_{\mathbf K},H^\prime] = -\epsilon_k  \alpha^\dagger_{\mathbf K}$.  We compute
\stepcounter{equation}
\begin{equation}
[c_{\mathbf{R}},H^\prime] = \sum_{\mathbf{R}^{\prime}} h^\prime_{\mathbf{R},{\mathbf{R}^\prime} } c_{\mathbf{R}^{\prime}}+\Delta_{\mathbf{R},{\mathbf{R}^\prime} } c^\dagger_{\mathbf{R}^{\prime}}
 \tag{S\arabic{equation}}\end{equation}
and substitute in the Bogoliubov transformation (\ref{eq:Bogoliubov}).  Comparing the coefficients of  $\alpha_{\mathbf K}$ and $\alpha^\dagger_{\mathbf K}$  on each side, we obtain the two BdG equations
\stepcounter{equation}
\begin{align*}
\epsilon_k U_{\mathbf{R},\mathbf{K}} & =  \sum_{\mathbf{R}^{\prime}} h^\prime_{\mathbf{R},{\mathbf{R}}^\prime }U_{\mathbf{R}^\prime,\mathbf{K}}  + \Delta_{\mathbf{R},{\mathbf{R}^\prime} } V_{\mathbf{R}^\prime,\mathbf{K}} \tag{S\arabic{equation}} \label{eq:BdG}\\
\epsilon_k V_{\mathbf{R},\mathbf{K}} & =  \sum_{\mathbf{R}^{\prime}} (-h^\prime_{\mathbf{R},{\mathbf{R}^\prime} })^*V_{\mathbf{R}^\prime,\mathbf{K}}  + (\Delta_{\mathbf{R},{\mathbf{R}^\prime} })^*  U_{\mathbf{R}^\prime,\mathbf{K}}.
\end{align*}
Substituting the Bogoliubov transformations in the definitions of $h^\prime_{{\mathbf R}^\prime,{\mathbf R}}$ and $\Delta_{{\mathbf R}^\prime,{\mathbf R}}$  and using the fact that $\alpha_{\mathbf K}$ annihilates the ground state, we obtain
\stepcounter{equation}
\begin{align*}
h^\prime_{{\mathbf R}^\prime,{\mathbf R}}  = &  h_{{\mathbf R}^\prime,{\mathbf R}} + \delta_{{\mathbf R}^\prime,{\mathbf R}} \sum_{{\mathbf R}^{\prime\prime}}  W({\mathbf R},{\mathbf R}^{\prime\prime}) \sum_{\mathbf K} V_{{\mathbf R}^{\prime\prime},{\mathbf K}}  V^*_{{\mathbf R}^{\prime\prime},{\mathbf K}} \\
&- W({\mathbf R},{\mathbf R}^\prime) \sum_{\mathbf K} V_{{\mathbf R},{\mathbf K}}  V^*_{{\mathbf R}^\prime,{\mathbf K}}
\tag{S\arabic{equation}} \label{eq:hprime}
\end{align*}
and
\stepcounter{equation}
\begin{equation}
\Delta_{{\mathbf R}^\prime,{\mathbf R}} = W({\mathbf R},{\mathbf R}^\prime) \sum_{\mathbf K} U_{{\mathbf R},{\mathbf K}} V^*_{{\mathbf R}^\prime,{\mathbf K}}. \label{eq:Delta}
 \tag{S\arabic{equation}}\end{equation}
The BdG equations are to be solved self-consistently with these expressions for $h^\prime_{{\mathbf R}^\prime,{\mathbf R}}$ and $\Delta_{{\mathbf R}^\prime,{\mathbf R}}$.

It is sometimes useful to express the BdG many-body ground state in Thouless form  \cite{Ring2004}.  To do so, we define $Z \equiv (V U^{-1})^*$ and write
\stepcounter{equation}
\begin{equation}
{|0\rangle} = {\mathcal N} \exp \left[\frac{1}{2}\sum_{{\mathbf R},{\mathbf R}^\prime} c^\dagger_{{\mathbf R}} Z_{{\mathbf R},{\mathbf R}^\prime} c^\dagger_{{\mathbf R}^\prime}\right] \ket{\mathrm{vac}}.
\label{eq:thouless0}
\tag{S\arabic{equation}}
\end{equation}
Here, ${\mathcal N}$ is a normalization constant.  This is a generalization of the standard BCS ground state wavefunction ${\mathcal N} \exp \left[\sum_{\mathbf k} (v_{\mathbf k}/u_{\mathbf k}) c^\dagger_{{\mathbf k},1} c^\dagger_{-{\mathbf k},-1}  \right] \ket{\mathrm{vac}}$.  It is more general because it does not assume the Cooper pairs form specifically in the momentum basis.

To show that (\ref{eq:thouless0}) is the many-body ground state of the BdG equations, we need to demonstrate that it is annihilated by $\alpha_\mathbf{K}$.  
Using the inverse Bogoliubov transformation (\ref{eq:Bogoliubovinverse}), we compute
\stepcounter{equation}
\begin{equation}
\left[\frac{1}{2}\sum_{{\mathbf R},{\mathbf R}^\prime} c^\dagger_{{\mathbf R}} Z_{{\mathbf R},{\mathbf R}^\prime} c^\dagger_{{\mathbf R}^\prime}, \alpha_\mathbf{K} \right]  =\sum_{\mathbf R} V^*_{\mathbf{R},\mathbf{K}}c^\dagger_{\mathbf R}.
 \tag{S\arabic{equation}}\end{equation}
 The argument uses (\ref{eq:Bogoliubovidentities}).
Note that the right hand side commutes with $\frac{1}{2}\sum_{{\mathbf R},{\mathbf R}^\prime} c^\dagger_{{\mathbf R}} Z_{{\mathbf R},{\mathbf R}^\prime} c^\dagger_{{\mathbf R}^\prime}$.  Now, according to a well-known lemma,
\stepcounter{equation}
\begin{equation}
e^X Y e^{-X}  = Y  + \left[Y,X\right] 
 \tag{S\arabic{equation}}\label{eq:commutelemma}\end{equation}
when $[X,[X,Y]]=0$.  (This is proven straightforwardly by differentiating $e^{sX} Y e^{-sX}$ with respect to $s$.)
Thus, we deduce that
\stepcounter{equation}
\begin{align*}
\exp &\left[-\frac{1}{2}\sum_{{\mathbf R},{\mathbf R}^\prime} c^\dagger_{{\mathbf R}} Z_{{\mathbf R},{\mathbf R}^\prime} c^\dagger_{{\mathbf R}^\prime}\right] \alpha_\mathbf{K} \exp \left[\frac{1}{2}\sum_{{\mathbf R},{\mathbf R}^\prime} c^\dagger_{{\mathbf R}} Z_{{\mathbf R},{\mathbf R}^\prime} c^\dagger_{{\mathbf R}^\prime}\right] \\
& = \alpha_\mathbf{K} -\sum_{\mathbf R} V^*_{\mathbf{R},\mathbf{K}}c^\dagger_{\mathbf R} = \sum_{\mathbf R} U^*_{\mathbf{R},\mathbf{K}}c_{\mathbf R}.
\tag{S\arabic{equation}} \label{eq:commutator}
\end{align*}
It follows that
\stepcounter{equation}
\begin{align*}
  \alpha_\mathbf{K} & \exp \left[\frac{1}{2}\sum_{{\mathbf R},{\mathbf R}^\prime} c^\dagger_{{\mathbf R}} Z_{{\mathbf R},{\mathbf R}^\prime} c^\dagger_{{\mathbf R}^\prime}\right] \ket{\mathrm{vac}} \tag{S\arabic{equation}} \label{eq:groundstateproof} \\
 & =  \exp \left[\frac{1}{2}\sum_{{\mathbf R},{\mathbf R}^\prime} c^\dagger_{{\mathbf R}} Z_{{\mathbf R},{\mathbf R}^\prime} c^\dagger_{{\mathbf R}^\prime}\right] \sum_{\mathbf R} U^*_{\mathbf{R},\mathbf{K}}c_{\mathbf R}  \ket{\mathrm{vac}} = 0.
\end{align*}
We see that (\ref{eq:thouless0}) is indeed the many-body ground state of the BdG equations.

The states $| \theta \rangle$ are defined in the text using the gauge transformation $U_{{\mathbf R},{\mathbf K}} \rightarrow e^{i \theta \Theta_{\mathbf r}/2} U_{{\mathbf R},{\mathbf K}} $ and $V_{{\mathbf R},{\mathbf K}} \rightarrow e^{-i \theta \Theta_{\mathbf r}/2} V_{{\mathbf R},{\mathbf K}}$.  Given the form (\ref{eq:thouless0}) for $|0\rangle$, this implies
\stepcounter{equation}
\begin{equation}
{|\theta\rangle} = {\mathcal N} \exp \left[\frac{1}{2}\sum_{{\mathbf R},{\mathbf R}^\prime} e^{i \theta (\Theta_{\mathbf r}+ \Theta_{{\mathbf r}^\prime})/2} c^\dagger_{{\mathbf R}}  Z_{{\mathbf R},{\mathbf R}^\prime} c^\dagger_{{\mathbf R}^\prime}\right] \ket{\mathrm{vac}}.
\tag{S\arabic{equation}}\label{eq:thoulesstheta}
\end{equation}

\subsection*{Computational Solution of BdG equations}

Our formalism is compatible with detailed first principles approaches to solving the BdG equations.  However, in this paper, for simplicity we model the superconducting qubits using a rectangular lattice with lattice vectors ${\mathbf a}_x$, ${\mathbf a}_y$, and ${\mathbf a}_z$.   The number of tight-binding sites in the $x$, $y$, and $z$ directions are $N_x$, $N_y$, and $N_z$ respectively, giving rise to lengths $L_x = N_x |{\mathbf a}_x|$, $L_y = N_y |{\mathbf a}_y|$, and $L_z = N_z |{\mathbf a}_z|$.  Symmetries can facilitate solution of the BdG equations.  We employ periodic boundary conditions in the $x$ and $y$ directions; as a result of the translational symmetry, our self-consistent quantities (\ref{eq:hprime}) and (\ref{eq:Delta}) are homogeneous in the $x$ and $y$ directions.  

We choose a spin-independent interaction of the form
\stepcounter{equation}
\begin{align*}
& W({\mathbf r},\sigma;{\mathbf r}^\prime,\sigma^\prime)  \\
& =w(x-x^\prime,y-y^\prime,z-z^\prime)= \left\{ \begin{array}{cc}  -g & {\mathbf r} = {\mathbf r}^\prime \\ \frac{\lambda}{|{\mathbf r} - {\mathbf r}^\prime |} & {\mathbf r} \ne {\mathbf r}^\prime \end{array} \right.
 \tag{S\arabic{equation}}
 \label{eq:formofW}
 \end{align*}
 where the attractive Hubbard interaction at ${\mathbf r} = {\mathbf r}^\prime$ gives rise to superconductivity.
The BdG equations can be written as
\stepcounter{equation}
\begin{align*}
\epsilon_k U_{m_x,m_y,z,\sigma,\mathbf{K}} & =  \sum_{z^\prime,\sigma^\prime} h^\prime_{m_x,m_y,z,\sigma, z^\prime,\sigma^\prime}
U_{m_x,m_y, z^\prime,\sigma^\prime,\mathbf{K}}  \nonumber \\
&\!\!\!\! + \Delta_{m_x,m_y, z,\sigma,z^\prime,\sigma^\prime} V_{m_x,m_y, z^\prime,\sigma^\prime,\mathbf{K}} \\
\epsilon_k  V_{m_x,m_y, z,\sigma,\mathbf{K}} & =  \sum_{z^\prime,\sigma^\prime} (-h^\prime_{-m_x, -m_y, z,\sigma,z^\prime,\sigma^\prime})^*V_{m_x,m_y, z^\prime,\sigma^\prime,\mathbf{K}} \nonumber \\
&\!\!\!\! + (\Delta_{-m_x,-m_y, z,\sigma,z^\prime,\sigma^\prime})^*  U_{m_x,m_y, z^\prime,\sigma^\prime,\mathbf{K}}
\tag{S\arabic{equation}}
\label{eq:BdGmxmy}
\end{align*}
where the transverse momenta $\frac{2 \pi m_x}{L_x}$ and $\frac{2 \pi m_y}{L_y}$ are good quantum numbers and \stepcounter{equation}
\begin{align*}
U_{m_x,m_y,z,\sigma,\mathbf{K}}  & = \frac{1}{\sqrt{N_x N_y}} \sum_{x,y} e^{-i\frac{2\pi m_x x}{L_x}-i\frac{2\pi m_y y}{L_y}}U_{x,y,z,\sigma,{\mathbf K}}\\
V_{m_x,m_y,z,\sigma,\mathbf{K}}  & = \frac{1}{\sqrt{N_x N_y}} \sum_{x,y} e^{-i\frac{2\pi m_x x}{L_x}-i\frac{2\pi m_y y}{L_y}}V_{x,y,z,\sigma,{\mathbf K}}
\tag{S\arabic{equation}}
\end{align*}
with inverse relation
\begin{align*}
U_{x,y,z,\sigma,{\mathbf K}}  & = \frac{1}{\sqrt{N_x N_y}} \sum_{m_x,m_y} e^{i\frac{2\pi m_x x}{L_x}+i\frac{2\pi m_y y}{L_y}}U_{m_x,m_y,z,\sigma,\mathbf{K}}\\
V_{x,y,z,\sigma,{\mathbf K}}  & = \frac{1}{\sqrt{N_x N_y}} \sum_{m_x,m_y} e^{i\frac{2\pi m_x x}{L_x}+i\frac{2\pi m_y y}{L_y}}V_{m_x,m_y,z,\sigma,\mathbf{K}}.
\tag{S\arabic{equation}}
\end{align*}

To specify the forms of the operators on the right hand side of (\ref{eq:BdGmxmy}), we define 
\stepcounter{equation}
\begin{align*}
\frac{1}{N_x N_y} & \sum_{x,y,x^\prime,y^\prime}  e^{-i \frac{2 \pi \ell_x}{L_x} x - i\frac{2 \pi \ell_y}{L_y}y } e^{i \frac{2 \pi m_x}{L_x} x^\prime - i\frac{2 \pi m_y}{L_y}y^\prime }  \\
& \hspace{1.0in} h_{x,y,z,\sigma,x^\prime,y^\prime,z,\sigma^\prime}  \\
& \equiv \delta_{\ell_x,m_x}\delta_{\ell_y,m_y}h_{m_x,m_y,z,\sigma,z^\prime,\sigma^\prime}
\tag{S\arabic{equation}}
\end{align*}
where
\stepcounter{equation}
\begin{align*}
&h_{m_x,m_y,z,\sigma,z^\prime,\sigma^\prime} = \\
&\sum_{x,y} e^{-i \frac{2 \pi m_x}{L_x} (x-x^\prime) - i\frac{2 \pi m_y}{L_y}(y-y^\prime) } h_{x,y,z,\sigma,x^\prime,y^\prime,z,\sigma^\prime}.
\tag{S\arabic{equation}}
\end{align*}
This definition is reasonable because $h_{x,y,z,\sigma,x^\prime,y^\prime,z,\sigma^\prime}$ depends on $x$, $x^\prime$, $y$, and $y^\prime$ only via $x-x^\prime$ and $y-y^\prime$.  (The use of periodic boundary conditions in the $x$ and $y$ directions is essential for this translational invariance.) 
Similarly, we define
\stepcounter{equation}
\begin{align*}
\frac{1}{N_x N_y} & \sum_{x,y,x^\prime,y^\prime}  e^{-i \frac{2 \pi \ell_x}{L_x} x - i\frac{2 \pi \ell_y}{L_y}y } e^{i \frac{2 \pi m_x}{L_x} x^\prime - i\frac{2 \pi m_y}{L_y}y^\prime } \\
& \hspace{1.0in} w(x-x^\prime,y-y^\prime,z-z^\prime)  \\
& \equiv \delta_{\ell_x,m_x}\delta_{\ell_y,m_y}\tilde{w}(m_x,m_y,z-z^\prime).
\tag{S\arabic{equation}}
\end{align*}
Here, we have
\stepcounter{equation}
\begin{align*}
&\tilde{w}(m_x,m_y,z-z^\prime) = \\
&\sum_{x,y} e^{-i \frac{2 \pi m_x}{L_x} (x-x^\prime) - i\frac{2 \pi m_y}{L_y}(y-y^\prime) } w(x-x^\prime,y-y^\prime,z-z^\prime)  \\
& =N_x N_y s(m_x,m_y) e^{-2 \pi \sqrt{(m_x/L_x)^2 +(m_y/L_y)^2}|z-z^\prime|} \\
& - \sum_{n_x,n_y} s(n_x,n_y)\delta_{z,z^\prime}  - g \delta_{z,z^\prime} .
\tag{S\arabic{equation}}
\end{align*}
 with
\stepcounter{equation}
\begin{align*}
 s(m_x,m_y)  & =   \tag{S\arabic{equation}} \\
  \frac{\lambda}{L_xL_y} &  \left\{ \begin{array}{cc}  2 \sqrt{\pi L_x L_y} & m_x = m_y = 0 \\
 \frac{1}{\sqrt{(m_x/L_x)^2 +(m_y/L_y)^2}} & \text{ otherwise}
 \end{array} \right. 
\end{align*}
Then, 
\stepcounter{equation}
\begin{align*}
h^\prime_{m_x, m_y, z^\prime,\sigma^\prime,z,\sigma}  = & h_{m_x, m_y, z^\prime,\sigma^\prime,z,\sigma}  \nonumber \\
 + \delta_{z^\prime,z} \delta_{\sigma^\prime,\sigma} \sum_{z^{\prime\prime},\sigma^{\prime\prime}} &  \tilde{w}(0,0,z-z^{\prime\prime})  \times \nonumber \\
 \frac{1}{N_x N_y} \sum_{\ell_x,\ell_y,\mathbf K} & V_{\ell_x,\ell_y,z^{\prime\prime},\sigma^{\prime\prime},{\mathbf K}}  V^*_{\ell_x,\ell_y,z^{\prime\prime},\sigma^{\prime\prime},{\mathbf K}}  \nonumber \\
 -  \sum_{\ell_x,\ell_y}   \tilde{w} (m_x-\ell_x&,m_y-\ell_y,z-z^{\prime})  \times  \nonumber \\
 \frac{1}{N_x N_y} \sum_{\mathbf K} & V_{\ell_x,\ell_y,z,\sigma,{\mathbf K}}  V^*_{\ell_x,\ell_y,z^{\prime},\sigma^{\prime},{\mathbf K}}  
\tag{S\arabic{equation}}
\end{align*}
and
\stepcounter{equation}
\begin{align*}
\Delta_{m_x,m_y, z,\sigma,z^\prime,\sigma^\prime} & =   \sum_{\ell_x,\ell_y}   \tilde{w} (m_x+\ell_x ,m_y+\ell_y,z-z^{\prime})  \times  \nonumber \\
 \frac{1}{N_x N_y} \sum_{\mathbf K} & U_{\ell_x,\ell_y,z^{\prime},\sigma^{\prime},{\mathbf K}}  V^*_{\ell_x,\ell_y,z,\sigma,{\mathbf K}}.
 \tag{S\arabic{equation}}
\end{align*}
Alternatively, we also have
\stepcounter{equation}
\begin{align*}
 \Delta_{m_x,m_y, z,\sigma,z^\prime,\sigma^\prime} & =  - \sum_{\ell_x,\ell_y}   \tilde{w} (m_x-\ell_x ,m_y-\ell_y,z-z^{\prime})  \times  \nonumber \\
 \frac{1}{N_x N_y} \sum_{\mathbf K} & U_{\ell_x,\ell_y,z,\sigma,{\mathbf K}}  V^*_{\ell_x,\ell_y,z^{\prime},\sigma^{\prime},{\mathbf K}}.
\tag{S\arabic{equation}}
\end{align*}
The total energy of the system is given by
\stepcounter{equation}
\begin{align*}
\langle H \rangle = &\!\!\!\!\!\!\!\!\!\! \sum_{\smash{\ell_x,\ell_y,z,\sigma,z^\prime,\sigma^\prime}} \Big[ h_{\ell_x, \ell_y,z,\sigma, z^\prime,\sigma^\prime}  \rho_{\ell_x,\ell_y,z,\sigma,z^{\prime},\sigma^{\prime}} \nonumber \\
 + & \frac{1}{2 N_x N_y} \sum_{\smash{m_x,m_y}}\!\!\!     \tilde{w}(0,0,z-z^{\prime})  \nonumber \\
&  \rho_{\ell_x,\ell_y,z,\sigma,z,\sigma} \rho_{m_x,m_y,z^{\prime},\sigma^{\prime},z^{\prime},\sigma^{\prime}}  \nonumber \\
& -   \tilde{w}(\ell_x-m_x,\ell_y-m_y,z-z^{\prime})  \nonumber \\
&  \rho_{\ell_x,\ell_y,z^{\prime},\sigma^{\prime},z,\sigma} \rho_{m_x,m_y,z,\sigma,z^{\prime},\sigma^{\prime}} \nonumber \\
& +   \tilde{w}(\ell_x-m_x,\ell_y-m_y,z-z^{\prime})  \nonumber \\
&  \kappa^*_{\ell_x,\ell_y,z,\sigma,z^{\prime},\sigma^{\prime}} \kappa_{m_x,m_y,z,\sigma,z^{\prime},\sigma^{\prime}}  \Big].
\tag{S\arabic{equation}}
\label{eq:totalenergy}
\end{align*}
Here,
\stepcounter{equation}
\begin{align*}
& \rho_{x,y,z,\sigma,x^{\prime},y^{\prime},z^{\prime},\sigma^{\prime}} =\langle c^\dagger_{{\mathbf R}^\prime} c_{\mathbf R}\rangle \\
& \hspace{0.5in} = \sum_{\mathbf K}  V_{x^\prime,y^\prime,z^\prime,\sigma^\prime,{\mathbf K}}  V^*_{x,y,z,\sigma,{\mathbf K}} 
 \tag{S\arabic{equation}}
 \label{eq:rhodef}
\end{align*}
leading to
\stepcounter{equation}
\begin{align*}
& \rho_{\ell_x,\ell_y,z,\sigma,z^{\prime},\sigma^{\prime}}  \\
= & \sum_{x,y } e^{-i\frac{2\pi \ell_x}{L_x}(x-x^\prime) -i\frac{2\pi \ell_y}{L_y}(y-y^\prime)}  \rho_{x,y,z,\sigma,x^{\prime},y^{\prime},z^{\prime},\sigma^{\prime}}
\\
& =\sum_{\mathbf K}  V_{-\ell_x,-\ell_y,z^\prime,\sigma^\prime,{\mathbf K}}  V^*_{-\ell_x,-\ell_y,z,\sigma,{\mathbf K}}
 \tag{S\arabic{equation}}
 \end{align*}
with the minus signs in front of $\ell_x$ and $\ell_y$ in the final expression resulting from the flip in the order of the primed and unprimed variables in (\ref{eq:rhodef}).  Similarly,
\stepcounter{equation}
\begin{align*}
& \kappa_{x,y,z,\sigma,x^{\prime},y^{\prime},z^{\prime},\sigma^{\prime}} =\langle c_{{\mathbf R}^\prime} c_{\mathbf R}\rangle \\
& \hspace{0.5in} = \sum_{\mathbf K}  U_{x^\prime,y^\prime,z^\prime,\sigma^\prime,{\mathbf K}}  V^*_{x,y,z,\sigma,{\mathbf K}} 
 \tag{S\arabic{equation}}
 \label{eq:kappadef}
\end{align*}
and
\stepcounter{equation}
\begin{align*}
& \kappa_{\ell_x,\ell_y,z,\sigma,z^{\prime},\sigma^{\prime}}  \\
= & \sum_{x,y } e^{-i\frac{2\pi \ell_x}{L_x}(x-x^\prime) -i\frac{2\pi \ell_y}{L_y}(y-y^\prime)} 
\kappa_{x,y,z,\sigma,x^{\prime},y^{\prime},z^{\prime},\sigma^{\prime}}
\\
& =\sum_{\mathbf K}  U_{-\ell_x,-\ell_y,z^\prime,\sigma^\prime,{\mathbf K}}  V^*_{-\ell_x,-\ell_y,z,\sigma,{\mathbf K}}.
 \tag{S\arabic{equation}}
 \end{align*}
 
 In the paper, we perform computations on rectangular lattices with $N_x=N_y=5$ and $N_z=220$ sites.  In the charge qubit case, the tunneling matrix elements in (\ref{eq:T}) are set to $t_{{\mathbf r}+{\mathbf a},{\mathbf r}}=t$, defining the energy scale of the Hamiltonian.  In the RF squid qubit case, a phase is included in $t_{{\mathbf r}+{\mathbf a_z},{\mathbf r}} = t e^{-i \pi/2(220)}$ since a magnetic field threads the squid loop.  In both cases, at a plane in the middle of the lattice of sites, reduced tunneling matrix elements in the $z$ direction to give rise to the Josephson junction.  The reduced value is $t_{{\mathbf r}+{\mathbf a_z},{\mathbf r}}  = 0.1 t$ in the charge qubit case and $t_{{\mathbf r}+{\mathbf a_z},{\mathbf r}}  = 0.3te^{-i \pi/2(220)}$ in the RF squid qubit case.  The potential is set to $v({\mathbf r})=0$, and the chemical potential is set to $\mu = -3.42t$ in (\ref{eq:P}).    In (\ref{eq:formofW}), we insert a strong attractive interaction $g=2.25t$ to ensure a superconducting gap  of reasonable magnitude (see  Fig. \ref{fig:chargequbit}c and Fig. \ref{fig:RFsquidqubit}c) despite the nearly 1-dimensional geometry of the lattice.  The repulsive interaction was taken small enough that the results were insensitive to its specific value ($\lambda = 5 \times 10^{-5} t$).
 
\subsection{Derivation of Matrix Elements}

As described in the text, we need to compute the matrix elements that appear in the Schrodinger equations (\ref{eq:Hthetathetaprime}) and (\ref{eq:H2by2}).  For concreteness, we will focus on the charge qubit case (\ref{eq:Hthetathetaprime}); the RF squid case is very similar.  The desired matrix elements take the form ${\langle  \theta^{\prime\prime} | \mathcal{O} |\theta^\prime \rangle}={\langle  \theta^{\prime\prime} | \mathcal{O}(c^\dagger_{\mathbf{R}_1},c_{\mathbf{R}_2}) |\theta^\prime \rangle}$ where the right hand side makes explicit the dependence of $\mathcal{O}$ on the operators $c^\dagger_{\mathbf{R}_1}$ and  $c_{\mathbf{R}_2}$.  To compute these matrix elements, we follow the derivation in Ref. \cite{Ring2004}.

The states $|\theta^\prime \rangle$ and $|\theta^{\prime\prime} \rangle$ are defined in the text by performing gauge transformations from the BdG solutions of (\ref{eq:BdG}):
\stepcounter{equation}
\begin{align*}
U^\prime_{{\mathbf R},{\mathbf K}}  = e^{i \theta^\prime \Theta_{\mathbf r}/2} U_{{\mathbf R},{\mathbf K}} \hspace{0.2in} &
V^\prime_{{\mathbf R},{\mathbf K}} = e^{-i \theta^\prime \Theta_{\mathbf r}/2} V_{{\mathbf R},{\mathbf K}} \\
U^{\prime\prime}_{{\mathbf R},{\mathbf K}}  = e^{i \theta^{\prime\prime} \Theta_{\mathbf r}/2} U_{{\mathbf R},{\mathbf K}} \hspace{0.2in} &
V^{\prime\prime}_{{\mathbf R},{\mathbf K}} = e^{-i \theta^{\prime\prime} \Theta_{\mathbf r}/2} V_{{\mathbf R},{\mathbf K}}   \tag{S\arabic{equation}}.
\end{align*}
These equations imply that  $|\theta^\prime \rangle$  is annihilated by the quasiparticle operator $\beta_{\mathbf K}$, where
\stepcounter{equation}
\begin{align*}
c_{\mathbf{R}} = \sum_{\mathbf L} U^\prime_{\mathbf{R},{\mathbf L}}\beta_{\mathbf L} + (V^\prime)^* _{\mathbf{R},{\mathbf L}}\beta^\dagger_{\mathbf L} \\
c^\dagger_{\mathbf{R}} = \sum_{\mathbf L} (U^\prime)^*_{\mathbf{R},{\mathbf L}} \beta^\dagger_{\mathbf L} + V^\prime_{\mathbf{R},{\mathbf L}}\beta_{\mathbf L},
 \tag{S\arabic{equation}}
 \label{eq:Bogoliubovcbeta}
 \end{align*}
 and $|\theta^{\prime\prime} \rangle$ is annihilated by the quasiparticle operator $\gamma_{\mathbf K}$, where
 \stepcounter{equation}
\begin{align*}
c_{\mathbf{R}} = \sum_{\mathbf L} U^{\prime\prime}_{\mathbf{R},{\mathbf L}}\gamma_{\mathbf L} + (V^{\prime\prime})^* _{\mathbf{R},{\mathbf L}}\gamma^\dagger_{\mathbf L} \\
c^\dagger_{\mathbf{R}} = \sum_{\mathbf L} (U^{\prime\prime})^*_{\mathbf{R},{\mathbf L}} \gamma^\dagger_{\mathbf L} + V^{\prime\prime}_{\mathbf{R},{\mathbf L}}\gamma_{\mathbf L}.
 \tag{S\arabic{equation}}
 \label{eq:Bogoliubovcgamma}
 \end{align*}
The inverse of this last definition is
\stepcounter{equation}
\begin{align*}
& \gamma_\mathbf{K}  = \sum_\mathbf{R}  (U^{\prime\prime})^*_{\mathbf{R},\mathbf{K}}c_{\mathbf{R}} + (V^{\prime\prime})^*_{\mathbf{R},\mathbf{K}}c^\dagger_{\mathbf{R}} \\
& \gamma^\dagger_\mathbf{K}  = \sum_\mathbf{R}  U^{\prime\prime}_{\mathbf{R},\mathbf{K}}c^\dagger_{\mathbf{R}} + V^{\prime\prime}_{\mathbf{R},\mathbf{K}}c_{\mathbf{R}};
 \tag{S\arabic{equation}} \label{eq:gammabeta}
 \end{align*}
combined with (\ref{eq:Bogoliubovcbeta}), this implies
\stepcounter{equation}
\begin{align*}
&\gamma_\mathbf{K}  = \sum_\mathbf{L}  {\mathcal U}^*_{{\mathbf L},{\mathbf K}}\beta_{\mathbf{L}} + {\mathcal V}^*_{{\mathbf L},{\mathbf K}}\beta^\dagger_{\mathbf{L}}  \\
& \gamma^\dagger_\mathbf{K}  = \sum_\mathbf{L}  {\mathcal U}_{{\mathbf L},{\mathbf K}}\beta^\dagger_{\mathbf{L}} + {\mathcal V}_{{\mathbf L},{\mathbf K}}\beta_{\mathbf{L}} 
 \tag{S\arabic{equation}}
 \end{align*}
where $\mathcal{U} = (U^\prime)^\dagger U^{\prime\prime} + (V^\prime)^\dagger V^{\prime\prime}$ and $\mathcal{V} = (V^\prime)^T U^{\prime\prime}+ (U^\prime)^T V^{\prime\prime}$.

To proceed, we use Thouless's theorem, which asserts that Bogoliubov states $|\theta^{\prime\prime} \rangle$ and $|\theta^\prime\rangle$ are related by
\stepcounter{equation}
\begin{equation}
|\theta^{\prime\prime} \rangle = \langle\theta^\prime |\theta^{\prime\prime} \rangle  e^{\sum_{{\mathbf K},{\mathbf K}^\prime} \beta^\dagger_{{\mathbf K}} \mathcal{Z}_{{\mathbf K},{\mathbf K}^\prime}  \beta^\dagger_{{\mathbf K}^\prime}/2} |\theta^\prime \rangle.
 \tag{S\arabic{equation}}\label{eq:thetaprimeprime}
 \end{equation}
where $\mathcal{Z} = (\mathcal{V} \mathcal{U}^{-1})^*$.  The theorem assumes that $ \langle\theta^\prime |\theta^{\prime\prime} \rangle \ne 0$, which we will see implies that $\mathcal{U}^{-1}$, and therefore $\mathcal{Z}$, exists.  To prove Thouless's theorem, one needs to show that $\gamma_\mathbf{K}$ annihilates the expression on the right hand side of (\ref{eq:thetaprimeprime}).  The proof closely parallels the argument leading up to (\ref{eq:groundstateproof}) with $\alpha_\mathbf{K}$, $c_\mathbf{R}$, and $U_{\mathbf{R},\mathbf{K}}$, $V_{\mathbf{R},\mathbf{K}}$, and $Z_{\mathbf{R},\mathbf{R}^\prime}$ replaced by $\gamma_\mathbf{K}$, $\beta_\mathbf{L}$, ${\mathcal U}_{\mathbf{L},\mathbf{K}}$, ${\mathcal V}_{\mathbf{L},\mathbf{K}}$. and ${\mathcal Z}_{\mathbf{K},\mathbf{K}^\prime}$ respectively.  In the first step of the proof, we calculate
\stepcounter{equation}
\begin{equation}
e^{ \sum_{{\mathbf K},{\mathbf K}^\prime} \beta_{{\mathbf K}^\prime} \mathcal{Z}^*_{{\mathbf K},{\mathbf K}^\prime}  \beta_{{\mathbf K}}/2} \beta_{\mathbf L} e^{ -\sum_{{\mathbf K},{\mathbf K}^\prime} \beta_{{\mathbf K}^\prime} \mathcal{Z}^*_{{\mathbf K},{\mathbf K}^\prime}  \beta_{{\mathbf K}}/2} = \beta_{\mathbf L}
 \tag{S\arabic{equation}} \label{eq:betacommutator}
 \end{equation}
and
\stepcounter{equation}
\begin{align*}
& e^{ \sum_{{\mathbf K},{\mathbf K}^\prime} \beta_{{\mathbf K}^\prime} \mathcal{Z}^*_{{\mathbf K},{\mathbf K}^\prime}  \beta_{{\mathbf K}}/2} \beta^\dagger_{\mathbf L} e^{ -\sum_{{\mathbf K},{\mathbf K}^\prime} \beta_{{\mathbf K}^\prime} \mathcal{Z}^*_{{\mathbf K},{\mathbf K}^\prime}  \beta_{{\mathbf K}}/2} \\
& = \beta^\dagger_{\mathbf L} + \Big[\sum_{{\mathbf K},{\mathbf K}^\prime} \beta_{{\mathbf K}^\prime} \mathcal{Z}^*_{{\mathbf K},{\mathbf K}^\prime}  \beta_{{\mathbf K}}/2,\beta^\dagger_{\mathbf L}\Big] \\
& = \beta^\dagger_{\mathbf L}  - \sum_{{\mathbf K}^\prime} \beta_{{\mathbf K}^\prime}  \mathcal{Z}^*_{{\mathbf K}^\prime,{\mathbf L}}.
\tag{S\arabic{equation}} \label{eq:betadaggercommutator}
\end{align*}
by applying lemma (\ref{eq:commutelemma}).  Then, we employ the definition (\ref{eq:gammabeta}) to obtain the analogues of (\ref{eq:commutator}) and (\ref{eq:groundstateproof}).  Note that (\ref{eq:thetaprimeprime}) reduces to (\ref{eq:thouless0}) when we set $U^\prime$ equal to the identity matrix and $V^\prime$ equal to the zero matrix.

Thouless's theorem allows us to evaluate the overlap $\langle\theta^\prime |\theta^{\prime\prime} \rangle$.  To do this, we simplify our expression for $|\theta^{\prime\prime} \rangle$ into a BCS-like form.  The theorem of Bloch and Messiah states there are unitary matrices $C$ and $D$ for which $\bar{\mathcal U} = D^\dagger {\mathcal U}C^\dagger$ and $\bar{\mathcal V} = D^T  {\mathcal V} C^\dagger$ are real and take simple block-diagonal forms.  The matrix $\bar{\mathcal U}$ has $2 \times 2$ blocks of the form 
$u_{\mathbf J} \begin{bmatrix}1& 0 \\ 0 & 1 \end{bmatrix}$ in which the 2 states ${\mathbf J}$ and ${\bar{\mathbf J}}$ are paired with $u_{\mathbf J} = u_{\bar{\mathbf J}}$.  In the usual BCS case in which pairing occurs in the momentum basis, we would have ${\mathbf J} = {\mathbf k},\sigma$ and ${\bar{\mathbf J}}= -{\mathbf k},-\sigma$.  The matrix $\bar{\mathcal V}$ has $2\times 2$ blocks of the form  $v_{\mathbf J} \begin{bmatrix}0& 1 \\ -1 & 0 \end{bmatrix}$, involving the same 2 states ${\mathbf J}$ and ${\bar{\mathbf J}}$.  The coefficients satisfy the normalization condition  $u_{\mathbf J}^2 + v_{\mathbf J}^2 = 1$.  Defining  $\hat{\beta}^\dagger_{\mathbf J} = \sum_{\mathbf K} \beta^\dagger_{\mathbf K} D_{{\mathbf K},{\mathbf J}}$ in terms of the matrix $D$, we find that $\sum_{{\mathbf K},{\mathbf K}^\prime} \beta^\dagger_{{\mathbf K}} \mathcal{Z}_{{\mathbf K},{\mathbf K}^\prime}  \beta^\dagger_{{\mathbf K}^\prime}/2 = \sum_{{\mathbf J}} \hat{\beta}^\dagger_{{\mathbf J}} \frac{v_{\mathbf J}}{u_{\mathbf J}}  \hat{\beta}^\dagger_{\bar{\mathbf J}}$.  Thus,
\stepcounter{equation}
\begin{align*}
|\theta^{\prime\prime} \rangle & = \langle\theta^\prime |\theta^{\prime\prime} \rangle  \exp \Big( \sum_{{\mathbf J}} \hat{\beta}^\dagger_{{\mathbf J}} \frac{v_{\mathbf J}}{u_{\mathbf J}}  \hat{\beta}^\dagger_{\bar{\mathbf J}}\Big) |\theta^\prime \rangle \\
&=  \langle\theta^\prime |\theta^{\prime\prime} \rangle  \Pi_{\mathbf J}  \Big(1+ \hat{\beta}^\dagger_{{\mathbf J}} \frac{v_{\mathbf J}}{u_{\mathbf J}}  \hat{\beta}^\dagger_{\bar{\mathbf J}}\Big) |\theta^\prime \rangle \\
& = \langle\theta^\prime |\theta^{\prime\prime} \rangle  \Pi_{\mathbf J} \frac{1}{u_{\mathbf J}}      \Pi_{\mathbf J}  \Big(u_{\mathbf J}+ \hat{\beta}^\dagger_{{\mathbf J}} v_{\mathbf J}  \hat{\beta}^\dagger_{\bar{\mathbf J}}\Big) |\theta^\prime \rangle.
 \tag{S\arabic{equation}}
 \end{align*}
The normalization condition $\langle\theta^{\prime\prime} |\theta^{\prime\prime} \rangle=1$ then implies $\left|\langle \theta^\prime|\theta^{\prime\prime}\rangle\right|^2 = \Pi_{\mathbf J} u_{\mathbf J}^2= \mathrm{det }\,\, {\mathcal U}$.  The Onishi formula $\left|\langle \theta^\prime|\theta^{\prime\prime}\rangle\right|^2 = \mathrm{det }\,\, {\mathcal U}$ leaves the phase of $\langle \theta^\prime|\theta^{\prime\prime}\rangle$ undetermined; by choosing the phases of $U_{{\mathbf R},{\mathbf K}}$ and $V_{{\mathbf R},{\mathbf K}}$ appropriately, we can ensure that $\langle \theta^\prime|\theta^{\prime\prime}\rangle$ is real. Then, the correct sign of $\langle \theta^\prime|\theta^{\prime\prime}\rangle = \pm\sqrt{ \det U}$ is fixed using continuity in $\theta^{\prime\prime}$ starting with $\langle \theta^ {\prime}|\theta^{\prime\prime}\rangle = 1$ at $\theta^{\prime\prime} = \theta^{\prime}$.

Using Thouless's theorem, we can also calculate matrix elements of the form
\stepcounter{equation}
\begin{align*}
\langle  & \theta^{\prime\prime} | \mathcal{O} |\theta^\prime \rangle/ \langle\theta^{\prime\prime} |\theta^\prime \rangle = \langle  \theta^{\prime\prime} | \mathcal{O}(c^\dagger_{\mathbf{R}_1},c_{\mathbf{R}_2}) |\theta^\prime \rangle/ \langle\theta^{\prime\prime} |\theta^\prime \rangle = \\
& \langle \theta^\prime | e^{ \sum_{{\mathbf K},{\mathbf K}^\prime} \beta_{{\mathbf K}^\prime} \mathcal{Z}^*_{{\mathbf K},{\mathbf K}^\prime}  \beta_{{\mathbf K}}/2} \mathcal{O}(c^\dagger_{\mathbf{R}_1},c_{\mathbf{R}_2}) |\theta^\prime\rangle = \\
&  \langle \theta^\prime | e^{ \sum_{{\mathbf K},{\mathbf K}^\prime} \beta_{{\mathbf K}^\prime} \mathcal{Z}^*_{{\mathbf K},{\mathbf K}^\prime}  \beta_{{\mathbf K}}/2} \mathcal{O}(c^\dagger_{\mathbf{R}_1},c_{\mathbf{R}_2}) \\
& \hspace{1.25in} e^{ - \sum_{{\mathbf K},{\mathbf K}^\prime} \beta_{{\mathbf K}^\prime} \mathcal{Z}^*_{{\mathbf K},{\mathbf K}^\prime}  \beta_{{\mathbf K}}/2} |\theta^\prime\rangle =\\
&  \langle \theta^\prime |  \mathcal{O}(e^{ \sum_{{\mathbf K},{\mathbf K}^\prime} \beta_{{\mathbf K}^\prime} \mathcal{Z}^*_{{\mathbf K},{\mathbf K}^\prime}  \beta_{{\mathbf K}}/2} c^\dagger_{\mathbf{R}_1} e^{ -\sum_{{\mathbf K},{\mathbf K}^\prime} \beta_{{\mathbf K}^\prime} \mathcal{Z}^*_{{\mathbf K},{\mathbf K}^\prime}  \beta_{{\mathbf K}}/2}, \\
& \hspace{0.2in} e^{ \sum_{{\mathbf K},{\mathbf K}^\prime} \beta_{{\mathbf K}^\prime} \mathcal{Z}^*_{{\mathbf K},{\mathbf K}^\prime}  \beta_{{\mathbf K}}/2} c_{\mathbf{R}_2} e^{ -\sum_{{\mathbf K},{\mathbf K}^\prime} \beta_{{\mathbf K}^\prime} \mathcal{Z}^*_{{\mathbf K},{\mathbf K}^\prime}  \beta_{{\mathbf K}}/2}) |\theta^\prime\rangle  \tag{S\arabic{equation}}
\end{align*}
where we have inserted $e^{ \sum_{{\mathbf K},{\mathbf K}^\prime} \beta_{{\mathbf K}^\prime} \mathcal{Z}^*_{{\mathbf K},{\mathbf K}^\prime}  \beta_{{\mathbf K}}/2} e^{-\sum_{{\mathbf K},{\mathbf K}^\prime} \beta_{{\mathbf K}^\prime} \mathcal{Z}^*_{{\mathbf K},{\mathbf K}^\prime}  \beta_{{\mathbf K}}/2}=1$ as needed between adjacent operators in $\mathcal{O}$ to derive the final line.

To proceed, we insert (\ref{eq:betacommutator}) and (\ref{eq:betadaggercommutator}) into the definitions of the Bogoliubov operators (\ref{eq:Bogoliubovcbeta}), obtaining
\stepcounter{equation}
\begin{align*}
& e^{ \sum_{{\mathbf K},{\mathbf K}^\prime} \beta_{{\mathbf K}^\prime} \mathcal{Z}^*_{{\mathbf K},{\mathbf K}^\prime}  \beta_{{\mathbf K}}/2} c_{\mathbf{R}}e^{ -\sum_{{\mathbf K},{\mathbf K}^\prime} \beta_{{\mathbf K}^\prime} \mathcal{Z}^*_{{\mathbf K},{\mathbf K}^\prime}  \beta_{{\mathbf K}}/2} \\
& = \sum_{\mathbf L} U^\prime_{\mathbf{R},{\mathbf L}} \beta_{\mathbf L} + (V^\prime)^* _{\mathbf{R},{\mathbf L}} (\beta^\dagger_{\mathbf L}  - \sum_{{\mathbf K}^\prime}\beta_{{\mathbf K}^\prime}  \mathcal{Z}^*_{{\mathbf K}^\prime,{\mathbf L}})
\tag{S\arabic{equation}}
\end{align*}
and
\stepcounter{equation}
\begin{align*}
& e^{ \sum_{{\mathbf K},{\mathbf K}^\prime} \beta_{{\mathbf K}^\prime} \mathcal{Z}^*_{{\mathbf K},{\mathbf K}^\prime}  \beta_{{\mathbf K}}/2} c^\dagger_{\mathbf{R}}e^{ -\sum_{{\mathbf K},{\mathbf K}^\prime} \beta_{{\mathbf K}^\prime} \mathcal{Z}^*_{{\mathbf K},{\mathbf K}^\prime}  \beta_{{\mathbf K}}/2} \\
& = \sum_{\mathbf L} (U^\prime)^*_{\mathbf{R},{\mathbf L}} (\beta^\dagger_{\mathbf L}  -  \sum_{{\mathbf K}^\prime} \beta_{{\mathbf K}^\prime}  \mathcal{Z}^*_{{\mathbf K}^\prime,{\mathbf L}})+ V^\prime _{\mathbf{R},{\mathbf L}} \beta_{\mathbf L}. 
\tag{S\arabic{equation}}
\end{align*}

It follows that 
\stepcounter{equation}
\begin{align*}
\langle  & \theta^{\prime\prime} | c^\dagger_{\mathbf{R}_1} c_{\mathbf{R}_2} |\theta^\prime \rangle/ \langle\theta^{\prime\prime} |\theta^\prime \rangle \\
& = \sum_{{\mathbf L},{\mathbf L}^\prime} (V^\prime _{\mathbf{R}_1,{\mathbf L}} \delta_{{\mathbf L},{\mathbf L}^\prime}-(U^\prime)^*_{\mathbf{R}_1,{\mathbf L}} \mathcal{Z}^*_{{\mathbf L}^\prime,{\mathbf L}})(V^\prime)^* _{\mathbf{R}_2,{\mathbf L}^\prime} \\
& = (V^\prime (V^\prime)^\dagger - (U^\prime)^* \mathcal{Z}^\dagger (V^\prime)^\dagger)_{\mathbf{R}_1,\mathbf{R}_2} \\
& = (V^\prime \mathcal{U} \frac{1}{\mathcal{U}}  (V^\prime)^\dagger + (U^\prime)^* \mathcal{V} \frac{1}{\mathcal{U}}  (V^\prime)^\dagger)_{\mathbf{R}_1,\mathbf{R}_2} \\
& = (V^\prime((U^\prime)^\dagger U^{\prime\prime} + (V^\prime)^\dagger V^{\prime\prime}) \frac{1}{\mathcal{U}} (V^\prime)^\dagger \\
& \hspace{0.25in} + (U^\prime)^*((V^\prime)^T U^{\prime\prime}+ (U^\prime)^T V^{\prime\prime}) \frac{1}{\mathcal{U}}  (V^\prime)^\dagger)_{\mathbf{R}_1,\mathbf{R}_2} \\
&= ((V^\prime)^* \frac{1}{\mathcal{U^T}}(V^{\prime\prime})^T)_{\mathbf{R}_2,\mathbf{R}_1}.
\tag{S\arabic{equation}}
\end{align*}
The third equality uses $\mathcal{Z} = - \mathcal{Z}^T$, and the final equality is derived using the orthogonality relations $(U^\prime)^*(V^\prime)^T + V^\prime (U^\prime)^\dagger = 0$ and $(V^\prime)^*(V^\prime)^T+U^\prime (U^\prime)^\dagger = I$.
Similarly, 
\stepcounter{equation}
\begin{align*}
\langle  & \theta^{\prime\prime} | c^\dagger_{\mathbf{R}_1} c^\dagger_{\mathbf{R}_2} |\theta^\prime \rangle/ \langle\theta^{\prime\prime} |\theta^\prime \rangle = -\langle  \theta^{\prime\prime} | c^\dagger_{\mathbf{R}_2} c^\dagger_{\mathbf{R}_1} |\theta^\prime \rangle/ \langle\theta^{\prime\prime} |\theta^\prime \rangle \\
 & = -(V^\prime (U^\prime)^\dagger - (U^\prime)^* \mathcal{Z}^\dagger (U^\prime)^\dagger)_{\mathbf{R}_2,\mathbf{R}_1} \\
 & = - ((U^\prime)^* \frac{1}{\mathcal{U}^T} (V^{\prime\prime})^T)_{\mathbf{R}_1,\mathbf{R}_2}  \tag{S\arabic{equation}}
\end{align*}
and
\stepcounter{equation}
\begin{align*}
\langle  & \theta^{\prime\prime} | c_{\mathbf{R}_3} c_{\mathbf{R}_4} |\theta^\prime \rangle/ \langle\theta^{\prime\prime} |\theta^\prime \rangle \\
 & = (U^\prime (V^\prime)^\dagger - (V^\prime)^* \mathcal{Z}^\dagger (V^\prime)^\dagger)_{\mathbf{R}_3,\mathbf{R}_4} \\
 & = ((V^\prime)^* \frac{1}{\mathcal{U}^T} (U^{\prime\prime})^T)_{\mathbf{R}_4,\mathbf{R}_3}. \tag{S\arabic{equation}}
\end{align*}
Finally, we have
\stepcounter{equation}
\begin{align*}
& \langle  \theta^{\prime\prime} | c^\dagger_{\mathbf{R}_1}c^\dagger_{\mathbf{R}_2} c_{\mathbf{R}_3}c_{\mathbf{R}_4} |\theta^\prime \rangle/ \langle\theta^{\prime\prime} |\theta^\prime \rangle \\
& = (V^\prime (V^\prime)^\dagger - (U^\prime)^* \mathcal{Z}^\dagger (V^\prime)^\dagger)_{\mathbf{R}_1,\mathbf{R}_4} \\
&\hspace{0.4in}(V^\prime (V^\prime)^\dagger - (U^\prime)^* \mathcal{Z}^\dagger (V^\prime)^\dagger)_{\mathbf{R}_2,\mathbf{R}_3}\\
&  - (V^\prime (V^\prime)^\dagger - (U^\prime)^* \mathcal{Z}^\dagger (V^\prime)^\dagger)_{\mathbf{R}_1,\mathbf{R}_3}\\
& \hspace{0.4in}(V^\prime (V^\prime)^\dagger - (U^\prime)^* \mathcal{Z}^\dagger (V^\prime)^\dagger)_{\mathbf{R}_2,\mathbf{R}_4}\\
& + (V^\prime (U^\prime)^\dagger - (U^\prime)^* \mathcal{Z}^\dagger (U^\prime)^\dagger)_{\mathbf{R}_1,\mathbf{R}_2} \\
&\hspace{0.4in} (U^\prime (V^\prime)^\dagger - (V^\prime)^* \mathcal{Z}^\dagger  (V^\prime)^\dagger)_{\mathbf{R}_3,\mathbf{R}_4} \\
& = ((V^\prime)^* \frac{1}{\mathcal{U^T}}(V^{\prime\prime})^T)_{\mathbf{R}_4,\mathbf{R}_1} ((V^\prime)^* \frac{1}{\mathcal{U^T}}(V^{\prime\prime})^T)_{\mathbf{R}_3,\mathbf{R}_2}\\
&  - ((V^\prime)^* \frac{1}{\mathcal{U^T}}(V^{\prime\prime})^T)_{\mathbf{R}_3,\mathbf{R}_1} ((V^\prime)^* \frac{1}{\mathcal{U^T}}(V^{\prime\prime})^T)_{\mathbf{R}_4,\mathbf{R}_2}\\
& - ((U^\prime)^* \frac{1}{\mathcal{U^T}}(V^{\prime\prime})^T)_{\mathbf{R}_1,\mathbf{R}_2}  ((V^\prime)^* \frac{1}{\mathcal{U^T}}(U^{\prime\prime})^T)_{\mathbf{R}_4,\mathbf{R}_3}.
\tag{S\arabic{equation}}
\end{align*}

To compute $\langle  \theta^{\prime\prime} |H | \theta^\prime \rangle/\langle \theta^{\prime\prime}|\theta^\prime\rangle$, we use  expression (\ref{eq:totalenergy}) but replace $\rho$ with 
\stepcounter{equation}
\begin{align*}
& \bar{\rho}_{\ell_x,\ell_y,z,\sigma,z^{\prime},\sigma^{\prime}}   = \tag{S\arabic{equation}}\\
 & \sum_{{\mathbf K},{\mathbf K}^\prime}  V^{\prime\prime}_{-\ell_x,-\ell_y,z^\prime,\sigma^\prime,{\mathbf K}^\prime} \left( \frac{1}{\mathcal{U}}\right)_{{\mathbf K}^\prime,{\mathbf K}} \Big(V^\prime_{-\ell_x,-\ell_y,z,\sigma,{\mathbf K}}\Big)^*
\end{align*}
and $\kappa$ with 
\stepcounter{equation}
\begin{align*}
& \bar{\kappa}_{\ell_x,\ell_y,z,\sigma,z^{\prime},\sigma^{\prime}}   = \tag{S\arabic{equation}}\\
 & \sum_{{\mathbf K},{\mathbf K}^\prime}  U^{\prime\prime}_{-\ell_x,-\ell_y,z^\prime,\sigma^\prime,{\mathbf K}^\prime} \left( \frac{1}{\mathcal{U}}\right)_{{\mathbf K}^\prime,{\mathbf K}} \Big(V^\prime_{-\ell_x,-\ell_y,z,\sigma,{\mathbf K}}\Big)^*.
\end{align*}

\subsection{Fixing the Order Parameter Phase Additive Constant}

Note that any self-consistent solution of the Bogoliubov-de Gennes equations remains a self-consistent solution under the transformation
\begin{align*}
\Delta_{{\mathbf R},{\mathbf R}^\prime} \rightarrow e^{i \xi}  \Delta_{{\mathbf R},{\mathbf R}^\prime}\\
U_{{\mathbf R},{\mathbf K}} \rightarrow e^{i \xi/2} U_{{\mathbf R},{\mathbf K}} \\
V_{{\mathbf R},{\mathbf K}} \rightarrow e^{-i \xi/2} U_{{\mathbf R},{\mathbf K}},
\end{align*}
where $\xi$ is a real constant.
Because $Z \equiv (V U^{-1})^*$, the ground state (\ref{eq:thouless0}) becomes
\stepcounter{equation}
\begin{equation}
{|0,\xi\rangle} = {\mathcal N} \exp \left[\frac{1}{2}\sum_{{\mathbf R},{\mathbf R}^\prime} e^{i\xi} c^\dagger_{{\mathbf R}} Z_{{\mathbf R},{\mathbf R}^\prime} c^\dagger_{{\mathbf R}^\prime}\right] \ket{\mathrm{vac}}.
\label{eq:thouless0xi}
\tag{S\arabic{equation}}
\end{equation}
Expanding the exponential, one sees that ${|0,\xi\rangle}$ is a superposition
\begin{align*}
{|0,\xi\rangle} = \sum_{\mathfrak n} e^{i{\mathfrak n} \xi} {|0,{\mathfrak n}\rangle}
\end{align*}
where ${|0,{\mathfrak n}\rangle} =  \int_0^{2 \pi} \frac{d\xi}{2\pi} e^{-in \xi} {|0,\xi\rangle} $ is an unnormalized state with ${\mathfrak n}$ total pairs ($2{\mathfrak n}$ total particles) in the system.  Clearly, $\xi$ shows up as the relative phase between states with different numbers of particles.  If a superconducting system with a fixed number of pairs $\bar{\mathfrak n}$ is modeled using the Bogoliubov-de Gennes equations, the state of the system can be described by one of the terms in the sum, i.e. ${|0,\bar{\mathfrak n}\rangle}/\sqrt{{\langle 0,\bar{\mathfrak n}}|{0,\bar{\mathfrak n}\rangle}}$.  For mathematical convenience, we often calculate physical properties of the system using the entire superposition ${|0,\xi\rangle}$.  However, correct physical predictions about the system cannot depend on the value of $\xi$.

It therefore is jarring that adding a constant to the rescaled order parameter phase (\ref{eq:Thetachargequbit}) does in fact alter the predictions of (\ref{eq:Hthetathetaprime}).  Indeed, taking $\Theta_{\mathbf r} \rightarrow \Theta_{\mathbf r} + \xi$ multiplies each $c^\dagger_{\mathbf R}$ in (\ref{eq:thoulesstheta}) by an extra factor of $e^{i \theta \xi/2}$.  Comparing the definition (\ref{eq:thouless0xi}), we see that $| \theta\rangle$ has changed to $|\theta, \theta \xi  \rangle$.   If we expand
\begin{align*}
{|\theta,\xi \theta\rangle} = \sum_{\mathfrak n} e^{i{\mathfrak n}\theta \xi} {|\theta,{\mathfrak n}\rangle},
\end{align*}
the matrix elements in (\ref{eq:Hthetathetaprime}) depend on $\xi$ as 
\stepcounter{equation}
\begin{align*}
\langle \theta,\theta \xi| \theta^\prime,\theta^\prime\xi  \rangle =  \sum_{\mathfrak n} e^{i {\mathfrak n} (\theta-\theta^\prime)\xi} {\langle \theta,{\mathfrak n}|}{\theta^\prime,{\mathfrak n}\rangle} \\
\langle \theta,\theta \xi |H| \theta^\prime,\theta^\prime \xi \rangle =  \sum_{\mathfrak n} e^{i {\mathfrak n} (\theta-\theta^\prime)\xi} {\langle \theta,{\mathfrak n}|}H{|\theta^\prime,{\mathfrak n}\rangle}.
\label{eq:sumovern}
\tag{S\arabic{equation}}
\end{align*}
In light of this expansion, one realizes that, for a system with a fixed number of pairs $\bar{\mathfrak n}$, one would actually like to solve a modified version of (\ref{eq:Hthetathetaprime}) with fixed particle number
\stepcounter{equation}
\begin{equation}
\sum_{\theta} \langle \theta, \bar{\mathfrak n} | H | \theta^\prime, \bar{\mathfrak n} \rangle \psi(\theta^\prime) = E \sum_{\theta} \langle \theta, \bar{\mathfrak n} | \theta^\prime, \bar{\mathfrak n} \rangle \psi(\theta^\prime).
\label{eq:Hthetathetaprimefixedn}
\tag{S\arabic{equation}}
\end{equation}
This equation does not depend on $\xi$, as required physically.

\begin{figure}
\begin{tabular}{cc}
(a) \begin{tabular}{c} \includegraphics[width=1.5in]{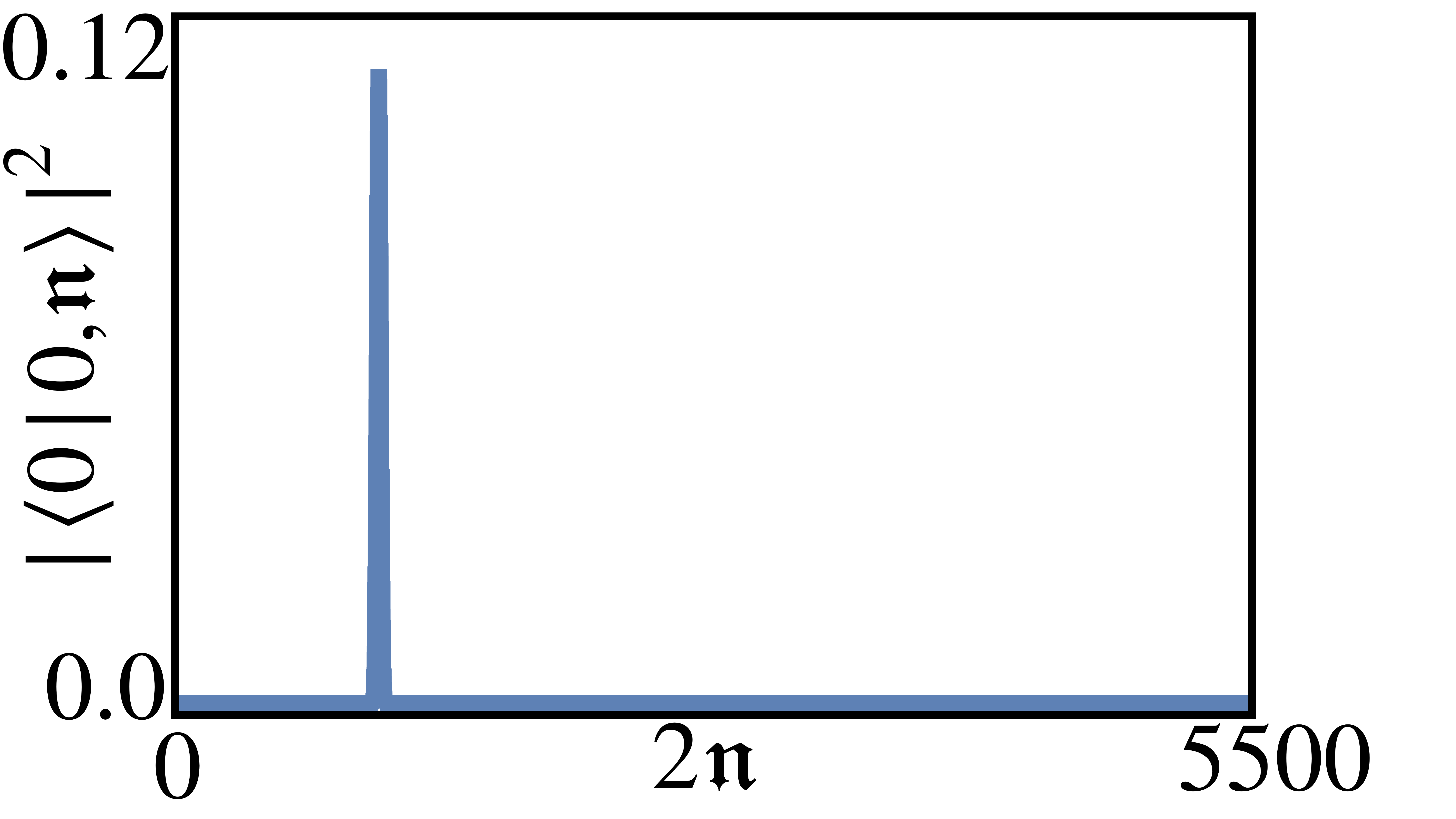} \end{tabular}&
(b) \begin{tabular}{c} \includegraphics[width=1.5in]{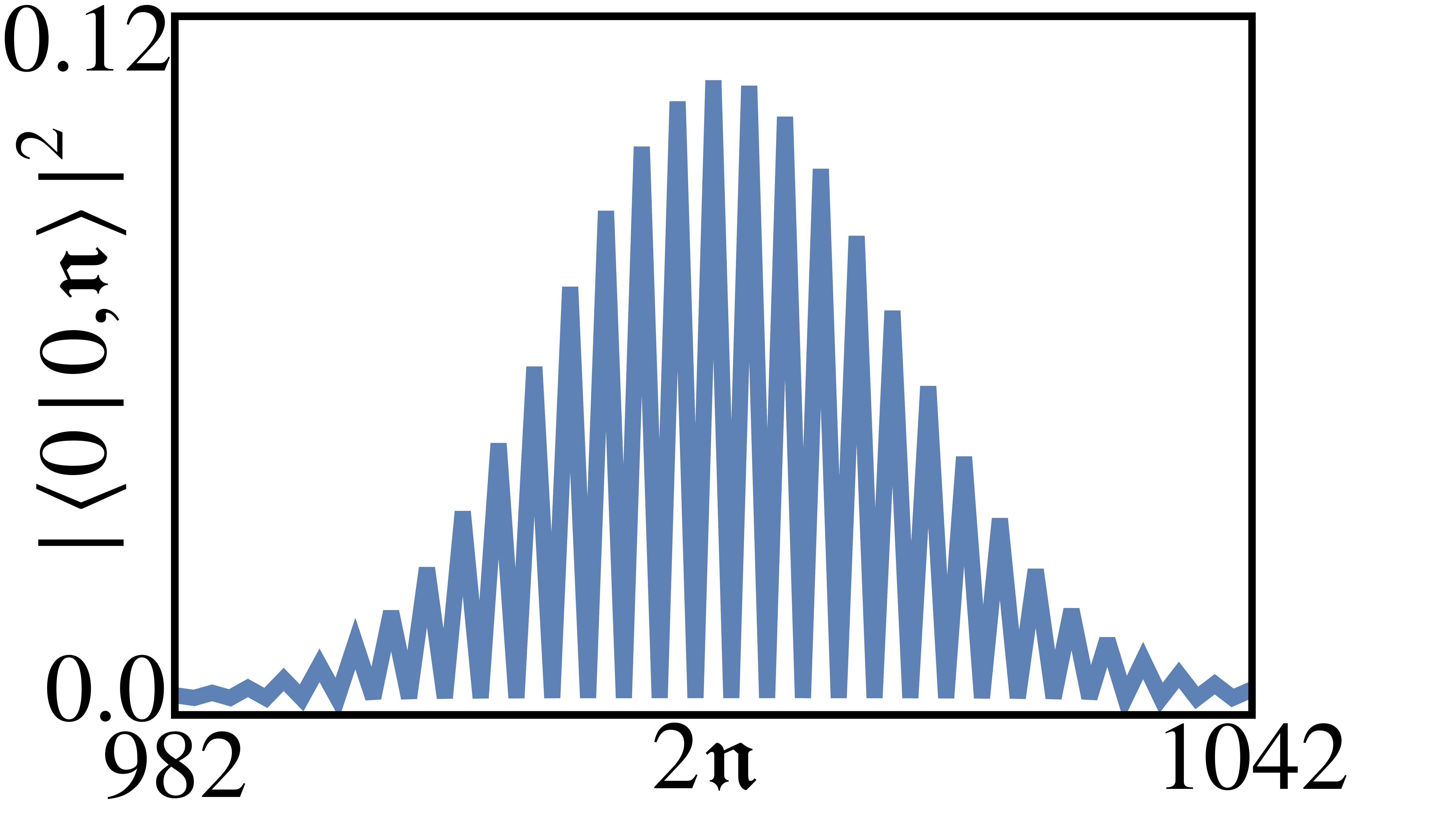} \end{tabular}
\end{tabular}
\caption{(a) Plot of overlap $|{\langle 0|}{0,{\mathfrak n}\rangle}|^2$ versus number of particles $2{\mathfrak n}$ for charge qubit parameters detailed in main text.  Strong peak is evident near $2{\mathfrak n} = 1012$ particles.  (b) Expanded plot in the range $2{\mathfrak n} = 982$ to $1042$ particles shows shape of peak.}
 \label{fig:overlapversusnumber}
\end{figure}

Nevertheless, the original equation (\ref{eq:Hthetathetaprime}) can be used provided that we choose a specific value of $\xi$ satisfying
\stepcounter{equation}
\begin{align*}
\langle \theta,\theta \xi| \theta^\prime,\theta^\prime\xi  \rangle & \approx \frac{e^{i \bar{\mathfrak n} (\theta-\theta^\prime)\xi}{\langle \theta,\bar{\mathfrak n}|}{\theta^\prime,\bar{\mathfrak n}\rangle}}{\sqrt{{\langle \theta,\bar{\mathfrak n}|}{\theta,\bar{\mathfrak n}\rangle} {\langle \theta^\prime,\bar{\mathfrak n}|}{\theta^\prime,\bar{\mathfrak n}\rangle}}} \\
& = \frac{e^{i \bar{\mathfrak n} (\theta-\theta^\prime)\xi}{\langle \theta,\bar{\mathfrak n}|}{\theta^\prime,\bar{\mathfrak n}\rangle}}{\langle 0,\bar{\mathfrak n}|0,\bar{\mathfrak n}\rangle}\\
\langle \theta,\theta \xi|H| \theta^\prime,\theta^\prime\xi  \rangle & \approx \frac{e^{i \bar{\mathfrak n} (\theta-\theta^\prime)\xi}{\langle \theta,\bar{\mathfrak n}|}H{|\theta^\prime,\bar{\mathfrak n}\rangle}}{\sqrt{{\langle \theta,\bar{\mathfrak n}|}{\theta,\bar{\mathfrak n}\rangle} {\langle \theta^\prime,\bar{\mathfrak n}|}{\theta^\prime,\bar{\mathfrak n}\rangle}}} \\
& = \frac{e^{i \bar{\mathfrak n} (\theta-\theta^\prime)\xi}{\langle \theta,\bar{\mathfrak n}|}H{|\theta^\prime,\bar{\mathfrak n}\rangle}}{\langle 0,\bar{\mathfrak n}|0,\bar{\mathfrak n}\rangle},
\label{eq:condition}
\tag{S\arabic{equation}}
\end{align*}
for $\bar{\mathfrak n}$ equal to the number of pairs in the system.  If this condition is satisfied, (\ref{eq:Hthetathetaprime}) approximately reduces to (\ref{eq:Hthetathetaprimefixedn}).  To satisfy (\ref{eq:condition}), in this paper we choose $\xi$ such that (\ref{eq:Thetachargequbit}) is an antisymmetric function of $z$, as in Fig. \ref{fig:chargequbit}c.  This implies that $e^{i {\mathfrak n} (\theta-\theta^\prime)\xi}{\langle \theta,{\mathfrak n}|}{\theta^\prime,{\mathfrak n}\rangle}$ is real for all ${\mathfrak n}$:
\begin{align*}
\Big(e^{i {\mathfrak n} (\theta-\theta^\prime)\xi}{\langle \theta,{\mathfrak n}|}{\theta^\prime,{\mathfrak n}\rangle}\Big)^* & = e^{- i {\mathfrak n} \theta\xi}{\langle -\theta,{\mathfrak n}|}{-\theta^\prime,{\mathfrak n}\rangle}e^{i {\mathfrak n} \theta^\prime \xi} \\
& \hspace{-0.25in} = e^{i {\mathfrak n} \theta\xi} {\langle \theta,{\mathfrak n}|} \hat{R}^\dagger \hat{R}{|\theta^\prime,{\mathfrak n}\rangle}e^{-i {\mathfrak n} \theta^\prime \xi} \\
& \hspace{-0.25in} = e^{i {\mathfrak n} (\theta-\theta^\prime)\xi}{\langle \theta,{\mathfrak n}|}{\theta^\prime,{\mathfrak n}\rangle}.
\end{align*}
In the second equality, we introduced an operator $\hat{R}$ that takes $z$ to $-z$, and we used the $z\leftrightarrow -z$ reflection symmetry of the problem about the Josephson junction.  The third equality uses the fact that two reflections $\hat{R}^\dagger \hat{R}$ in succession produce an identity operation.  Similarly, our choice of $\xi$ ensures that $e^{i {\mathfrak n} (\theta-\theta^\prime)\xi}{\langle \theta,{\mathfrak n}|}H|{\theta^\prime,{\mathfrak n}\rangle}$ is real for all ${\mathfrak n}$:
\begin{align*}
&\Big(e^{i {\mathfrak n} (\theta-\theta^\prime)\xi}{\langle \theta,{\mathfrak n}|}H|{\theta^\prime,{\mathfrak n}\rangle}\Big)^* \\
&= e^{i {\mathfrak n} (-\theta+\theta^\prime)\xi}{\langle -\theta,{\mathfrak n}|}H^*|{-\theta^\prime,{\mathfrak n}\rangle} \\
& = e^{i {\mathfrak n} (\theta-\theta^\prime)\xi}{\langle \theta,{\mathfrak n}|}\hat{R}^\dagger H^* \hat{R}|{\theta^\prime,{\mathfrak n}\rangle} \\
& = e^{i {\mathfrak n} (\theta-\theta^\prime)\xi}{\langle \theta,{\mathfrak n}|}H|{\theta^\prime,{\mathfrak n}\rangle}.
\end{align*}
Now, numerically, we find that the terms in (\ref{eq:sumovern}) are strongly peaked as a function of ${\mathfrak n}$ about ${\mathfrak n}=\bar{\mathfrak n}$.  A sample calculation, performed using the charge qubit parameters described in the main text, is shown in Fig. \ref{fig:overlapversusnumber} for $e^{i {\mathfrak n} (\theta-\theta^\prime)\xi}{\langle \theta,{\mathfrak n}|}{\theta^\prime,{\mathfrak n}\rangle}$ in the case $\theta = \theta^\prime=0$.  Since they are real and do not have a rapidly varying phase, the terms with ${\mathfrak n}$ close to $\bar{\mathfrak n}$ add constructively and determine the value of the sum.  As a result, (\ref{eq:condition}) is satisfied: we have numerically verified agreement to within $\sim1\%$ for the overlap $\langle \theta,\theta \xi| \theta^\prime,\theta^\prime\xi  \rangle$ and for the one-body operators in the Hamiltonian $\langle \theta,\theta \xi|c^\dagger_{\mathbf R} c_{{\mathbf R}^\prime} | \theta^\prime,\theta^\prime\xi  \rangle$ at a few choices of $\theta$, $\theta^\prime$, ${\mathbf R}$, and ${{\mathbf R}^\prime}$.  (We expect that (\ref{eq:condition}) would also be satisfied for the two-body operators in the Hamiltonian but did not check this numerically.). Had we chosen  $\xi$ injudiciously, there would have been differences in phase leading to cancellations within the sums (\ref{eq:sumovern}).  As a result, (\ref{eq:condition}) would not be well satisfied.

Although this discussion has focused on the case of the charge qubit, analogous remarks apply to the RF squid qubit.  The rescaled phase function (\ref{eq:Theta}) is chosen so it is antisymmetric on $z$, as shown in Fig. \ref{fig:RFsquidqubit}c.  This allows us to get physically relevant results by solving (\ref{eq:H2by2}) without projecting on to states of fixed particle number.

\subsection{Definition of Orthonormal Basis}

In the limit of large systems, the states $|\theta\rangle$ approach orthonormality.  This follows from the Onishi formula $ \left|\langle \theta|\theta^\prime\rangle\right|^2 = \mathrm{det }\,\, {\mathcal U}$ in the limit in which the matrix (\ref{eq:Uchargequbit}) becomes large.  To solve (\ref{eq:Hthetathetaprime}) for a finite system, though, it is convenient to construct an orthonormal basis explicitly.  Our non-orthonormal basis, defined in the main text, comprises states $|\theta\rangle$, where $-\theta_{max} < \theta \le \theta_{max}$.   The value of $\theta_{max}$ is determined by $|\theta\rangle = |\theta + 2 \theta_{max}\rangle$; based on (\ref{eq:thoulesstheta}) a sufficient condition is 
\stepcounter{equation}  
\begin{align*}
e^{i 2 \theta_{max}(\Theta_{\mathbf r}+\Theta_{\mathbf r^\prime})/2}= 1 \hspace{0.1in} \forall \hspace{0.1in} {\mathbf r} \text{ and } {\mathbf r^\prime}.
 \tag{S\arabic{equation}}
 \label{eq:thetamax}
\end{align*}
For instance, if we have a charge qubit in which $\Theta_{\mathbf r}/2$ changes from $-1/4$ to $1/4$ abruptly across the Josephson junction from one lattice site to the next, then we have $\theta_{max} = 2 \pi$.  If instead $\Theta_{\mathbf r}/2$ changes from $-1/4$ to $1/4$ linearly over $\Delta z$ lattice sites, $\Delta z$ referring to the thickness of the Josephson junction, then $\theta_{max} = 2 \pi \Delta z$.  

We define orthonormal states $|n\rangle$ as follows.
Using expression (\ref{eq:thoulesstheta}) for $|\theta^\prime\rangle$ and defining 
\stepcounter{equation}  
\begin{align*}
\bar{c}_{\mathbf R}^\dagger =  e^{i\theta^\prime \Theta_{\mathbf R}/2}c _{\mathbf R}^\dagger,
 \tag{S\arabic{equation}}
 \label{eq:barcdagger}
\end{align*}
one sees that the overlap matrix elements satisfy $\langle \theta|\theta^\prime \rangle = \langle \theta-\theta^\prime|0 \rangle$.  Fourier transforming, we define $o_n$ by
\stepcounter{equation}  
\begin{align*}
&\frac{1}{2 \theta_{max}}\smashoperator{\int_{-\theta_{max}}^{\theta_{max}}} d\theta  e^{i \theta n} \frac{1}{2 \theta_{max}}\smashoperator{\int_{-\theta_{max}}^{\theta_{max}}} d\theta^\prime  e^{-i \theta^\prime n^\prime} \langle \theta-\theta^\prime|0 \rangle \\
&= \left\{ \begin{array}{ll} o_n & n = n^\prime\\ 0 & \text{otherwise} \end{array}\right.
 \tag{S\arabic{equation}}
 \label{eq:overlapmatrix}
\end{align*}
where 
\stepcounter{equation}  
\begin{align*}
n=2\pi m/2 \theta_{max}
\tag{S\arabic{equation}}
 \label{eq:nval}
 \end{align*}
 for $m$ an integer in the range $-M/2, \dots, M/2$ with $M$ an even integer.
It follows that the states 
\stepcounter{equation}  
\begin{align*}
| n \rangle = \frac{1}{\sqrt{o_n}} \frac{1}{2 \theta_{max}}\smashoperator{\int_{-\theta_{max}}^{\theta_{max}}} d\theta e^{-i \theta n} |\theta\rangle
 \tag{S\arabic{equation}}
 \label{eq:ndef}
\end{align*}
satisfy the orthonormality condition 
\stepcounter{equation}  
\begin{align*}
\langle n|n^\prime\rangle 
= \left\{ \begin{array}{ll} 1 & n = n^\prime\\ 0 & \text{otherwise} \end{array}\right.
 \tag{S\arabic{equation}}
\end{align*}
Some eigenvalues $o_n$ approach zero because the basis of $| \theta \rangle$ states is overcomplete; we exclude the corresponding states $|n \rangle$ from our basis.

Given the definition of $|n\rangle$, it is possible to perform a unitary transformation to a coarse-grained $\bar{\theta}$ basis defined by an inverse Fourier transform over the valid  $|n\rangle$ states: 
 \stepcounter{equation}
\begin{align*}
|\bar{\theta}\rangle =  \sum_{n} e^{i \bar{\theta} n} |n\rangle/\sqrt{M+1}.
 \tag{S\arabic{equation}}
 \label{eq:coarsetheta}
\end{align*} 
The number of such states is $M+1$, and they lie in value between $-\theta_{max}$ and $\theta_{max}$, so their allowed values are $\bar{\theta} = 2 \theta_{max} m/(M+1)$ with $m=-M/2,\dots,M/2$ integral.

An upper bound on $M$ is given by $\theta_{max}N_{pairs}/\pi$, where $N_{pairs}$ is the total number of pairs occupying the system.  This estimate is obtained by applying the theorem of Bloch and Messiah to (\ref{eq:thoulesstheta}).  As noted previously, this theorem states there are unitary matrices $C$ and $D$ for which $\bar{U} = D^\dagger U  C^\dagger$ and $\bar{V} = D^T U C^\dagger$ are real and take simple block-diagonal forms.  The matrix $\bar{U}$ has $2 \times 2$ blocks of the form 
$u_{\mathbf J} \begin{bmatrix}1& 0 \\ 0 & 1 \end{bmatrix}$, while these blocks take the form  $v_{\mathbf J} \begin{bmatrix}0& 1 \\ -1 & 0 \end{bmatrix}$ in the case of $\bar{V}$.  Here, the $2\times 2$ block involving state ${\mathbf J}$ involves a paired state that we label $\bar{\mathbf J}$.  Defining
\stepcounter{equation}  
\begin{align*}
d^\dagger_{\mathbf J} = \sum_{\mathbf R} c^\dagger_{\mathbf R} e^{i \theta \Theta_{\mathbf r}/2}   D_{{\mathbf R},{\mathbf J}},
 \tag{S\arabic{equation}}
 \label{eq:ddef}
\end{align*}
we find that $\frac{1}{2} \sum_{{\mathbf R},{\mathbf R}^\prime} e^{i \theta (\Theta_{\mathbf r}+ \Theta_{{\mathbf r}^\prime})/2} c^\dagger_{{\mathbf R}}  Z_{{\mathbf R},{\mathbf R}^\prime} c^\dagger_{{\mathbf R}^\prime}= \sum_{{\mathbf J}} d^\dagger_{{\mathbf J}} \frac{v_{\mathbf J}}{u_{\mathbf J}} d^\dagger_{\bar{\mathbf J}}$.  Inserting this into (\ref{eq:thoulesstheta}) yields
\stepcounter{equation}
\begin{align*}
|\theta \rangle & =  {\mathcal N}   \exp \Big( \sum_{{\mathbf J}}d^\dagger_{{\mathbf J}} \frac{v_{\mathbf J}}{u_{\mathbf J}}  d^\dagger_{\bar{\mathbf J}}\Big) \ket{\mathrm{vac}} \\
&=  {\mathcal N}  \Pi_{\mathbf J}  \Big(1+ d^\dagger_{{\mathbf J}} \frac{v_{\mathbf J}}{u_{\mathbf J}}  d^\dagger_{\bar{\mathbf J}}\Big) \ket{\mathrm{vac}} \\
&\approx  {\mathcal N} \smashoperator{\prod_{{\mathbf J}  \sim \text{Fermi sea}}} \Big(1+ d^\dagger_{{\mathbf J}} \frac{v_{\mathbf J}}{u_{\mathbf J}}  d^\dagger_{\bar{\mathbf J}}\Big) \ket{\mathrm{vac}} \\
&=  {\mathcal N}  \Big(1+\smashoperator{\sum_{{\mathbf J}  \sim \text{Fermi sea}}}d^\dagger_{{\mathbf J}} \frac{v_{\mathbf J}}{u_{\mathbf J}}  d^\dagger_{\bar{\mathbf J}}+ \dots + \smashoperator{\prod_{{\mathbf J}  \sim \text{Fermi sea}}}d^\dagger_{{\mathbf J}} \frac{v_{\mathbf J}}{u_{\mathbf J}}  d^\dagger_{\bar{\mathbf J}}\Big) \ket{\mathrm{vac}} 
 \tag{S\arabic{equation}}
 \end{align*}
The notation ${{\mathbf J}  \sim \text{Fermi sea}}$ indicates states that are within the Fermi sea or not too far above the Fermi surface.  The approximate equality in the third line uses that fact that high-energy states well above the Fermi surface have $v_{\mathbf J} \rightarrow 0$.  The final equality comes from multiplying out the product into a sum of terms.  Inserting this expression into (\ref{eq:ndef}) yields
 \stepcounter{equation}
\begin{align*}
|n\rangle & \approx \frac{\mathcal N}{\sqrt{o_n}}\frac{1}{2 \theta_{max}}\smashoperator{\int_{-\theta_{max}}^{\theta_{max}}} d\theta e^{-i n \theta}  \Big(1+ \dots + \smashoperator{\prod_{{\mathbf J}  \sim \text{Fermi Sea}}} d^\dagger_{{\mathbf J}} \frac{v_{\mathbf J}}{u_{\mathbf J}}  d^\dagger_{\bar{\mathbf J}}\Big) \ket{\mathrm{vac}} \\
& = \frac{\mathcal N}{\sqrt{o_n}}\frac{1}{2 \theta_{max}}\smashoperator{\int_{-\theta_{max}}^{\theta_{max}}} d\theta e^{-i n \theta}  \Big(1+ \dots + \smashoperator{\prod_{{\mathbf J}  \sim \text{Fermi sea}}} \\
& \hspace{0.1in}   \sum_{{\mathbf R},{\mathbf R^\prime}} c^\dagger_{\mathbf R} e^{i \theta \Theta_{\mathbf r}/2}   D_{{\mathbf R},{\mathbf J}} \frac{v_{\mathbf J}}{u_{\mathbf J}}  c^\dagger_{\mathbf R^\prime} e^{i \theta \Theta_{\mathbf r^\prime}/2}   D_{{\mathbf R^\prime},\bar{\mathbf J}}\Big) \ket{\mathrm{vac}} 
 \tag{S\arabic{equation}}
 \label{eq:ncalc}
\end{align*}
where (\ref{eq:ddef}) has been used on the final line.
If $n$ is too large, the right hand side will vanish.  To see this, note that
 \stepcounter{equation}
\begin{align*}
& \frac{1}{2 \theta_{max}}\smashoperator{\int_{-\theta_{max}}^{\theta_{max}}}d\theta  e^{-i n \theta} e^{i\theta(\Theta_{\mathbf r_1} + \dots+ \Theta_{\mathbf r_{2T}})/2} \\
& = \left\{\begin{array}{ll} 1 & n = (\Theta_{\mathbf r_1} + \dots+ \Theta_{\mathbf r_{2T}})/2 \\ 0 & \text{otherwise}\end{array}\right.
 \tag{S\arabic{equation}}
 \label{eq:nproj}
\end{align*}
using (\ref{eq:thetamax}).
Assuming that $\Theta_{\mathbf r}$ is at most $1/2$, the largest possible value of $(\Theta_{\mathbf r_1} + \dots+ \Theta_{\mathbf r_{2T}})/2$ is $T/2$; the Kronecker delta then enforces $n = T/2$.  Thus, the largest possible value of $n$ that could possibly give a non-vanishing result for (\ref{eq:ncalc}) is $n = N_{pairs}/2$, where $N_{pairs}$ is the number of occupied pairs appearing in $\smashoperator{\prod_{{\mathbf J}  \sim \text{Fermi sea}}}$.  Similarly, the smallest possible value of $n$ is $-N_{pairs}/2$.  Given (\ref{eq:nval}), we find that $2\pi (M/2)/2 \theta_{max} \le N_{pairs}/2$, or $M \le N_{pairs}\theta_{max}/\pi$ as claimed above.

In the case of a charge qubit composed of two weakly coupled superconducting islands with a relatively abrupt phase change, as in Fig. \ref{fig:chargequbit}c, we can find a better estimate of $M+1$.  The basis size $M+1$ turns out to scale as the number of pairs $N_{shared}$ in $|\theta\rangle$ that are shared on both islands, which is roughly the number of pairs near the chemical potential of the system.  To show this, we write
\stepcounter{equation}
\begin{align*}
|\theta\rangle=  {\mathcal N}&  \smashoperator{\prod_{{\mathbf J}  \sim \text{Fermi sea}}} \Big(1+ d^\dagger_{{\mathbf J}} \frac{v_{\mathbf J}}{u_{\mathbf J}}  d^\dagger_{\bar{\mathbf J}}\Big) \ket{\mathrm{vac}} \\
=   {\mathcal N}& \smashoperator{\prod_{{\mathbf J}  \text{ left}}} \Big(1+ d^\dagger_{{\mathbf J}} \frac{v_{\mathbf J}}{u_{\mathbf J}}  d^\dagger_{\bar{\mathbf J}}\Big) \times \\
 &  \smashoperator{\prod_{{\mathbf J}  \text{ right}}} \Big(1+ d^\dagger_{{\mathbf J}} \frac{v_{\mathbf J}}{u_{\mathbf J}}  d^\dagger_{\bar{\mathbf J}}\Big)\smashoperator{\prod_{{\mathbf J}  \text{ shared}}} \Big(1+ d^\dagger_{{\mathbf J}} \frac{v_{\mathbf J}}{u_{\mathbf J}}  d^\dagger_{\bar{\mathbf J}}\Big) \ket{\mathrm{vac}} \\
&  \approx   {\mathcal N^\prime}\smashoperator{\prod_{{\mathbf J}  \text{ left}}} d^\dagger_{{\mathbf J}} d^\dagger_{\bar{\mathbf J}} \smashoperator{\prod_{{\mathbf J}  \text{ right}}} d^\dagger_{{\mathbf J}} d^\dagger_{\bar{\mathbf J}}\smashoperator{\prod_{{\mathbf J}  \text{ shared}}} \Big(1+ d^\dagger_{{\mathbf J}} \frac{v_{\mathbf J}}{u_{\mathbf J}}  d^\dagger_{\bar{\mathbf J}}\Big) \ket{\mathrm{vac}}.
 \tag{S\arabic{equation}}
\end{align*}
where ${{\mathbf J}  \text{ left}}$ and ${{\mathbf J}  \text{ right}}$ refer to states that are localized well within the Fermi sea of the left or right island respectively.  These states have $v_{\mathbf J}/u_{\mathbf J}$ very large, permitting the final approximate equality.  Let $N_{left}$ be the number of pairs in the ${{\mathbf J}  \text{ left}}$ product, $N_{right}$ be the number of pairs in the ${{\mathbf J}  \text{ right}}$ product, and $N_{shared}$ be the number of pairs in the ${{\mathbf J}  \text{ shared}}$ product.  We are considering the case of an abrupt phase change, in which the electrons in the system have $\Theta_{\mathbf R} = -1/2$ or $1/2$ depending on whether they inhabit the left island or the right island.  Thus, when we calculate (\ref{eq:ncalc}), the integrals take the form
\begin{align*}
\frac{1}{2 \theta_{max}}\smashoperator{\int_{-\theta_{max}}^{\theta_{max}}}d\theta  e^{-i n \theta} e^{i  \theta (N_{right}-N_{left})/2} e^{i\theta(\Theta_{\mathbf r_1} + \dots+ \Theta_{\mathbf r_{2T}})/2}
\end{align*}
with the largest possible value of $T$ given by the number of pairs $N_{shared}$ that can wander from the left island to the right island.   We see that $n$ ranges from $(N_{right}-N_{left})/2 - N_{shared}/2$ to $(N_{right}-N_{left})/2 + N_{shared}/2$.  Recalling (\ref{eq:nval}), we see that the size of the basis is determined by $2\pi (M/2)/2\theta_{max} \le N_{shared}/2$, so that $M \le N_{shared} \theta_{max}/\pi$.  Physically, $|n\rangle$ is a state with $2n$ more pairs on the right island than on the left island.

The construction of an orthonormal basis for (\ref{eq:H2by2}) closely parallels the construction for (\ref{eq:Hthetathetaprime}).  In the limit of large systems, the states ${|\!\!\circlearrowright,\theta\rangle}$ and ${|\!\!\circlearrowleft,\theta\rangle}$ approach orthonormality.  For smaller systems, we construct an orthonormal basis explicitly as follows.  The original, non-orthogonal basis is composed of ${|\!\!\circlearrowright,\theta\rangle}$ and ${|\!\!\circlearrowleft,\theta\rangle}$ with $-\theta_{max} < \theta \le \theta_{max}$ and $\theta_{max}$ fixed by (\ref{eq:thetamax}).  For an RF squid qubit, $\theta_{max}$ will approach $2 \pi N_z$, with $N_z$ the number of sites around the circumference of the system.  This greatly exceeds the value of $\theta_{max}$ in the charge qubit case, consistent with the fact that $\theta$ is unbounded in LE theory in the case of an RF squid.   The overlap matrix is composed of $2\times 2$ blocks
\stepcounter{equation}  
\begin{align*}
& \begin{bmatrix} \langle \circlearrowright,\theta|\circlearrowright,\theta^\prime\rangle & \langle \circlearrowright,\theta|\circlearrowleft,\theta^\prime\rangle \\ \langle \circlearrowleft,\theta|\circlearrowright,\theta^\prime\rangle & \langle \circlearrowleft,\theta|\circlearrowleft,\theta^\prime\rangle \end{bmatrix} \\
& = \begin{bmatrix} \langle \circlearrowright,\theta-\theta^\prime|\circlearrowright,0\rangle & \langle \circlearrowright,\theta-\theta^\prime|\circlearrowleft,0\rangle \\ \langle \circlearrowleft,\theta-\theta^\prime|\circlearrowright,0\rangle & \langle \circlearrowleft,\theta-\theta^\prime|\circlearrowleft,0\rangle \end{bmatrix} 
\tag{S\arabic{equation}}
 \label{eq:overlapmatrix}
\end{align*}
for $-\theta_{max} < \theta,\theta^\prime \le \theta_{max}$.  After a Fourier transform, the overlap matrix becomes block diagonal with $2 \times 2$ blocks of the form
\stepcounter{equation}  
\begin{align*}
& \smashoperator{\int_{-\theta_{max}}^{\theta_{max}}} d\theta \smashoperator{\int_{-\theta_{max}}^{\theta_{max}}} d\theta^\prime  \frac{e^{i\theta n-i \theta^\prime n^\prime}}{(2 \theta_{max})^2} \begin{bmatrix} \langle \circlearrowright,\theta|\circlearrowright,\theta^\prime\rangle & \langle \circlearrowright,\theta|\circlearrowleft,\theta^\prime\rangle \\ \langle \circlearrowleft,\theta|\circlearrowright,\theta^\prime\rangle & \langle \circlearrowleft,\theta|\circlearrowleft,\theta^\prime\rangle \end{bmatrix}\\
&= \left\{ \begin{array}{ll} O_n & n = n^\prime\\ 0 & \text{otherwise} \end{array}\right.
 \tag{S\arabic{equation}}
 \label{eq:overlapmatrixn}
\end{align*}
where $O_n$ is a $2 \times 2$ matrix and $n,n^\prime$ satisfy (\ref{eq:nval}).  The eigenvalues of $O_n$, labeled $o_{1,n}$ and $o_{2,n}$, have corresponding eigenvectors
\begin{align*}
O_{n} \begin{bmatrix} a_{i,n} \\ b_{i,n} \end{bmatrix} = o_{i,n}   \begin{bmatrix} a_{i,n} \\ b_{i,n} \end{bmatrix}
\end{align*}
for $i=1,2$.
We can define orthonormal states in terms of these eigenvectors by
\stepcounter{equation}  
\begin{align*}
|i,n\rangle =   \frac{1}{\sqrt{o_{i,n}}} &  \frac{1}{2 \theta_{max}}\smashoperator{\int_{-\theta_{max}}^{\theta_{max}}} d\theta e^{-i \theta n}(a_{i,n} {|\!\!\circlearrowright,\theta\rangle} +   b_{i,n} {|\!\!\circlearrowleft,\theta\rangle})
 \tag{S\arabic{equation}}
 \label{eq:inbasis}
\end{align*}
for $i=1,2$.  They satisfy 
\stepcounter{equation}  
\begin{align*}
\langle i,n|i^\prime,n^\prime\rangle = \left\{ \begin{array}{ll} 1& i = i^\prime \text{ and } n = n^\prime\\ 0 & \text{otherwise} \end{array}\right.
 \tag{S\arabic{equation}}
\end{align*}
  Because the original, non-orthogonal basis is overcomplete, some eigenvalues $o_{i,n}$ approach zero; the corresponding states $|i,n\rangle$ are omitted from our basis.  We denote by $2(M+1)$ the number of valid orthogonal basis states $|i,n\rangle$. 

Given this definition of $|i,n\rangle$, it is possible to perform a unitary transformation to a coarse-grained phase basis defined by an inverse Fourier transform over the valid  $|i,n\rangle$ states: 
 \stepcounter{equation}
\begin{align*}
|i,\bar{\theta}\rangle =  \sum_{n=-M/2}^{M/2} e^{i \bar{\theta} n} |i,n\rangle/\sqrt{M+1}.
 \tag{S\arabic{equation}}
 \label{eq:coarsethetai}
\end{align*} 
Since the number of such states for a given $i$ is $M+1$, the spacing between them is $\Delta \bar{\theta} = 2 \theta_{max}/(M+1)$.  As $M+1$ grows with system size, this spacing shrinks to zero.

Applying the theorem of Bloch and Messiah as above to ${|\!\!\circlearrowright,\theta\rangle}$ and ${|\!\!\circlearrowleft,\theta\rangle}$, we can argue that a rough upper bound on $M+1$  is given by the number of occupied pairs in the system.  It is important to emphasize that the quantum number $n$ in $|i,n\rangle$ does not admit a simple interpretation in terms of the positions of Cooper pairs in the system.

\subsection{Derivation of Lumped Element Equations}

In the coarse-grained $\bar{\theta}$ basis, the Schrodinger equation (\ref{eq:Hthetathetaprime}) becomes an $(M+1) \times (M+1)$ matrix equation
 \stepcounter{equation}
\begin{align*}
\sum_{\bar{\theta}^\prime}\langle \bar{\theta}|{\mathcal H}|\bar{\theta}^\prime\rangle \psi(\bar{\theta}^\prime) = E \psi(\bar{\theta}^\prime)
 \tag{S\arabic{equation}}
 \label{eq:coarsethetaH}
\end{align*}
where
\stepcounter{equation}
\begin{align*}
\langle \bar{\theta}|{\mathcal H}|\bar{\theta}^\prime\rangle & = \frac{1}{M+1}  \frac{1}{(2 \theta_{max})^2} \sum_{n=-M/2}^{M/2}  \frac{e^{-i \bar{\theta} n}}{\sqrt{o_n}} \smashoperator{\int_{-\theta_{max}}^{\theta_{max}}} d\theta e^{i \theta n} \\
& \sum_{n^\prime=-M/2}^{M/2} \frac{e^{i \bar{\theta}^\prime n^\prime} }{\sqrt{o_{n^\prime}}} \smashoperator{\int_{-\theta_{max}}^{\theta_{max}}} d\theta e^{-i \theta^\prime n^\prime}
 \langle \theta|H| \theta^\prime \rangle.
 \tag{S\arabic{equation}}
 \label{eq:coarsethetaHmatrixelement}
\end{align*}
As the size of the system, and the basis size $M+1$, grows, the overlap matrix elements $\langle \theta | \theta^\prime \rangle$ decay rapidly for $\theta\ne \theta^\prime$ as a consequence of (\ref{eq:Uchargequbit}).  The original $|\theta\rangle$ basis tends toward orthonormality, and one expects the orthonormal basis $|\bar{\theta}\rangle$ that we constructed to approach $|\theta\rangle|_{\theta=\bar{\theta}}$, the original basis state with $\theta$ evaluated at $\bar{\theta}$.  Thus, in this limit we approximate $\langle \bar{\theta}|{\mathcal H}|\bar{\theta}^\prime\rangle \approx  \langle \theta|H| \theta^\prime \rangle|_{\theta=\bar{\theta},\theta^\prime=\bar{\theta}^\prime}$.  Recall the decomposition (\ref{eq:H}), $\langle \theta|H| \theta^\prime \rangle = \langle \theta|T+P+W| \theta^\prime \rangle$.   Employing (\ref{eq:barcdagger}), we can show that the potential and interaction energies satisfy $\langle \theta|P+W| \theta^\prime \rangle = \langle \theta-\theta^\prime|P+W|0\rangle$.
Therefore, along the diagonal of the Hamiltonian matrix in (\ref{eq:coarsethetaH}), $\langle \theta|P+W| \theta \rangle=\langle 0|P+W| 0 \rangle$ contributes an overall constant that simply shifts $E$ in (\ref{eq:coarsethetaH}).  Only $\langle \theta|T| \theta \rangle$ depends upon $\theta$; this dependence is physically important and should not be neglected.  However, in the off-diagonal elements, we will approximate $\langle \theta|T| \theta^\prime \rangle$ as $\langle \theta-\theta^\prime|T|0\rangle$, which is reasonable if the error incurred thereby is small: $|\langle \theta-\theta^\prime|T|0\rangle -\langle \theta|T| \theta^\prime \rangle|\ll |\langle \theta-\theta^\prime|T+P+W|0\rangle|$.  Summarizing, we have
\stepcounter{equation}
\begin{align*}
\langle \bar{\theta}|{\mathcal H}|\bar{\theta}^\prime\rangle & \approx \delta_{\theta,\theta^\prime} (\langle \theta|T|\theta\rangle -\langle 0|T|0\rangle)|_{\theta=\bar{\theta},\theta^\prime=\bar{\theta}^\prime}  \\
& + \langle \theta-\theta^\prime|T+P+W|0\rangle|_{\theta=\bar{\theta},\theta^\prime=\bar{\theta}^\prime}
\tag{S\arabic{equation}}
 \label{eq:coarsethetaHmatrixelementapproximation}
\end{align*}
Clearly, the first line is diagonal in the phase basis.  The second line, since it depends only on $\theta-\theta^\prime$, becomes diagonal in the basis of $|n\rangle$ states (\ref{eq:ndef}).  

We can develop the analysis further in the case of a charge qubit composed of two weakly coupled superconducting islands with a relatively abrupt phase change, as in Fig. \ref{fig:chargequbit}c.  The matrix element of the tunneling Hamiltonian is
\stepcounter{equation}
\begin{align*}
 \langle \theta|T| \theta^\prime \rangle =  -  \langle \theta-\theta^\prime | & \sum_{{\mathbf R}}  \sum_{\smash{{\mathbf a}=\pm{\mathbf a}_x,\pm{\mathbf a}_y,\pm{\mathbf a}_z}} \!\!\!\!\!\!\!\!\!\! t_{{\mathbf r}+{\mathbf a},{\mathbf r}} \\
&  \,\,\, e^{i\theta^\prime( \Theta_{\mathbf r}-\Theta_{{\mathbf r}+{\mathbf a}})/2} c^\dagger_{{\mathbf r}+{\mathbf a},\sigma}  c^{ }_{{\mathbf r},\sigma} | 0 \rangle
 \tag{S\arabic{equation}}
 \label{eq:Tmatrixelement}
\end{align*}
where we have used (\ref{eq:barcdagger}).
It is useful to decompose $T = T_{near} + T_{far}$, where $T_{near}$ contains the small fraction of terms in which ${\mathbf r}$ is right near the junction and $T_{far}$ contains all other tunneling terms.  As a result of the form of Fig. \ref{fig:chargequbit}c, far from the junction, $\Theta_{\mathbf r} \approx \Theta_{{\mathbf r}+{\mathbf a}}$, so that $e^{i\theta^\prime( \Theta_{\mathbf r}-\Theta_{{\mathbf r}+{\mathbf a}})/2} \approx 1$.  The expression (\ref{eq:Tmatrixelement}) then implies $\langle \theta|T_{far}| \theta^\prime \rangle \approx \langle \theta-\theta^\prime|T_{far}| 0\rangle$.  So, neglecting the contribution of $T_{near}$, we conclude that $T_{far}+P+W$ is diagonal in the basis  (\ref{eq:ndef}).  A reasonable approximation is $\langle n|T_{far}|n^\prime\rangle = E_{seas}n^2\delta_{n,n^\prime}$, $\langle n|W|n^\prime\rangle = 4E_C n^2\delta_{n,n^\prime}$ and $\langle n|P|n^\prime\rangle = - \tilde{\mu} n\delta_{n,n^\prime}$.  Here, $\tilde{\mu}$ denotes the electrochemical potental difference between the islands.  The capacitive energy, $4E_C = (2 e)^2/2C$ in terms of an appropriate capacitance $C$, is familiar from LE theory.  The unfamiliar coefficient $E_{seas}$ is determined by the total energy of the two Fermi seas residing in the two superconducting islands. For simplicity, assume that each of the islands has volume $L^3$.   Let the total number of electrons in the system be $2N_{tot}$, with $N_{tot}+2n$ residing on one island and $N_{tot}-2n$ residing on the other island.  Then total energy of the two Fermi seas is $\frac{3}{5}\frac{\hbar^2}{2mL^{2}}(3\pi^2)^{2/3}((N_{tot}+2n)^{5/3} + (N_{tot}-2n)^{5/3})$.  Taylor expanding this expression to second order, we deduce that $E_{seas} = \frac{\hbar^2}{2mL^{2}}(3\pi^2)^{2/3}\frac{8}{3}N_{tot}^{-1/3}=\frac{8E_F}{3N_{tot}}$ where $E_F$ is the Fermi energy of each island when $n=0$.  For physical devices, $E_{seas}$ is much smaller than $4E_C$.  However, in simulations such as the one performed in the text, where the Coulomb interaction is replaced with a short-range Hubbard interaction, $E_{seas}$ becomes the important energy.

The contribution of $T_{near}$ to the diagonal matrix elements takes the form $\langle \theta|T_{near}|\theta\rangle \approx - E_J \cos \theta/2$.  Here, we have introduced the parameter $E_J$ and have used the fact that $\Theta_{\mathbf r} - \Theta_{{\mathbf r}+{\mathbf a}} \approx  1$ when ${\mathbf r}$ and ${\mathbf r} + {\mathbf a}$ stand on opposite sides of the junction, implying $e^{i\theta^\prime( \Theta_{\mathbf r}-\Theta_{{\mathbf r}+{\mathbf a}})/2} \approx e^{i\theta^\prime/2}$ in (\ref{eq:Tmatrixelement}).

Assembling these results, we make a continuum approximation to our matrix equation (\ref{eq:coarsethetaH}) for $M+1$ large, obtaining the lumped element equation
\stepcounter{equation}
\begin{align*}
4E_C \Big(-i\frac{d}{d\bar{\theta}}-n_0\Big)^2  \psi(\bar{\theta})+ E_J \Big(1-\cos \frac{\bar{\theta}}{2}\Big) \psi(\bar{\theta})= E  \psi(\bar{\theta})
\tag{S\arabic{equation}}
 \label{eq:coarsethetaHLEpreliminary}
\end{align*}
where $n_0 =\tilde{\mu}/8E_C$.  This should be compared with the standard lumped element expression (\ref{eq:lumped}).  There is a factor of 2 difference in the potential stemming from the fact that $-2\pi < \bar{\theta}\le 2\pi$ in our theory.  Setting $\tilde{\theta} = \bar{\theta}/2$, $\tilde{E}_C = E_C/4$ and $\tilde{n}_0 = 2n_0$, we have
\stepcounter{equation}
\begin{align*}
4\tilde{E}_C \Big(-i\frac{d}{d\tilde{\theta}}-\tilde{n}_0\Big)^2  \psi(\tilde{\theta})+ E_J \Big(1-\cos \tilde{\theta}\Big) \psi(\tilde{\theta})= E  \psi(\tilde{\theta})
\tag{S\arabic{equation}}
 \label{eq:coarsethetaHLE}
\end{align*}
with $-\pi < \tilde{\theta} \le \pi$.  This agrees with  (\ref{eq:lumped}).

We can make a similar argument in the case of an RF squid qubit.  The transformation to the orthogonal basis changes (\ref{eq:H2by2}) to the matrix equation 
\begin{align*}
\sum_{\bar{\theta}^\prime} \begin{bmatrix} \langle 1,\bar{\theta}|{\mathcal H}|1,\bar{\theta}^\prime\rangle & \langle 1,\bar{\theta}|{\mathcal H}|2,\bar{\theta}^\prime\rangle \\ \langle 2,\bar{\theta}|{\mathcal H}|1,\bar{\theta}^\prime\rangle & \langle 2,\bar{\theta}|{\mathcal H}|2,\bar{\theta}^\prime\rangle \end{bmatrix} \begin{bmatrix}  \psi(1,\bar{\theta}^\prime) \\  \psi(2,\bar{\theta}^\prime)\end{bmatrix} = E \begin{bmatrix}  \psi(1,\bar{\theta}) \\  \psi(2,\bar{\theta})\end{bmatrix}
\end{align*}
where $ \langle i,\bar{\theta}|{\mathcal H}|i^\prime,\bar{\theta}^\prime\rangle$ satisfies a definition analogous to (\ref{eq:coarsethetaHmatrixelement})
\stepcounter{equation}
\begin{align*}
\langle i, \bar{\theta}|&{\mathcal H}|i^\prime,\bar{\theta}^\prime\rangle = \frac{1}{M+1}  \frac{1}{(2 \theta_{max})^2} \sum_{n=-M/2}^{M/2}  \frac{e^{-i \bar{\theta} n}}{\sqrt{o_n}} \smashoperator{\int_{-\theta_{max}}^{\theta_{max}}} d\theta e^{i \theta n} \\
& \sum_{n^\prime=-M/2}^{M/2} \frac{e^{i \bar{\theta}^\prime n^\prime} }{\sqrt{o_{n^\prime}}} \smashoperator{\int_{-\theta_{max}}^{\theta_{max}}} d\theta^\prime e^{-i \theta^\prime n^\prime}
\\
 & \begin{bmatrix}a_{i,n} \\ b_{i,n}\end{bmatrix}^\dagger  \begin{bmatrix} \langle \circlearrowright,\theta|H|\!\!\circlearrowright, \theta^\prime \rangle & \langle \circlearrowright,\theta|H|\!\!\circlearrowleft, \theta^\prime \rangle \\
  \langle \circlearrowleft,\theta|H|\!\!\circlearrowright, \theta^\prime \rangle & \langle \circlearrowleft,\theta|H|\!\!\circlearrowleft, \theta^\prime \rangle  \end{bmatrix} \begin{bmatrix}a_{i^\prime,n^\prime} \\ b_{i^\prime,n^\prime}\end{bmatrix} 
 \tag{S\arabic{equation}}
 \label{eq:coarsethetaH2by2matrixelement}
\end{align*}

For large systems, the overlap matrix (\ref{eq:overlapmatrix}) tends to that of an orthonormal set of states, so we can approximate ${|1,\bar{\theta}\rangle = |\!\!\circlearrowright,\theta \rangle|_{\theta = \bar{\theta}}}$ and ${|2,\bar{\theta}\rangle = |\!\!\circlearrowleft,\theta \rangle|_{\theta = \bar{\theta}}}$.  
Then, our matrix equation becomes
\stepcounter{equation}
\begin{align*}
\sum_{\bar{\theta}^\prime} &\left. \begin{bmatrix} \langle \circlearrowright,\theta|H|\!\!\circlearrowright,\theta^\prime\rangle & \langle \circlearrowright,\theta|H|\!\!\circlearrowleft,\theta^\prime\rangle \\ \langle \circlearrowleft,\theta|H|\!\!\circlearrowright,\theta^\prime\rangle & \langle \circlearrowleft,\theta|H|\!\!\circlearrowleft,\theta^\prime\rangle \end{bmatrix}\right|_{\theta=\bar{\theta},\theta^\prime=\bar{\theta}^\prime} \begin{bmatrix}  \psi(\circlearrowright,\bar{\theta}^\prime) \\  \psi(\circlearrowleft,\bar{\theta}^\prime) \end{bmatrix} \\
& = E \begin{bmatrix}  \psi(\circlearrowright,\bar{\theta}) \\  \psi(\circlearrowleft,\bar{\theta}) \end{bmatrix}. 
 \tag{S\arabic{equation}}
 \label{eq:coarsethetaH2by2}
\end{align*}

The potential and interaction matrix elements satisfy
\begin{align*}
&\begin{bmatrix} \langle \circlearrowright,\theta|P+W|\!\!\circlearrowright, \theta^\prime \rangle & \langle \circlearrowright,\theta|P+W|\!\!\circlearrowleft, \theta^\prime \rangle \\
  \langle \circlearrowleft,\theta|P+W|\!\!\circlearrowright, \theta^\prime \rangle & \langle \circlearrowleft,\theta|P+W|\!\!\circlearrowleft, \theta^\prime \rangle  \end{bmatrix} \\
  &
  = \begin{bmatrix} \langle \circlearrowright,\theta-\theta^\prime|P+W|\!\!\circlearrowright, 0 \rangle & \langle \circlearrowright,\theta-\theta^\prime|P+W|\!\!\circlearrowleft, 0 \rangle \\
  \langle \circlearrowleft,\theta-\theta^\prime|P+W|\!\!\circlearrowright, 0 \rangle & \langle \circlearrowleft,\theta-\theta^\prime|P+W|\!\!\circlearrowleft, 0 \rangle  \end{bmatrix},
\end{align*}
which can be verified using a substitution like (\ref{eq:barcdagger}).

For $\theta \ne \theta^\prime$, we adopt the approximation
\begin{align*}
&\begin{bmatrix} \langle \circlearrowright,\theta|T|\!\!\circlearrowright, \theta^\prime \rangle & \langle \circlearrowright,\theta|T|\!\!\circlearrowleft, \theta^\prime \rangle \\
  \langle \circlearrowleft,\theta|T|\!\!\circlearrowright, \theta^\prime \rangle & \langle \circlearrowleft,\theta|T|\!\!\circlearrowleft, \theta^\prime \rangle  \end{bmatrix} \\
  &
  \approx \begin{bmatrix} \langle \circlearrowright,\theta-\theta^\prime|T|\!\!\circlearrowright, 0 \rangle & \langle \circlearrowright,\theta-\theta^\prime|T|\!\!\circlearrowleft, 0 \rangle \\
  \langle \circlearrowleft,\theta-\theta^\prime|T|\!\!\circlearrowright, 0 \rangle & \langle \circlearrowleft,\theta-\theta^\prime|T|\!\!\circlearrowleft, 0 \rangle  \end{bmatrix}
\end{align*}
for the tunneling matrix elements.  This relies on the approximation $e^{i\theta^\prime( \Theta_{\mathbf r}-\Theta_{{\mathbf r}+{\mathbf a}})/2} \approx 1$, which is roughly true since $\Theta_{\mathbf r}$ changes gradually along the long circumference of the RF squid qubit (see Fig. \ref{fig:RFsquidqubit}c).  When $\theta=\theta^\prime$, on the diagonal of the Hamiltonian matrix, we retain the explicit $\theta$ dependence seen in the double-well potential in Fig. \ref{fig:RFsquidqubit}d.  This leads to
\begin{widetext}
\stepcounter{equation}
\begin{align*}
\begin{bmatrix} \langle 1,\bar{\theta}|{\mathcal H}|1,\bar{\theta}^\prime\rangle & \langle 1,\bar{\theta}|{\mathcal H}|2,\bar{\theta}^\prime\rangle \\ \langle 2,\bar{\theta}|{\mathcal H}|1,\bar{\theta}^\prime\rangle & \langle 2,\bar{\theta}|{\mathcal H}|2,\bar{\theta}^\prime\rangle \end{bmatrix} & \approx  \left. \delta_{\theta,\theta^\prime}  \begin{bmatrix} \langle \circlearrowright,\theta|T|\!\!\circlearrowright, \theta \rangle-\langle \circlearrowright,0|T|\!\!\circlearrowright, 0 \rangle & \\
  & \langle \circlearrowleft,\theta|T|\!\!\circlearrowleft, \theta \rangle-\langle \circlearrowleft,0|T|\!\!\circlearrowleft, 0 \rangle  \end{bmatrix}\right|_{\theta=\bar{\theta},\theta^\prime=\bar{\theta}^\prime} \\
  &+  \left.\begin{bmatrix} \langle \circlearrowright,\theta-\theta^\prime|T+P+W|\!\!\circlearrowright, 0 \rangle & \langle \circlearrowright,\theta-\theta^\prime|T+P+W|\!\!\circlearrowleft, 0 \rangle \\
  \langle \circlearrowleft,\theta-\theta^\prime|T+P+W|\!\!\circlearrowright, 0 \rangle & \langle \circlearrowleft,\theta-\theta^\prime|T+P+W|\!\!\circlearrowleft, 0 \rangle  \end{bmatrix}\right|_{\theta=\bar{\theta},\theta^\prime=\bar{\theta}^\prime} 
\tag{S\arabic{equation}}
\label{eq:coarsethetaH2by2matrixelementapproximation}
\end{align*}
\end{widetext}

In the first term on the right hand side, we approximate
\begin{align*}
\langle \circlearrowright,\theta|T|\!\!\circlearrowright, \theta \rangle-\langle \circlearrowright,0|T|\!\!\circlearrowright, 0 \rangle  = & -E_J \cos\frac{1}{2}(\theta-\theta^\circlearrowright) \\
&+ \frac{1}{2} E_L (\theta-\theta^\circlearrowright)^2\\
\langle \circlearrowleft,\theta|T|\!\!\circlearrowleft, \theta \rangle -\langle \circlearrowleft,0|T|\!\!\circlearrowleft, 0 \rangle  = & -E_J \cos\frac{1}{2}(\theta-\theta^\circlearrowleft) \\
& +\frac{1}{2} E_L (\theta-\theta^\circlearrowleft)^2.
\end{align*}  In each equation, the contribution proportional to $E_L$ comes from tunneling terms in (\ref{eq:T}) distant from the Josephson junction, while the contribution proportional to $E_J$ comes from tunneling through the Josephson junction.  

The second line on the right hand side of (\ref{eq:coarsethetaH2by2matrixelementapproximation}) depends only on $\theta-\theta^\prime$.   It therefore simplifies in the basis (\ref{eq:inbasis}).  As the system size grows and the overlap matrix (\ref{eq:overlapmatrix}) tends to that of an orthonormal set of states, the matrix $O_n$ becomes nearly diagonal: its eigenvectors tend to $a_{i,n} = 1$, $b_{i_n} =0$ and $a_{i,n} = 0$, $b_{i_n} =1$.  Considering the expression (\ref{eq:inbasis}), it becomes reasonable to adopt the notation ${|\!\!\circlearrowright,n\rangle} = |1,n\rangle$ and  ${|\!\!\circlearrowleft,n\rangle} = |2,n\rangle$.  In this basis, the second line on the right hand side of (\ref{eq:coarsethetaH2by2matrixelementapproximation})
is block diagonal with $2\times 2$ blocks
\stepcounter{equation}
\begin{align*}
\begin{bmatrix}E_{\circlearrowright,n;\circlearrowright,n} & E_{\circlearrowright,n;\circlearrowleft,n} \\ E_{\circlearrowleft,n;\circlearrowright,n} &  E_{\circlearrowleft,n;\circlearrowleft,n} \end{bmatrix}.
\tag{S\arabic{equation}}
\label{eq:coarsethetaH2by2n}
\end{align*}
Now, the matrix elements in the second line of (\ref{eq:coarsethetaH2by2matrixelementapproximation}) can be assumed real since one can introduce phases if necessary into the state definitions.  (It is true that conditions such as $|\!\!\circlearrowright, \theta_{max} \rangle = |\!\!\circlearrowright, -\theta_{max} \rangle$ could prevent the introduction of such phases consistently for all $\theta-\theta^\prime$ in (\ref{eq:coarsethetaH2by2matrixelementapproximation}).  However, since $\theta_{max}$ is so large for an RF squid qubit, states such as $|\!\!\circlearrowright,\theta_{max}\rangle$ have extremely high energy and play no role in the accessible energy eigenstates of the system.  Therefore, the matrix elements in the second line of (\ref{eq:coarsethetaH2by2matrixelementapproximation}) can be written as a real part plus a correction that vanishes for energetically accessible states.)  Because the second line of (\ref{eq:coarsethetaH2by2matrixelementapproximation}) should decay with $|\theta-\theta^\prime|$, we make a tight-binding approximation, retaining neighbors with $\theta-\theta^\prime|_{\theta=\bar{\theta},\theta^\prime=\bar{\theta}^\prime} = 0,\pm \Delta \bar{\theta}$.   (Recall that $\Delta \bar{\theta} = 2 \theta_{max}/(M+1)$ as argued after (\ref{eq:coarsethetai}).)  Then, each of the 4 functions in (\ref{eq:coarsethetaH2by2n}) equals a constant plus a term proportional to $\cos n \Delta \bar{\theta}$.  As system size increases and $\Delta \bar{\theta}$ shrinks, we can truncate the cosine at second order to obtain
\stepcounter{equation}
\begin{align*}
\begin{bmatrix}E_{\circlearrowright} & \\  &  E_{\circlearrowright} \end{bmatrix}+\begin{bmatrix}E_{\circlearrowright;\circlearrowright} & E_{\circlearrowright;\circlearrowleft} \\ E_{\circlearrowright;\circlearrowleft} &  E_{\circlearrowright;\circlearrowright} \end{bmatrix} n^2.
\tag{S\arabic{equation}}
\label{eq:coarsethetaH2by2napprox}
\end{align*}
This result may be reminiscent of the energy proportional to $n^2$ obtained in the case of a charge qubit with an abrupt junction, where $|n\rangle$ is a state with $2n$ extra Cooper pairs on one side of the Josephson junction.  (See discussion below (\ref{eq:Tmatrixelement}).)  However, it has been derived quite differently here; in particular, we avoided any claim that the states $|i,n\rangle$ have a simple physical interpretation in terms of the positions of Cooper pairs.

Assembling our results, we find the continuum limit of our matrix equation (\ref{eq:coarsethetaH2by2})
\stepcounter{equation}
\begin{align*}
-&\begin{bmatrix} E_{\circlearrowright;\circlearrowright} & E_{\circlearrowright;\circlearrowleft} \\ E_{\circlearrowleft;\circlearrowright} &  E_{\circlearrowleft;\circlearrowleft} \end{bmatrix} \frac{d^2}{d\bar{\theta}^2} \begin{bmatrix}  \psi(\circlearrowright,\bar{\theta}) \\  \psi(\circlearrowleft,\bar{\theta}) \end{bmatrix} \\
&+  \begin{bmatrix} (E_{\circlearrowright} - E_J\cos\frac{1}{2}(\bar{\theta}-\theta^\circlearrowright)+ \frac{1}{2} E_L (\bar{\theta}-\theta^\circlearrowright)^2)  \psi(\circlearrowright,\bar{\theta}) \\  (E_{\circlearrowleft}-E_J\cos\frac{1}{2}(\bar{\theta}-\theta^\circlearrowleft)+ \frac{1}{2} E_L (\bar{\theta}-\theta^\circlearrowleft)^2) \psi(\circlearrowleft,\bar{\theta}) \end{bmatrix} \\
& = E  \begin{bmatrix}  \psi(\circlearrowright,\bar{\theta}) \\  \psi(\circlearrowleft,\bar{\theta}) \end{bmatrix}
\tag{S\arabic{equation}}.
\label{eq:continuumlimitH2by2}
\end{align*}
An approximate one-component equation can be obtained by adopting the ansatz
\stepcounter{equation}
\begin{align*}
 \begin{bmatrix}  \psi(\circlearrowright,\bar{\theta}) \\  \psi(\circlearrowleft,\bar{\theta}) \end{bmatrix} & \approx \left\{\begin{array}{cc}\begin{bmatrix}  \psi(\bar{\theta})  \\ 0  \end{bmatrix}  &\bar{\theta} \le \bar{\theta}_m \\ \\
\begin{bmatrix}  0  \\ \psi(\bar{\theta})  \end{bmatrix}  &\bar{\theta} > \bar{\theta}_m
 \end{array}\right. = \begin{bmatrix}  1-f(\bar{\theta})  \\ f(\bar{\theta}) \end{bmatrix} \psi(\bar{\theta}).
\tag{S\arabic{equation}}
\label{eq:ansatzH2by2}
\end{align*}
Here $f(\bar{\theta})$ is a step function that increases from $0$ to $1$ when $\bar{\theta}$ transitions through the local maximum $\bar{\theta}_m$ defined by $E_{\circlearrowright} - E_J\cos\frac{1}{2}(\bar{\theta}_m-\theta^\circlearrowright)+ \frac{1}{2} E_L (\bar{\theta}_m-\theta^\circlearrowright)^2 = E_{\circlearrowleft}-E_J\cos\frac{1}{2}(\bar{\theta}_m-\theta^\circlearrowleft)+ \frac{1}{2} E_L (\bar{\theta}_m-\theta^\circlearrowleft)^2$.  When the RF squid qubit is threaded by a half superconducting flux quantum, the symmetric double-well potential depicted in Fig. \ref{fig:RFsquidqubit}d has a local maximum at $\bar{\theta}_m=0$.
Inserting the ansatz, we are left with
\stepcounter{equation}
\begin{align*}
-4E_C(\bar{\theta}) \frac{d^2}{d\bar{\theta}^2}  \psi(\bar{\theta}) + E_I(\bar{\theta})  \psi(\bar{\theta}) = E  \psi(\bar{\theta})
\tag{S\arabic{equation}}
\label{eq:onecomponent}
\end{align*}
with
\begin{align*}
4E_C(\bar{\theta}) =  \left\{\begin{array}{cc}  E_{\circlearrowright;\circlearrowright} &\bar{\theta} \le \bar{\theta}_m \\
 E_{\circlearrowleft;\circlearrowleft} &\bar{\theta} > \bar{\theta}_m \end{array} \right.
\end{align*}
and
\begin{align*}
&E_I(\bar{\theta})  = \\
& \left\{\begin{array}{ccc} E_{\circlearrowright} - E_J\cos\frac{1}{2}(\bar{\theta}-\bar{\theta}^\circlearrowright)+ \frac{1}{2} E_L (\bar{\theta}-\bar{\theta}^\circlearrowright)^2 && \bar{\theta} \le \bar{\theta}_m \\
\\
E_{\circlearrowleft} - E_J\cos\frac{1}{2}(\bar{\theta}-\bar{\theta}^\circlearrowright)+ \frac{1}{2} E_L (\bar{\theta}-\bar{\theta}^\circlearrowright)^2& &\bar{\theta} < \bar{\theta}_m \end{array} \right. .
\end{align*}
 In this derivation, we neglect the off-diagonal values $E_{\circlearrowright,\circlearrowleft}$ and $E_{\circlearrowleft,\circlearrowright}$, assuming that they are small compared to $E_I(\bar{\theta})$.  We also neglect terms proportional to $\frac{d f(\bar{\theta})}{d \bar{\theta}} \frac{d \psi(\bar{\theta})}{d \bar{\theta}}$ or $\frac{d^2 f(\bar{\theta})}{d \bar{\theta}^2}  \psi(\bar{\theta})$.  This is justified if $\psi(\bar{\theta})$ and $ \frac{d \psi(\bar{\theta})}{d \bar{\theta}}$ are small at the local maximum $\bar{\theta}_m$, the only point at which $\frac{d f(\bar{\theta})}{d \bar{\theta}}$ and $\frac{d^2 f(\bar{\theta})}{d \bar{\theta}^2}$ do not vanish.  For instance, this approximation seems particularly appropriate for low-energy eigenstates of a double-well potential that nearly vanish inside the potential barrier (see Fig. \ref{fig:RFsquidqubit}d).

\subsection{Number of Entangled Electrons}

Computing the number of entangled electrons in an RF squid qubit is beyond the scope of LE theory.  One might attempt an answer within LE theory (see, e.g. Ref. [S1])  by working in the basis $|n\rangle$ and regarding $n$ as the number of Cooper pairs on the capacitors shunting the Josephson junction of the qubit.  For example, one might associate the uncertainty $\Delta n$ with the number of Cooper pairs participating in the supercurrent by flowing on and off the junction capacitance.  However, this association leads to unphysical conclusions.  For the parameters of the RF squid qubit experiment \cite{Friedman2000}, for example, one computes $\Delta n \sim 50$ pairs in the entangled state.   Traveling with speed $v$ around a ring of circumference $L_z$, they should produce a current $I$ satisfying $\Delta n \sim  I L_z/ 2e v$.  Inserting $L_z \sim 500 \mu$m and $I \sim \mu$A, we find agreement only if $v$ approaches the speed of light $c$.  But physically, the maximum plausible speed is the Fermi velocity $v_F \sim 0.01 c$.

Our theory provides a microscopic many-body quantum state of the RF squid qubit, which is not provided by LE theory.  We use this quantum state in the main text to evaluate (\ref{eq:DeltaN}).  In this section, we give a short derivation of an approximate expression \cite{Korsbakken2009,*Korsbakken2010} for (\ref{eq:DeltaN}) that assumes that the many-body state is a superposition of displaced Fermi seas counterpropagating in a ring of circumference $L_z$.  A displaced Fermi sea is a sphere of momentum eigenstates that is centered at a non-zero momentum.  Thus, the basis that diagonalizes the expression ${\langle\circlearrowright\!| c^\dagger_{\mathbf Q}c_{{\mathbf Q}^\prime}  |\!\circlearrowright\rangle} - {\langle\circlearrowleft\!| c^\dagger_{\mathbf Q}c_{{\mathbf Q}^\prime}  |\!\circlearrowleft\rangle}$ appearing in  (\ref{eq:DeltaN}) is given by momentum and spin: ${\mathbf Q} = ({\mathbf q},\sigma)$.

\begin{figure}
\begin{tabular}{cc}
(a) \begin{tabular}{c} \includegraphics[width=1.5in]{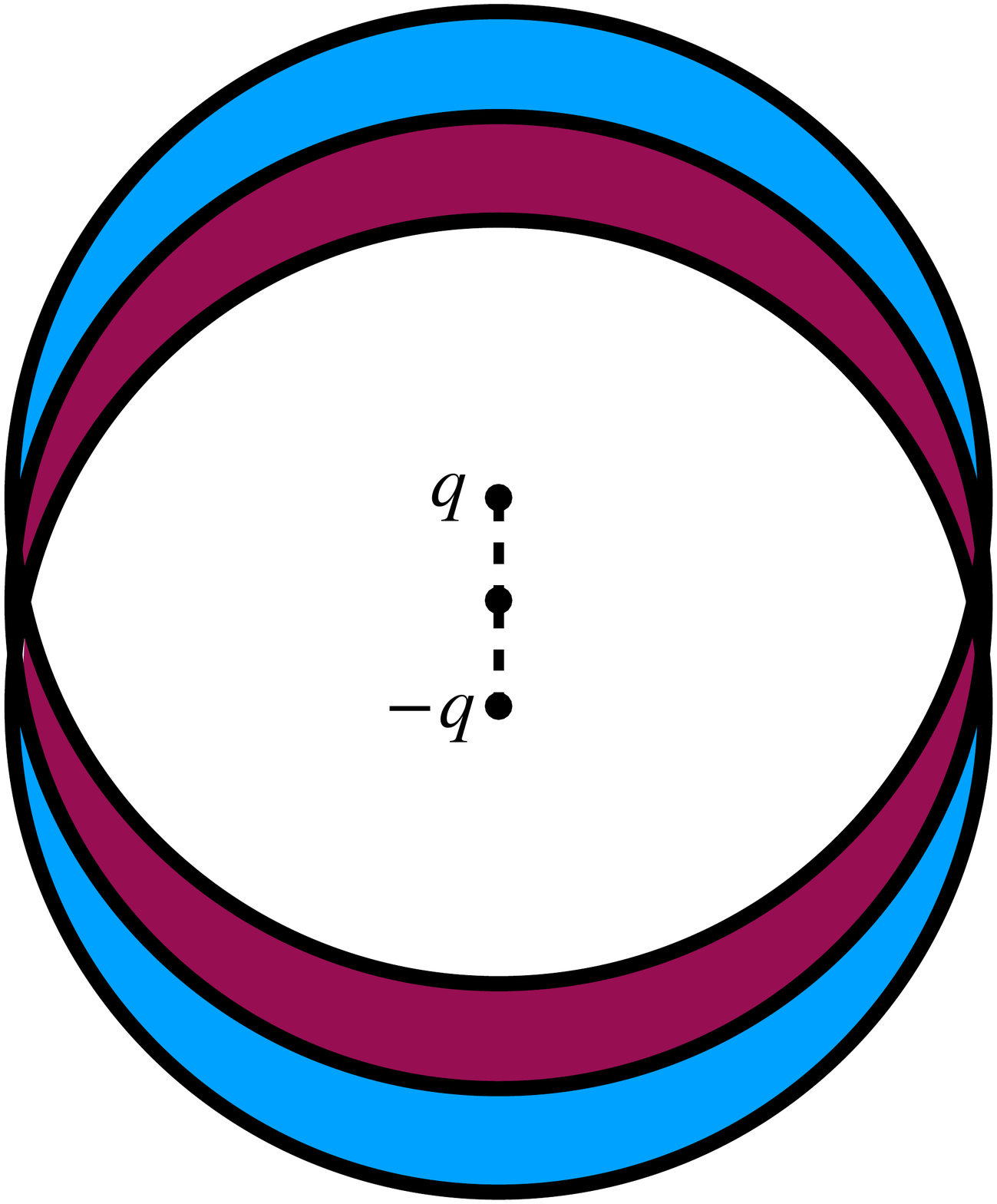} \end{tabular}&
(b) \begin{tabular}{c} \includegraphics[width=1.5in]{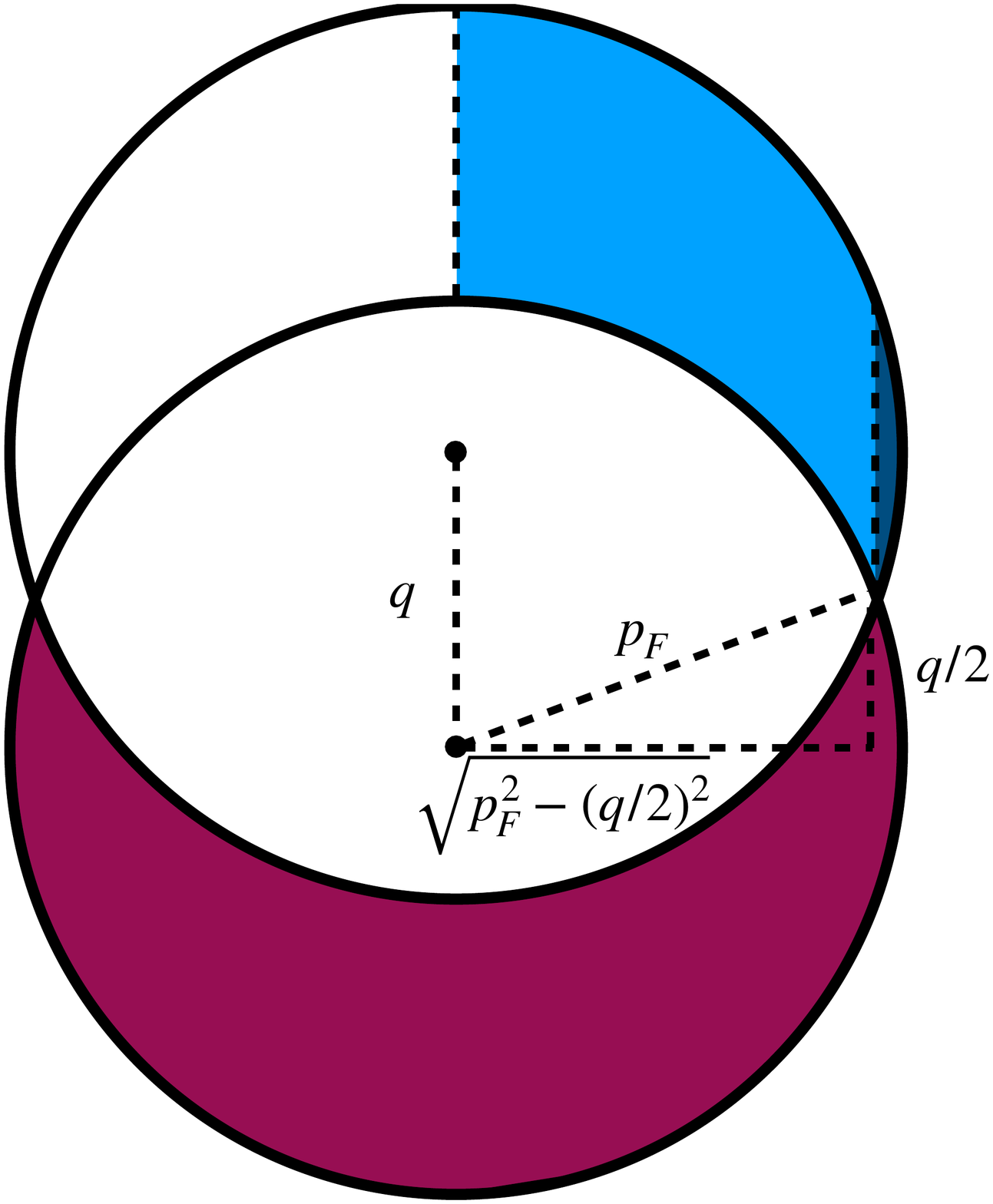} \end{tabular}
\end{tabular}
\caption{(a) Cross-section of three displaced Fermi seas, one shifted up by $q$,  one shifted down by $-q$, and one undisplaced.   If the blue slivers are exchanged and the purple approximate slivers are exchanged, the displaced Fermi seas are mapped into one other.  Thus, the number of entangled electrons is the number of electronic states occupying 1 blue sliver and 1 purple sliver.  (b) Diagram used to evaluate volume of sliver in momentum space.  Upper circle, centered at $q$, and middle circle, centered at $0$, from (a) are depicted.  Sliver is  decomposed into several regions (first term in (\ref{eq:sliverintegral}) proceeds over light blue part and second term in (\ref{eq:sliverintegral}) over the dark blue part).}
 \label{fig:sliverintegral}
\end{figure}

If the displaced Fermi seas carry current $\pm I/2$ and are centered at momentum $\pm q$, then the total number of electrons below the Fermi surface of each sea is $N \sim I L_z/2e(q/m)$, where $m$ is the electron mass.  However, not all $N$ electrons are entangled when we superpose displaced Fermi seas.  The core electronic states are occupied in both displaced Fermi seas; these core electronic states do not participate in the entanglement  \cite{Korsbakken2009,Korsbakken2010}, and the sum (\ref{eq:DeltaN}) is deliberately defined so that they do not contribute.   In Fig. \ref{fig:sliverintegral}a, which depicts the two displaced Fermi seas graphically, these core electrons occupy the white region.  Only the 4 colored slivers in Fig. \ref{fig:sliverintegral}a make contributions to (\ref{eq:DeltaN}).  And, recalling the factor of $1/2$ in (\ref{eq:DeltaN}), the quantity $\Delta N$ equals the number of electronic states in 1 blue sliver plus 1 purple sliver.  The slivers are approximately congruent, and we evaluate the volume of a blue sliver with the assistance of Fig. \ref{fig:sliverintegral}b.   If $p_F$ is the Fermi momentum, the volume of the blue sliver in momentum space is
\stepcounter{equation}
\begin{align*}
\int_{Sliver} d^3 p & = \Bigg[2 \pi  \int_0^{\sqrt{p_F^2 - (q/2)^2}} dp_r  p_r \int_{\sqrt{p_F^2 - p_r^2}}^{q+\sqrt{p_F^2 - p_r^2}} dp_z +\\
& 2 \pi \int_{\sqrt{p_F^2-(q/2)^2}}^{p_F} dp_r p_r \int_{q-\sqrt{p_F^2 - p_r^2}}^{q+\sqrt{p_F^2 - p_r^2}}  dp_z \Bigg] \\
& = \pi p_F^2 q - \frac{\pi}{24}q^3 \approx \pi p_F^2 q. 
\tag{S\arabic{equation}} \label{eq:sliverintegral}
\end{align*}
We have performed the integral using cylindrical coordinates.  In the final line, we assume $p_F \gg q$.  Intuitively, the light blue region of Fig. \ref{fig:sliverintegral}b, rotated around the z axis, approximately has the volume of cylinder of base $\pi p_F^2$ and height $q$.  The number of electronic states in 1 blue sliver plus 1 purple sliver in Fig.  \ref{fig:sliverintegral}a is $\Delta N \approx N (2 \pi p_F^2 q) /(4 \pi p_F^3 /3) = N 3 q/2 p_F$, using the fact that the Fermi sea has volume $4 \pi p_F^3 /3$ in momentum space.  Substituting in the expression for $N$ above, we find $\Delta N \sim 3 I L_z/4 e(p_F/m) =  3 I L_z/4 e v_F$.  

\noindent [S1] F. Marquardt, B. Abel, and J. von Delft, Phys. Rev. A {\bf 78}, 012109 (2008)

\end{document}